\documentclass[10pt,journal]{IEEEtran}
\usepackage{mathrsfs}
\usepackage{algorithm}
\usepackage{algpseudocode}
\usepackage{cite}
\usepackage{color}
\usepackage{psfrag}
\usepackage{subfigure}
\usepackage{amssymb}
\usepackage{epsfig}
\usepackage{pifont}
\usepackage{amsmath}
\usepackage{array}
\usepackage{pifont}
\usepackage{amsfonts}
\usepackage{fancyhdr}
\usepackage{amscd}
\usepackage{bm}
\usepackage{amssymb}
\usepackage{graphics,eepic,epic}
\usepackage{latexsym}
\usepackage{euscript}
\usepackage{graphics,eepic,epic,psfrag}
\newtheorem{theorem}{Theorem}
\newtheorem{proposition}{Proposition}

\newtheorem{lemma}{Lemma}

\newcommand{\eda}[1]{{\color{black}#1}}
\newcommand{\edb}[1]{{\color{black}#1}}
\begin{document}
\title{Phase Transition Analysis for Covariance Based Massive Random Access with Massive MIMO}
\author{Zhilin~Chen,~\IEEEmembership{Member,~IEEE,}
        Foad~Sohrabi,~\IEEEmembership{Member,~IEEE,}
        Ya-Feng~Liu,~\IEEEmembership{Senior Member,~IEEE,}
        and~Wei~Yu,~\IEEEmembership{Fellow,~IEEE}
\thanks{
Manuscript submitted to
\emph{IEEE Transactions on Information Theory}
on March 6, 2020, revised on July 2, 2021, and accepted on November 8, 2021.
The work of Zhilin Chen, Foad Sohrabi, and Wei Yu was supported by the Natural Sciences and Engineering Research Council (NSERC) of Canada.
The work of Ya-Feng Liu was supported by the National Natural Science Foundation of China (NSFC) under Grant 12022116 and Grant 12021001. \par
The materials in this paper have been presented in part at the IEEE International Conference on Communications (ICC), Shanghai, China, May 2019 \cite{Chen2019a}, and at the Asilomar Conference on Signals, Systems, and Computers, Pacific Grove, CA, USA, November 2019 \cite{Chen2019b}. \par
Zhilin Chen, Foad Sohrabi, and Wei Yu are with The Edward S. Rogers Sr. Department of Electrical and Computer Engineering, University of Toronto, Toronto, ON M5S 3G4, Canada (e-mails:\{zchen, fsohrabi, weiyu\}@comm.utoronto.ca). \par
Ya-Feng Liu is with the State Key Laboratory of Scientific and Engineering Computing, Institute of Computational Mathematics and Scientific/Engineering Computing, Academy of Mathematics and Systems Science, Chinese Academy of Sciences, Beijing, 100190, China (e-mail: yafliu@lsec.cc.ac.cn).}}
\maketitle

\begin{abstract}
This paper considers a massive random access problem in which a large number
of sporadically active devices wish to communicate with a base station (BS)
equipped with massive multiple-input multiple-output (MIMO) antennas. Each
device is preassigned a unique signature sequence, and the BS identifies the
active devices by detecting which sequences are transmitted. This device
activity detection problem can be formulated as a maximum likelihood estimation
(MLE) problem for which the sample covariance matrix of the received signal
is a sufficient statistic. The goal of this paper is to characterize the
feasible set of problem parameters under which this covariance based approach is
able to successfully recover the device activities in the massive MIMO regime.
Through an analysis of the asymptotic behaviors of MLE via its associated
Fisher information matrix, this paper derives a necessary and sufficient
condition on the Fisher information matrix to ensure a vanishing
probability of detection error as the number of antennas goes to infinity, based on
which a numerical phase transition analysis is obtained. This condition is
also examined from a perspective of covariance matching, which relates the
phase transition analysis to a recently derived scaling law. Further,
we provide a characterization of the distribution of the estimation error in
MLE, based on which the error probabilities in device activity detection can
be accurately predicted.  Finally, this paper studies a random access scheme
with joint device activity and data detection and analyzes its performance
in a similar way.
\end{abstract}
\IEEEpeerreviewmaketitle

\section{Introduction}

Uncoordinated random access is a challenging task for massive machine-type
communications (mMTC), in which a large number of sporadically active devices
attempt to communicate with the base station (BS) in the uplink
\cite{Hasan2013, Bockelmann2016, Liu2018b}. Conventional cellular systems
provide random access for human-type communications by employing a set of
orthogonal sequences, from which every active device randomly and independently
selects one sequence to transmit as a pilot for requesting access
\cite{Dahlman2013}. When the number of active devices is comparable to the
number of available orthogonal sequences, this uncoordinated random access
approach inevitably leads to collisions. A subsequent collision-resolution
mechanism is then needed, which introduces delay because of the required
multiple rounds of signaling. Such a scheme may not be suitable for mMTC due to
the fact that the delay caused by contention resolution can be severe
\cite{Dawy2017}.

The issue of collision in random access for mMTC can be avoided by using
non-orthogonal sequences \cite{Chen2018}. The basic idea is to use a large set
of non-orthogonal sequences and to preassign a unique pilot sequence to each
device, then let all the active devices transmit their pilots simultaneously as
identifiers. The BS can take advantage of the sporadic nature of the device
activity pattern and use a sparse recovery (i.e., compressed sensing) algorithm
to detect which sequences are transmitted, thereby identifying
the active devices.

The ability to perform sparse recovery can be greatly enhanced if the BS is
equipped with a large number of antennas. This is because the non-orthogonality
of the pilot sequences leads to significant interference between the pilots, and
a massive multiple-input multiple-output (MIMO) system is ideally suited for exploiting
the spatial dimensions for interference mitigation \cite{Liu2018}. The goal of
this paper is to understand the fundamental limit of sparse recovery for mMTC.
Specifically, we ask the following question. Given a pilot sequence length $L$
and assuming a fixed set of non-orthogonal pilot sequences, how many
simultaneously active users (i.e., $K$) can be identified out of a large number of $N$
potential users, when the number of antennas $M$ at the BS is large.

The answer to the above question depends on the way the problem is formulated.
One possible formulation is the following. Because the wireless channels are
time-varying and are not known precisely either at the transmitters or at the
receiver, one can formulate the problem as a joint device activity detection
and channel estimation problem. This approach is taken in \cite{Chen2018,
Liu2018}, where an approximate message-passing (AMP) algorithm is used for
sparse recovery. For the case where the BS has a single antenna, we generally
need $K<L$ for successful recovery. But interestingly, as pointed out in
\cite{Liu2018}, as the number of BS antennas $M$ goes to infinity, successful
sparse recovery may be possible even for $K \geq L$, although AMP would become
increasingly more difficult to converge at large $M$ \cite{Fengler2019a}.

The above AMP approach falls under the Bayesian framework, as it assumes the
knowledge of channel statistics and aims to estimate the instantaneous channel
state information (CSI). An alternative formulation is to forgo the estimation
of instantaneous CSI altogether, instead focusing on estimating the statistical
channel information (in particular, the large-scale fading), and to use the
estimated statistical information to determine whether a device is active or
not. This non-Bayesian approach is pioneered in \cite{Fengler2019a}
and is termed the \emph{covariance approach}, because a certain sample covariance
matrix of the received sequence is a sufficient statistic for this
estimation task.
This covariance approach is ideally suited for large $M$, because
the covariance can be accurately estimated using a large number of observation samples.
When $M$ is large, this covariance based approach
has the key advantage that it is capable of detecting a much
larger number of active devices, as observed in
\cite{Fengler2019a}.
In fact, accurate activity
detection is possible in the regime where $K=O(L^2/\log^2(N/L^2))$ for sufficiently large $M$.

The above scaling law is discovered in \cite{Fengler2019a}, which states a
relationship among parameters $K$, $L$, and $N$ and a condition on how
large $M$ has to be so that the probability of error would go down to zero
exponentially as a function of the pilot length $L$. It is proved
in \cite{Fengler2019a} that the scaling law holds rigorously for the
nonnegative least squares (NNLS) formulation of the problem, where the device activities along
with the large-scale fading coefficients are the unknown variables, and also a so-called ``restricted'' version of the maximum likelihood
estimation (MLE), where the large-scale fading \edb{coefficients are} assumed to be known and the device activity detection problem is formulated as a combinatorial problem over $\{0,1\}^N$. It is conjectured in \cite{Fengler2019a} that the same
scaling law also holds for the unrestricted MLE, where the device
activities and the large-scale fading coefficients are both
unknown variables as in the NNLS formulation. The scaling law in \cite{Fengler2019a} is
derived based on randomly generated pilot sequences so that the restricted
isometry property of certain measurement matrix holds in the compressed sensing
context.

In this paper, we revisit the issue of \edb{the} scaling law by studying the (unrestricted) MLE directly from an estimation theoretical perspective. Specifically, through an asymptotic performance analysis of the MLE in the regime where $M$ tends to infinity, this paper characterizes the values of $K$, $L$, and $N$ for arbitrary pilot sequences under which reliable
activity detection can be ensured. Note that unlike the scaling law in \cite{Fengler2019a} that relates parameters $K$, $L$, $N$, and $M$ which are all finite, our analysis considers finite $K$, $L$, and $N$ only, while letting $M$ tend to infinity. The asymptotic analysis in this paper leads to a numerical method for characterizing the phase transition as $M$ goes to infinity and an accurate prediction of the probability of detection error in
the regime where $M$ is large but finite. Moreover, via a re-examination of our analysis from the covariance matching point of view, a connection between our analysis and the above scaling law is established and the conjecture in \cite{Fengler2019a} is addressed in the asymptotic regime of $M$ going to infinity.

It is pertinent to note that, unlike the AMP approach, to use the covariance approach for efficient data transmission, a subsequent channel estimation stage would normally be needed, because the covariance approach does not reveal instantaneous CSI.
However, if each device only has a small amount of data to
transmit, it is possible to conceive a random access scheme in which each
device is preassigned multiple distinct sequences, and
the data bits are embedded in the choice of which sequence to transmit at each
device, so that the BS can perform joint device activity and data detection
\cite{Senel2018}.
The covariance approach is well suited for such a scenario, because of its
$O(L^2)$ scaling that allows many more sequences to be detected.
Our phase transition analysis of the covariance approach naturally carries
over to this case.

\subsection{Related Work}

The classical random access strategy originated from the ALOHA system
\cite{Abramson1970}, which further evolved into a variety of enhanced ALOHA
schemes \cite{Casini2007,Liva2011,Narayanan2012,Paolini2015a} some of which
employ iterative interference cancellation to resolve collision. The classical
ALOHA can be thought of as a strategy that uses orthogonal sequences for device
identification followed by collision resolution and retransmission.

Recently a number of non-orthogonal sequence based random access schemes for
mMTC have been proposed, e.g., the two-phase grant-free random access
\cite{Liu2018b}, the grant-free random access with data embedding \cite{Senel2018} or data spreading \cite{Han2019}.
The non-orthogonal sequences can be used as signatures for active device detection, e.g., 
\cite{Chen2018,Wang2019},
as codewords for data transmission, e.g., \cite{Polyanskiy2017,Amalladinne2018,Fengler2019}, or as both, e.g., \cite{Senel2018,Chen2019a}. By detecting which sequences are transmitted, the BS acquires the identification of the active devices and/or the data bits.

The sequence detection problem in random access is closely related to the compressed sensing problem due to the sporadic nature of the device activity, for which various sparse recovery techniques have been explored, e.g., orthogonal matching pursuit \cite{Dekorsy2013,Boljanovic2017}, basis pursuit denoising \cite{Wunder2014}, Bayesian sparse recovery \cite{Lau2015, Ahn2019}, and dimension reduction based optimization \cite{Shao2020}.
Specifically, the computationally efficient AMP algorithm is used for the device
activity detection problem in \cite{Goertz2015,Chen2017,Sun2019,Jiang2020}
for single-antenna systems, in \cite{Chen2018,Liu2018,Ke2019} for multi-antenna
systems, and in \cite{Simeone2016,Chen2019c} for multi-cell or cloud radio
access networks. An important feature of AMP is that the performance can be
analyzed via an analytical framework of state evolution \cite{Donoho2009},
based on which the detection error can be accurately predicted.

\eda{
As mentioned earlier, when the BS is equipped with a large number of antennas,
it is possible to detect the device activities by estimating the channel statistics
based on certain sample covariance of the received signal.
This covariance approach is proposed in
\cite{Fengler2019a}
for massive MIMO systems, where the sequence detection problem is formulated as either an MLE problem or an NNLS problem. 
As compared to the compressed sensing approach that aims to recover the instantaneous channel vectors, the covariance based method aims to estimate the large-scale fading coefficients \edb{of the channels} by averaging the received signal over multiple antennas, thus the channel hardening effect in the massive MIMO systems can be exploited.  
It is shown in \cite{Haghighatshoar2018,Fengler2019} that when the number of BS antennas is large, the covariance based method with the MLE formulation can outperform the AMP approach.
}

As already mentioned, the performance of the covariance based approach under the NNLS formulation has been analyzed in \cite{Fengler2019a}, where an error bound and an analytic scaling law on $K$, $L$, $N$, and $M$ are derived assuming a specific class of random pilot sequences.
In contrast to \cite{Fengler2019a}, this paper considers the generic MLE formulation with arbitrary pilot sequences, and derives a necessary and sufficient condition for reliable activity detection for $K$, $L$, and $N$ in the asymptotic regime where $M$ tends to infinity. Our main result is a numerical characterization of the phase transition.

Most of the above works, which use non-orthogonal sequences as pilots, take a
\emph{sourced} approach to massive connectivity. In contrast, an
\emph{unsourced} random access approach has been proposed in
\cite{Polyanskiy2017} and further developed in
\cite{Amalladinne2018,Fengler2019}, where the detection of device activities
amounts to determining a list of messages from the active devices without
identifying which message belongs to which device.  The device identification
information is embedded in the data payload. The detection problem for
this scenario is different from the one considered here.

\subsection{Main Contributions}

This paper studies the covariance based approach for device activity detection
with non-orthogonal sequences in a massive MIMO system. We adopt the MLE
formulation and characterize the conditions for successful detection when the
number of antennas at the BS tends to infinity. The main contributions are as
follows:

\begin{itemize}
\item

We study the performance of the device activity detection by analyzing
the asymptotic behaviors of MLE via its associated Fisher information matrix.
Given a device activity detection problem with finite $N$, $K$, and $L$, we derive a necessary and sufficient condition on the
Fisher information matrix under which a vanishing probability of detection error can be
ensured as $M$ tends to infinity. This condition involves solving a linear
programming (LP) problem, based on which a numerical phase transition analysis
can be obtained. As compared to the analytic scaling law in
\cite{Fengler2019a}, which is derived based on the NNLS \edb{formulation and the} restricted MLE formulation, our phase
transition analysis is numerical and is based on the unrestricted MLE formulation. Moreover, the scaling law in \cite{Fengler2019a} assumes
a specific class of signature sequences that are uniformly drawn from a sphere,
whereas our phase transition analysis applies to any arbitrary signature
sequences.
\item We provide an equivalent necessary and sufficient condition
from the perspective of covariance matching to allow a characterization of the phase
transition in $N$, $K$, and $L$, with $M$ tending to infinity. The new
condition reveals the connection between the phase transition analysis in this
paper and the analytic scaling law in \cite{Fengler2019a}, and addresses the conjecture in \cite{Fengler2019a} on the unrestricted MLE in the asymptotic regime of $M$.

\item 
We provide a way to accurately predict the error probabilities for
device activity detection under finite $M$.
This is accomplished by characterizing the distribution of the estimation error
of MLE. 
We show that the distribution of detection error can be obtained by
solving a quadratic programming (QP) problem involving the Fisher information matrix.
\item Finally, we
study the joint device activity and data detection for a random access
scheme where each device is associated with multiple distinct sequences to
convey a few data bits. We show that this joint device activity and data
detection problem can be formulated in a similar way, and the performance can
be analyzed accordingly.
\end{itemize}

\subsection{Paper Organization and Notation}
The reminder of this paper is organized as follows. Section~\ref{sec.system} introduces the system model. Section~\ref{sec.detection} studies the device activity detection problem. Section~\ref{sec.analysis} analyzes the asymptotic performance and presents a phase transition analysis. Section~\ref{sec.matching} examines the phase transition analysis from the covariance matching perspective.
Section~\ref{sec.joint} studies the joint device activity and data detection problem. Simulation results are provided in Section~\ref{sec.simu}. Conclusions are drawn in Section~\ref{sec.conclusion}.

Throughout this paper, lower-case, boldface lower-case, and boldface upper-case letters denote scalars, vectors, and matrices, respectively. Calligraphy letters denote sets. Superscripts $(\cdot)^{H}$, $(\cdot)^{T}$, $(\cdot)^{*}$, $(\cdot)^{-1}$, and $(\cdot)^{\dagger}$ denote conjugate transpose, transpose, conjugate, inverse, and Moore-Penrose inverse, respectively. Further, $\mathbf{I}$ denotes identity matrix with appropriate dimensions, $\mathbb{E}[\cdot]$ denotes expectation, $\operatorname{Var}[\cdot]$ denotes variance,
$\operatorname{Re}(\cdot)$ denotes real part, $\operatorname{Im}(\cdot)$ denotes imaginary part, $\mathrm{tr}(\mathbf{X})$ denotes the trace of $\mathbf{X}$, $\mathrm{diag}(x_1,\ldots,x_n)$ (or $\mathrm{diag}(\mathbf{X}_1,\ldots,\mathbf{X}_n)$) denotes a (block) diagonal matrix formed by $x_1,\ldots,x_n$ (or $\mathbf{X}_1,\ldots,\mathbf{X}_n$), $\triangleq$ denotes definition, $|\cdot|$ denotes the determinant of a matrix, $\|\mathbf{X}\|_F$ denotes the Frobenius norm of $\mathbf{X}$, $\|\mathbf{x}\|_2$ denotes the $\ell_2$ norm of $\mathbf{x}$, $\|\mathbf{x}\|_1$ denotes the $\ell_1$ norm of $\mathbf{x}$, $\|\mathbf{x}\|_0$ denotes the number of nonzero entries in $\mathbf{x}$, $\odot$ denotes element-wise product, and $\otimes$ denotes Kronecker product. Finally, $\mathcal{N}(\boldsymbol\mu, \mathbf{\Sigma})$ (or $\mathcal{CN}(\boldsymbol\mu, \mathbf{\Sigma})$) denotes a (complex) Gaussian distribution with mean $\boldsymbol\mu$ and covariance $\mathbf{\Sigma}$. {\color{black}Table \ref{tab.notation} summarizes the notations used in this paper.}
\begin{table}[h]
\centering
\caption{Summary of Notations}
        {\color{black}
        \begin{tabular}{|c|p{0.76\columnwidth}|}
        \hline
        Notation  & \hspace{70pt} Description \\ \hline
        $N$, $K$ & Total number of devices, number of active devices \\
        $L$ & Signature sequence length\\
        $M$      & Number of antennas at the BS\\
        $a_n$ & Activity indicator of device $n$\\
        $a_n^q$ & Sequence selection indicator for sequence $q$ of device $n$\\
        $g_n$& Channel large-scale fading coefficient of device $n$\\
        $\mathbf{y}_m$ & Received signal at the $m$-th antenna\\
        $\mathbf{s}_n$ & Signature sequence of device $n$\\
        $\mathbf{S}$ & Signature sequence matrix formed as $[\mathbf{s}_1,\ldots, \mathbf{s}_N]$\\
        $\widehat{\mathbf{S}}$ & Column-wise Kronecker product of $\mathbf{S}^*$ and $\mathbf{S}$\\
        $\widetilde{\mathbf{S}}$ & Signature sequence matrix in data embedding scheme\\
        $\mathbf{h}_n$ & Rayleigh fading component of device $n$\\
        $\mathbf{H}$ & Channel matrix, i.e., $[\mathbf{h}_1,\ldots, \mathbf{h}_N]^T$\\
        $\widetilde{\mathbf{H}}$ & Channel matrix in data embedding scheme\\
        $\gamma_n$ & Indicator of activity and large-scale fading of device $n$\\
        $\hat{\gamma}_n^{(M)}$, $\gamma_n^0$ & Maximum likelihood estimate of $\gamma_n$, true value of $\gamma_n$\\
        $\boldsymbol\gamma$ & Vector of large-scale fading coeffiecints  $[\gamma_1,\ldots,\gamma_N]^T$ \\
        $\hat{\boldsymbol\gamma}^{(M)}$, $\boldsymbol\gamma^0$ & Maximum likelihood estimate of $\boldsymbol\gamma$, true value of $\boldsymbol\gamma$\\
        $\widetilde{\boldsymbol\gamma}$ & Indicators of sequence selection and large-scale fading\\
        $\boldsymbol\Gamma$, $\boldsymbol\Gamma^0$ & Diagonal matrix formed by $\gamma_1,\ldots,\gamma_N$, true value of $\boldsymbol\Gamma$\\
        $\mathbf{J}(\gamma)$ & Fisher information matrix of $\gamma$\\
        $\boldsymbol\Sigma$ & Covariance matrix of the received signal at the BS\\
        $\hat{\boldsymbol\Sigma},\boldsymbol\Sigma^0$  & Sample covariance matrix, true value of $\boldsymbol\Sigma$\\
        $\widetilde{\boldsymbol\Sigma}$ & Covariance matrix in data embedding scheme\\
        $\mathcal{I}$& Set of the indices of inactive devices\\
        $\mathcal{I}^c$ & Complement of $\mathcal{I}$ with respect to $\{1,\ldots,N\}$\\
        $\mathcal{N}$ & Null space of $\mathbf{J}(\boldsymbol\gamma^0)$ in $\mathbb{R}^N$\\
        $\widetilde{\mathcal{N}}$ & Null space of $\widehat{\mathbf{S}}$ in $\mathbb{R}^N$\\
        $\mathcal{C}$ & Cone in $\mathbb{R}^N$ with entries indexed by $\mathcal{I}$ being nonnegative\\
        $b$ & Number of bits of the embedded data per device\\
        $Q$ & Number of sequences per device in data embedding
        \\\hline
        \end{tabular}\vspace{0.5em}
        }
\label{tab.notation}
\end{table}

\section{System Model}
\label{sec.system}

Consider an uplink single-cell massive random access scenario with $N$
single-antenna devices communicating with a BS equipped with $M$ antennas.
We primarily focus on the massive MIMO regime where $M$ is large. A block fading
channel model is assumed, i.e., the channel coefficients remain constant for a
coherence interval. We assume that the user traffic is sporadic, i.e., only  $K\ll N$
devices are active during each coherence interval. For the purpose of
device identification, each device $n$ is preassigned a unique signature
sequence $\mathbf{s}_n=[s_{1n},\ldots, s_{Ln}]^T\in \mathbb{C}^{L}$,
where $L$ is the sequence length which is assumed to be shorter than the
length of the coherence interval. In the pilot phase, we assume that
all the active devices transmit their signature sequences synchronously
at the same time. The objective is for the BS to detect which subset of
devices are active based on the received signal.

Let $a_{n}\in\{1,0\}$ denote the activity of device $n$ in a given coherence
interval, i.e., $a_n=1$ if the device is active and $a_n=0$ otherwise.
We model the channel vector between the BS and device $n$ as a random vector
$g_n\mathbf{h}_{n}$,
where $\mathbf{h}_{n}\in \mathbb{C}^{M}$ is the Rayleigh fading component that
has the distribution $\mathcal{CN}(\mathbf{0},\mathbf{I})$, and $g_{n}$ is the
large-scale fading component due to path-loss and shadowing. 
The received signal $\mathbf{Y}\in \mathbb{C}^{L\times M}$ at the BS in the pilot
phase can be expressed as
\begin{align}\label{eq.sys}
\mathbf{Y}&=\sum_{n=1}^{N}a_{n}\mathbf{s}_{n}g_n\mathbf{h}_{n}^T +\mathbf{W}\nonumber\\
&=
\begin{bmatrix}
\mathbf{s}_{1}&\ldots & \mathbf{s}_{N}
\end{bmatrix}
\begin{bmatrix}
a_{1}g_1 &  & \\
 &\ddots  & \\
 &  &a_{N}g_N
\end{bmatrix}
\begin{bmatrix}
\mathbf{h}_{1}^T\\
\vdots\\
\mathbf{h}_{N}^T
\end{bmatrix}
+\mathbf{W}\nonumber\\
&\triangleq\mathbf{S}\boldsymbol{\Gamma}^{\frac{1}{2}}\mathbf{H}+\mathbf{W},
\end{align}
where $\mathbf{S}\triangleq[\mathbf{s}_{1},\ldots,\mathbf{s}_{N}]\in\mathbb{C}^{L\times N}$ is the signature sequence matrix, $\boldsymbol{\Gamma}\triangleq\operatorname{diag}(\gamma_1,\ldots,\gamma_N)\in\mathbb{R}^{N \times N}$ with $\gamma_n=(a_ng_n)^2$ is a diagonal matrix indicating both the device activity $a_n$ and the large-scale fading component $g_n$, $\mathbf{H}\triangleq[\mathbf{h}_{1},\ldots,\mathbf{h}_{N}]^T\in\mathbb{C}^{N \times M}$ is the channel matrix, and
$\mathbf{W}\in \mathbb{C}^{L\times M}$ is the effective independent and identically distributed (i.i.d.) Gaussian noise with variance $\sigma_{w}^2$ normalized by the device transmit power for simplicity. We use $\boldsymbol{\gamma}\triangleq[\gamma_1,\ldots,\gamma_N]^T\in\mathbb{R}^{N}$ to denote the diagonal entries of $\boldsymbol{\Gamma}$.

The signature sequence matrix $\mathbf{S}$ is assumed to be known at the BS.
We identify the device activity pattern based on $\mathbf{Y}$
by exploiting the sparsity in $(a_1,\ldots,a_N)$.
One way of formulating this detection problem is to estimate the instantaneous
CSI $a_n g_n \mathbf{h}_n$ for all devices, as represented by the row sparse matrix
$\mathbf{X}\triangleq\boldsymbol{\Gamma}^{\frac{1}{2}}
\mathbf{H}\in \mathbb{C}^{N\times M}$.
The active devices are simply devices with nonzero effective instantaneous channels.
This is a compressed sensing problem of recovering
nonzero rows of the matrix $\mathbf{X}$ from the received signal
$\mathbf{Y}=\mathbf{S}\mathbf{X}+\mathbf{W}$. If we assume prior knowledge or
prior statistics of $g_n$, this problem can be solved under a Bayesian framework
using, e.g., the AMP algorithm \cite{Chen2018,Liu2018}.

An alternative formulation is to regard $\mathbf{h}_n$ as random, and to
detect the device activities by estimating only the $a_n g_n$ term for all devices.
The active devices are simply those whose effective large-scale fading
coefficients are nonzero. This is akin to estimating the activity indicator $a_n$
from the parameters of the channel statistics as represented by $\boldsymbol{\gamma}$.
In such a formulation, $\boldsymbol{\gamma}$ can be treated as a set of deterministic but unknown parameters, and $\mathbf{Y}$ is modeled as an observation that follows the
conditional distribution $p(\mathbf{Y}|\boldsymbol{\gamma})$ based on
the statistics of $\mathbf{h}_n$ and $\mathbf{W}$.  This method is called
\emph{the covariance approach} \cite{Fengler2019a},
because of the essential role played by the sample covariance of $\mathbf{Y}$
in the estimation process, as shown in the next section.

The key difference between the two approaches is that the estimation of
$\boldsymbol{\gamma}$ involves a much smaller number of unknown parameters
than the estimation of $\mathbf{X}$, so it is more efficient to detect
the device activities based on $\boldsymbol{\gamma}$. On the other hand, the estimation of the channel
statistics requires a large number of samples, so the covariance approach is
most effective in the massive MIMO regime, where the large number of antennas
provide many observation samples of the large-scale fading \edb{coefficients}. When the number of
BS antennas is small, the AMP-based approach may be preferable. This paper
focuses attention to the massive MIMO regime. The aim is to provide a tractable
performance analysis for the covariance based approach.

\section{Covariance Based Device Activity Detection}
\label{sec.detection}

\subsection{Problem Formulation}

Following the approach suggested in \cite{Fengler2019a}, we use MLE to estimate
$\boldsymbol{\gamma}$ from $\mathbf{Y}$, thereafter obtain the device activity
indicator $a_n$ from $\boldsymbol{\gamma}$. To compute the likelihood, we first observe from
\eqref{eq.sys} that given $\boldsymbol{\gamma}$, the columns of $\mathbf{Y}$,
denoted by $\mathbf{y}_m\in\mathbb{C}^{L}, 1\leq m\leq M $, are independent due
to the i.i.d.\ channel coefficients over the different antennas. Each column
follows a complex Gaussian distribution as
\begin{align}\label{eq.gauss}
\mathbf{y}_m \sim \mathcal{CN}\left(\mathbf{0},\boldsymbol\Sigma\right),
\end{align}
where $\boldsymbol\Sigma$ is the covariance matrix that can be computed as
\begin{align}\label{eq.gauss_cov}
\boldsymbol\Sigma&=\mathbb{E}\left[\mathbf{y}_m\mathbf{y}_m^H\right]\nonumber\\
&=\mathbf{S}\boldsymbol{\Gamma}\mathbf{S}^H+\sigma_w^2\mathbf{I}\nonumber\\
&=\sum_{n=1}^N \gamma_n\mathbf{s}_n\mathbf{s}_n^H + \sigma_w^2\mathbf{I}.
\end{align}
Due to the independence of the columns of $\mathbf{Y}$, the likelihood of $\mathbf{Y}$ is
\begin{align}\label{eq.likelihood}
p(\mathbf{Y}|\boldsymbol{\gamma})&=\prod_{m=1}^M\frac{1}{|\pi\boldsymbol\Sigma|}\exp{\left(-\mathbf{y}_m^H\boldsymbol\Sigma^{-1}\mathbf{y}_m\right)}\nonumber\\
&=\frac{1}{|\pi\boldsymbol\Sigma|^M}\exp{\left(-\operatorname{tr}\left(\boldsymbol\Sigma^{-1}\mathbf{Y}\mathbf{Y}^H\right)\right)}.
\end{align}
The maximization of $\log p(\mathbf{Y}|\boldsymbol{\gamma})$ can be cast as the minimization of $-\frac{1}{M}\log p(\mathbf{Y}|\boldsymbol{\gamma})$ formulated as
\begin{subequations}\label{eq.prob1}
\begin{alignat}{2}\label{eq.prob1.1}
&\underset{\boldsymbol\gamma}{\operatorname{minimize}}    &\quad& \log\left|\boldsymbol\Sigma\right|+ \operatorname{tr}\left(\boldsymbol\Sigma^{-1}\hat{\boldsymbol{\Sigma}}\right)\\
&\operatorname{subject\,to} &      &\, \boldsymbol{\gamma} \geq 0,
\end{alignat}
\end{subequations}
where
\begin{equation}
	\hat{\boldsymbol{\Sigma}}\triangleq \frac{1}{M}\mathbf{Y}\mathbf{Y}^H = \frac{1}{M}\sum_{m=1}^M \mathbf{y}_m\mathbf{y}_m^H
	\label{eq.sigma_hat}
\end{equation}
is the sample covariance matrix of the received signal averaged over different antennas, and $\boldsymbol{\gamma} \geq 0$ is due to the fact that $\gamma_n=(a_ng_n)^2\geq 0$. 

We observe from \eqref{eq.prob1} that the MLE problem depends on $\mathbf{Y}$
through the sample covariance matrix $\hat{\boldsymbol{\Sigma}}$. For this
reason, the approach based on solving the formulation in \eqref{eq.prob1} is
termed the covariance based approach in this paper.  As $M$ increases,
$\hat{\boldsymbol{\Sigma}}$ tends to the true covariance matrix of $\mathbf{Y}$,
but the size of the optimization problem does not change. Thus, the complexity
of solving \eqref{eq.prob1} does not scale with $M$. This is a desirable property
especially for massive MIMO systems. 

It is worth mentioning that the use of maximum likelihood for parameter estimation with multivariate Gaussian observations has appeared in various contexts. For example, a similar optimization problem is formulated in \cite{Ottersten1993} for the direction of arrival estimation.
Other related examples include sparse approximation \cite{Wipf2007}. 

\subsection{Algorithms}

The optimization problem \eqref{eq.prob1} is not convex due to the
fact that $\log|\boldsymbol\Sigma|$ is concave whereas
$\operatorname{tr}(\boldsymbol\Sigma^{-1}\hat{\boldsymbol{\Sigma}})$ is convex.
However, various algorithms have been shown to have excellent performance in
practice for solving \eqref{eq.prob1}. For example,
\cite{Wipf2007} proposes a multiple sparse Bayesian learning (M-SBL) algorithm
based on expectation maximization that estimates $\boldsymbol\gamma$
iteratively. \eda{Moreover, \cite{Fengler2019a}} suggests a coordinate
descent algorithm that randomly updates each coordinate of the estimate of
$\boldsymbol{\gamma}$ iteratively until convergence. Although the problem is
non-convex, there is evidence that M-SBL or coordinate descent may be able to
achieve global optimality if $\boldsymbol{\Gamma}^{\frac{1}{2}}\mathbf{H}$ or
$\mathbf{S}$ satisfies certain conditions; see \cite{Wipf2007} and
\cite{Fengler2019a}.

For numerical experiments, this paper adopts
the coordinate descent method from \cite{Fengler2019a}. Let {$\hat{\boldsymbol{\gamma}}^{\text{($M$)}}$} be the estimate of $\boldsymbol{\gamma}$ by the coordinate descent method {after convergence}, where the superscript $M$ indicates the number of the antennas. Once {$\hat{\boldsymbol{\gamma}}^{\text{($M$)}}$} is obtained, we employ the element-wise thresholding to determine $a_n$ from {$\hat{\gamma}_n^{\text{($M$)}}$,} the $n$-th entry of {$\hat{\boldsymbol{\gamma}}^{\text{($M$)}}$}, using a threshold $l_{th}$, i.e., $a_{n}=1$ if {$\hat{\gamma}_n^{\text{($M$)}}\geq l_{th}$} and $a_{n}=0$ otherwise.
The probabilities of missed detection and false alarm can be traded off by setting different values for $l_{th}$. A description of the coordinate descent algorithm is given in Algorithm~\ref{alg.cd}.

The complexity of the coordinate descent \edb{algorithm} is dominated by the matrix-vector multiplications in steps $5$--$7$, whose complexity is ${O}(L^2)$. As a result, the overall complexity is ${O}(TNL^2)$, where $T$ is the number of iterations.
As the complexity of the algorithm is linear in $N$ and quadratic in $L$, it is
suitable for scenarios with large $N$ and small $L$, which is often the case for low-latency mMTC.

\begin{algorithm}[t]
\caption{Coordinate descent to estimate $\boldsymbol{\gamma}$ }
\label{alg.cd}
\begin{algorithmic}[1]
\State{Initialize ${\hat{\boldsymbol{\gamma}}^{\text{($M$)}}}=\mathbf{0}$ and $ \hat{\boldsymbol{\Sigma}}=\sigma_w^2\mathbf{I}$.}
\For{$i = 1, 2,\ldots, T$}
\State{Randomly select a permutation $i_1,i_2,\ldots,i_{N}$ of the coordinate indices $\{1,2,\ldots,N\}$} of {$\hat{\boldsymbol{\gamma}}^{\text{($M$)}}$.}
\For{$n = 1 ~\mathrm{to}~ N$}
\State{$\delta = \max\Big\{
\frac{\mathbf{s}_{i_n}^H\hat{\boldsymbol{\Sigma}}^{-1}\frac{\mathbf{Y}\mathbf{Y}^H}{M}\hat{\boldsymbol{\Sigma}}^{-1}\mathbf{s}_{i_n}-\mathbf{s}_{i_n}^H\hat{\boldsymbol{\Sigma}}^{-1}\mathbf{s}_{i_n}}
{(\mathbf{s}_{i_n}^H\hat{\boldsymbol{\Sigma}}^{-1}\mathbf{s}_{i_n})^2},{-\hat{\gamma}_{i_n}^{\text{($M$)}}} \Big\}$}
\State{{$\hat{\gamma}_{i_n}^{\text{($M$)}}\leftarrow \hat{\gamma}_{i_n}^{\text{($M$)}} +\delta$}}
\State{$\hat{\boldsymbol{\Sigma}}^{-1}\leftarrow \hat{\boldsymbol{\Sigma}}^{-1} -\delta\frac{\hat{\boldsymbol{\Sigma}}^{-1}\mathbf{s}_{i_n}\mathbf{s}_{i_n}^H\hat{\boldsymbol{\Sigma}}^{-1}}{1+\delta\mathbf{s}_{i_n}^H\hat{\boldsymbol{\Sigma}}^{-1}\mathbf{s}_{i_n}}$}
 \EndFor
\EndFor
\State{Output ${\hat{\boldsymbol{\gamma}}^{\text{($M$)}}=[\hat{\gamma}_{1}^{\text{($M$)}},\ldots,\hat{\gamma}_{N}^{\text{($M$)}}]^T}$.}
\end{algorithmic}
\end{algorithm}

\section{Asymptotic Performance Analysis via Fisher Information Matrix}
\label{sec.analysis}

It is challenging to analyze the performance of specific algorithms for solving
the MLE problem (\ref{eq.prob1}), because most practical algorithms can only
guarantee local optimality. In this section, we assume instead that the MLE
problem (\ref{eq.prob1}) is solved to global optimality and analyze
the asymptotic properties of the true MLE solution {$\hat{\boldsymbol{\gamma}}^{\text{($M$)}}$}
in the regime $M\rightarrow \infty$.
Although the global minimizer of \eqref{eq.prob1} may not be easily found
in practice due to the computational complexity constraint, simulation results
show that the analysis still provides useful insights into the performance of
practical algorithms for solving (\ref{eq.prob1}).
The analysis hinges upon the Fisher information matrix associated with the MLE problem.

For notational clarity, let $\boldsymbol{\gamma}^0$ denote the true parameter to be estimated. We aim to study two questions: (i) What are the conditions on the system parameters $N, K,$ and $L$ such that the estimate {$\hat{\boldsymbol{\gamma}}^{\text{($M$)}}$} can approach the true parameter $\boldsymbol{\gamma}^0$ as $M\rightarrow \infty$? (ii) If these conditions are satisfied but if $M$ is finite, how is the estimation error ${\hat{\boldsymbol{\gamma}}^{\text{($M$)}}}-\boldsymbol{\gamma}^0$ distributed? The answer to the first question helps identify the desired operating regime in the space of $N, K,$ and $L$ for getting an accurate estimate {$\hat{\boldsymbol{\gamma}}^{\text{($M$)}}$} via MLE with massive MIMO, and the answer to the second question helps characterize the error probabilities for practical device activity detection settings.

\subsection{Asymptotic Properties of MLE}
We investigate the above two questions by exploiting the asymptotic properties of MLE: \emph{consistency} and \emph{asymptotic normality}. Recall from \eqref{eq.gauss} that the received signals $\mathbf{y}_m$ at different antennas can be seen as i.i.d. samples of the underlying channel distribution.
It is known from the standard estimation theory (e.g., \cite{Kay1993}) that under certain regularity conditions, the MLE 
is \emph{consistent}, i.e.,
\begin{align}\label{eq.asymp.p}
{\hat{\boldsymbol{\gamma}}^{\text{($M$)}}} \overset{P}{\rightarrow} \boldsymbol{\gamma}^0, \quad\operatorname{as}~ M\rightarrow \infty,
\end{align}
where $\overset{P}{\rightarrow}$ denotes convergence in probability. Furthermore, if the true parameter $\boldsymbol\gamma^0$ is an interior point in the parameter space of $\boldsymbol\gamma$, the estimation error $\sqrt{M}({\hat{\boldsymbol{\gamma}}^{\text{($M$)}}}-\boldsymbol{\gamma}^0)$ converges in distribution to a multivariate Gaussian distribution as the number of i.i.d. samples goes to infinity, i.e.,
\begin{align}\label{eq.asymp.d}
\sqrt{M}({\hat{\boldsymbol{\gamma}}^{\text{($M$)}}}-\boldsymbol{\gamma}^0)\overset{D}{\rightarrow} \mathcal{N}\left(0,M\mathbf{J}^{-1}(\boldsymbol\gamma^0)\right), ~\operatorname{as}~ M\rightarrow \infty.
\end{align}
Here, $\mathbf{J}(\boldsymbol\gamma)$ is the Fisher information matrix, whose $(i,j)$-th entry is defined as
\begin{align}\label{eq.fim.def}
[\mathbf{J}(\boldsymbol\gamma)]_{ij}=\mathbb{E}\left[\left(\frac{\partial \log p(\mathbf{Y}|\boldsymbol{\gamma})}{\partial \gamma_i}\right)\left(\frac{\partial \log p(\mathbf{Y}|\boldsymbol{\gamma})}{\partial \gamma_j}\right)\right],
\end{align}
where $p(\mathbf{Y}|\boldsymbol{\gamma})$ is given in \eqref{eq.likelihood}, and the expectation is taken with respect to $\mathbf{Y}$.

However, for the MLE problem considered in this paper, the results in \eqref{eq.asymp.p} and \eqref{eq.asymp.d} cannot be directly applied as the following two regularity conditions may not be satisfied:
\eda{
\begin{enumerate}
\item The consistency of MLE requires that the true parameter $\boldsymbol\gamma^0$ is \emph{identifiable}, i.e., there exists no other $\boldsymbol{\gamma}^\prime \neq \boldsymbol\gamma^0$ such that $p(\mathbf{Y}|\boldsymbol{\gamma}^\prime)=p(\mathbf{Y}|\boldsymbol{\gamma}^0)$. This is not guaranteed in our problem because the dimension of the parameter $\boldsymbol\gamma^0$, i.e., $N$, could be much larger than the dimensions of the sample covariance matrix $\hat{\boldsymbol{\Sigma}}$, i.e., $L\times L$, and therefore ambiguity may occur in the estimation of $\boldsymbol\gamma^0$.
\item The asymptotic normality of MLE requires that the true parameter $\boldsymbol\gamma^0$ {is} an interior point of its parameter space, {i.e., $[0,+\infty)^{N}$ in the problem under consideration.}
	However, in our problem, $\boldsymbol\gamma^0$ always lies on the boundary of $[0,+\infty)^{N}$, because most of the entries in $\boldsymbol\gamma^0$ are zero due to the inactive devices. For these entries, the estimation error ${\hat{\gamma}_n^{\text{($M$)}}}-\gamma_n^0$ is always nonnegative.
Thus, the estimation error ${\hat{\boldsymbol{\gamma}}^{\text{($M$)}}}-\boldsymbol{\gamma}^0$ cannot be Gaussian distributed.
\end{enumerate}
}

In this paper, we deal with the issue of consistency by proposing a new necessary and sufficient condition for the parameter identifiability, and deal with the asymptotic distribution of $\sqrt{M}({\hat{\boldsymbol{\gamma}}^{\text{($M$)}}}-\boldsymbol{\gamma}^0)$ by taking the boundary case into consideration. Since the Fisher information matrix $\mathbf{J}(\boldsymbol\gamma)$ plays a key role in our analysis, we first provide an explicit expression for $\mathbf{J}(\boldsymbol\gamma)$.
\begin{theorem}\label{prop.fim}
Consider the likelihood function in \eqref{eq.likelihood}, where $\boldsymbol{\gamma}$ is the parameter to be estimated. The associated $N\times N$ Fisher information matrix of $\boldsymbol{\gamma}$ is given by
\begin{align}\label{eq.fim}
\mathbf{J}(\boldsymbol{\gamma}) = M\left(\mathbf{P}\odot \mathbf{P}^*\right),
\end{align}
where $\mathbf{P}\triangleq \mathbf{S}^H\left(\mathbf{S}\boldsymbol\Gamma\mathbf{S}^H+\sigma_w^2\mathbf{I}\right)^{-1}\mathbf{S}$.
\end{theorem}
\begin{IEEEproof}
Please see Appendix~\ref{sec.apd.fisher}.
\end{IEEEproof}

Note that it is possible for $\mathbf{J}(\boldsymbol{\gamma})$ to be singular. This can be shown by using the fact that the rank of $\mathbf{J}(\boldsymbol{\gamma})$ must satisfy
\begin{align}\label{eq.rank}
\operatorname{Rank}(\mathbf{P} \odot \mathbf{P}^*)\overset{(a)}{\leq} \operatorname{Rank}(\mathbf{P})^2 \overset{(b)}{\leq} L^2,
\end{align}
where $(a)$ is due to $\operatorname{Rank}(\mathbf{U}\odot \mathbf{V})\leq\operatorname{Rank}(\mathbf{U})\operatorname{Rank}(\mathbf{V})$ for arbitrary matrices $\mathbf{U}$ and $\mathbf{V}$, and $(b)$ is based on $\operatorname{Rank}(\mathbf{P})\leq\min\{N,L\}$.
Since $\mathbf{P}\odot \mathbf{P}^*$ is of size $N\times N$, we can conclude from \eqref{eq.rank} that $\mathbf{J}(\boldsymbol{\gamma})$ is singular if $N>L^2$, i.e., the dimension of $\boldsymbol{\gamma}$ is larger than the size of the sample covariance matrix $\hat{\boldsymbol{\Sigma}}$ in \eqref{eq.sigma_hat}. The singularity of $\mathbf{J}(\boldsymbol{\gamma})$ complicates the analysis of the estimation problem. Our analysis below takes singular $\mathbf{J}(\boldsymbol{\gamma})$ into consideration.

\subsection{A Necessary and Sufficient Condition for Consistency of $\hat{\boldsymbol{\gamma}}$}
We first establish a necessary and sufficient condition on $\mathbf{J}(\boldsymbol{\gamma})$ such that {$\hat{\boldsymbol{\gamma}}^{\text{($M$)}}$} can approach $\boldsymbol{\gamma}^0$ in the large $M$ limit.
\begin{theorem}\label{prop.condition}
Consider the MLE problem in \eqref{eq.prob1} for device activity detection with given signature sequence matrix $\mathbf{S}\in\mathbb{C}^{L \times N}$ and noise variance $\sigma_w^2$, and let {$\hat{\boldsymbol{\gamma}}^{\text{($M$)}}$} be a sequence of solutions of \eqref{eq.prob1} as $M$ increases. Let $\boldsymbol{\gamma}^0$ be the true parameter whose $N-K$ zero entries are indexed by $\mathcal{I}$, i.e.,
\begin{align}
\mathcal{I}\triangleq\{i\mid\gamma_i^0=0\}.
\end{align}
Define
\begin{align}
\mathcal{N}&\triangleq\{\mathbf{x}\mid \mathbf{x}^T\mathbf{J}(\boldsymbol{\gamma}^0)\mathbf{x}=0, \mathbf{x}\in \mathbb{R}^{N}\},\label{eq.subspace}\\
\mathcal{C}&\triangleq\{\mathbf{x}\mid x_i\geq 0, i\in \mathcal{I}, \mathbf{x}\in \mathbb{R}^{N}\},\label{eq.cone}
\end{align}
where $x_i$ is the $i$-th entry of $\mathbf{x}$.
Then a necessary and sufficient condition for the consistency of {$\hat{\boldsymbol{\gamma}}^{\text{($M$)}}$}, i.e., ${\hat{\boldsymbol{\gamma}}^{\text{($M$)}}}\rightarrow \boldsymbol{\gamma}^0$ as $M\rightarrow\infty$, is that the intersection of $\mathcal{N}$ and $\mathcal{C}$ is the zero vector, i.e., $\mathcal{N}\cap\mathcal{C}=\{\mathbf{0}\}$.
\end{theorem}
\begin{IEEEproof}
Please see Appendix~\ref{sec.apd.condition}.
\end{IEEEproof}

\eda{
An interpretation of the sets $\mathcal{N}$ and $\mathcal{C}$ in Theorem~\ref{prop.condition} is as follows:
\begin{itemize}
\item $\mathcal{N}$ is the null space of $\mathbf{J}(\boldsymbol{\gamma}^0)$, which contains all directions $\mathbf{x}$ from $\boldsymbol{\gamma}^0$ along which the likelihood function stays unchanged, i.e., $ p(\mathbf{Y}|\boldsymbol{\gamma}^0)= p(\mathbf{Y}|\boldsymbol{\gamma}^0+t\mathbf{x})$ holds for small positive $t$ and any $\mathbf{x}\in \mathcal{N}$.
\item $\mathcal{C}$ is a cone, which contains vectors whose coordinates indexed by $\mathcal{I}$ are always nonnegative---in other words, directions $\mathbf{x}$ from $\boldsymbol{\gamma}^0$ along which $\boldsymbol{\gamma}^0+t\mathbf{x}\in [0,+\infty)^N$ holds for small positive $t$.
\end{itemize}
}

The condition $\mathcal{N}\cap\mathcal{C}=\{\mathbf{0}\}$ ensures that the likelihood function $p(\mathbf{Y}|\boldsymbol{\gamma})$ in the feasible neighborhood of $\boldsymbol{\gamma}^0$ is not identical to $p(\mathbf{Y}|\boldsymbol{\gamma}^0)$, so that the true parameter $\boldsymbol{\gamma}^0$ is uniquely identifiable around its neighborhood through the likelihood function. Such a property is often referred to as the local identifiability \cite{Rothenberg1971}, which is of course necessary in order to have ${\hat{\boldsymbol{\gamma}}^{\text{($M$)}}} \rightarrow\boldsymbol{\gamma}^0$. Otherwise, {$\hat{\boldsymbol{\gamma}}^{\text{($M$)}}$} may converge to other neighboring points that have the identical likelihood function as $p(\mathbf{Y}|\boldsymbol{\gamma}^0)$.

The local identifiability of $\boldsymbol\gamma^0$ establishes the necessary
part of the theorem. To prove the sufficiency, we need to show that the
true parameter $\boldsymbol\gamma^0$ is also \emph{globally} identifiable if
$\mathcal{N}\cap\mathcal{C}=\{\mathbf{0}\}$ holds. For general estimation
problems, it is usually difficult to examine the global identifiability based
on its associated Fisher information matrix since the Fisher information matrix
provides only local information of the likelihood function.
However, for the problem considered in this paper, by exploiting the Gaussian model for the observations and the fact that the covariance matrix is a linear function of $\boldsymbol\gamma$ as shown in \eqref{eq.gauss_cov}, it is possible to show that the local identifiability and the global identifiability are equivalent (see Appendix~\ref{sec.apd.condition}). For this reason, the condition $\mathcal{N}\cap\mathcal{C}=\{\mathbf{0}\}$ is both necessary and sufficient for the consistency of the MLE.

\eda{
As an illustration of the condition $\mathcal{N}\cap\mathcal{C}=\{\mathbf{0}\}$, Fig.~\ref{fig.intersection} shows two toy examples of $\mathcal{N}$ and $\mathcal{C}$ in $\mathbb{R}^3$, where the red circle represents $\boldsymbol\gamma^0$ and the shaded cube represents $\boldsymbol\gamma^0+\mathcal{C}$. Note that $\boldsymbol\gamma^0=[\gamma_1^0,\gamma_2^0,\gamma_3^0]^T \in \mathbb{R}^3$. On the left is an example in which $\boldsymbol\gamma^0$ is given by $\gamma_1^0=\gamma_2^0=0, \gamma_3^0>0$, so we have $\mathcal{I}=\{1,2\}$. In this case, $\mathcal{N}\cap\mathcal{C}=\{\mathbf{0}\}$ holds if $\mathcal{N}$ is a one-dimensional subspace such as the one represented by the yellow line. Note that $\mathcal{N}\cap\mathcal{C}=\{\mathbf{0}\}$ cannot hold if $\mathcal{N}$ is two dimensional. On the right is an example in which $\boldsymbol\gamma^0$ is given by $\gamma_1^0=\gamma_2^0=\gamma_3^0=0$, so we have  $\mathcal{I}=\{1,2,3\}$. In this case, $\mathcal{N}\cap\mathcal{C}=\{\mathbf{0}\}$ can hold for a two-dimensional subspace $\mathcal{N}$ such as the one represented by the yellow plane.

It can be observed from the above examples that whether or not $\mathcal{N}\cap\mathcal{C}=\{\mathbf{0}\}$ holds depends on the shapes of $\mathcal{N}$ and $\mathcal{C}$.
Specifically, $|\mathcal{I}|$ and the dimension of $\mathcal{N}$ place condition on whether $\mathcal{N}\cap\mathcal{C}=\{\mathbf{0}\}$ can hold. Intuitively, it is more difficult to satisfy the condition $\mathcal{N}\cap\mathcal{C}=\{\mathbf{0}\}$ if $|\mathcal{I}|$ is small, or if the dimension of $\mathcal{N}$ is large. In the following {proposition}, we state a necessary condition for $\mathcal{N}\cap\mathcal{C}=\{\mathbf{0}\}$ in terms of $|\mathcal{I}|$ and the dimension of $\mathcal{N}$.
\begin{proposition}\label{prop.dim_necessary}
A necessary condition for $\mathcal{N}\cap\mathcal{C}=\{\mathbf{0}\}$ to hold is $\rm{dim}(\mathcal{N})<|\mathcal{I}|$, where $\rm{dim}(\mathcal{N})$ is the {dimension} of $\mathcal{N}$.
\end{proposition}
\begin{IEEEproof}
Please see Appendix~\ref{sec.apd.dim_necessary}.
\end{IEEEproof}

Note that based on \eqref{eq.rank}, the dimension of $\mathcal{N}$ is $N-L^2$ with high probability if the entries of $\mathbf{S}$ are generated randomly. Since $|\mathcal{I}| = N-K$, from Proposition~\ref{prop.dim_necessary} we immediately have that $K<L^2$ is necessary for $\mathcal{N}\cap\mathcal{C}=\{\mathbf{0}\}$ to hold. This result is quite intuitive as $L^2$ can be seen as the number of effective (real-valued) observations from the sample covariance.

We can also use a dimension counting argument to establish the following sufficient condition for $\mathcal{N}\cap\mathcal{C}=\{\mathbf{0}\}$. Consider the special case where $\mathbf{J}(\boldsymbol{\gamma}^0)$ is non-singular, which is true with high probability if $N\leq L^2$ and the entries of $\mathbf{S}$ are generated randomly. We then have $\mathcal{N}=\{\mathbf{0}\}$, and the condition in Theorem~\ref{prop.condition} is immediately satisfied.

\begin{figure}
\centerline{\epsfig{figure=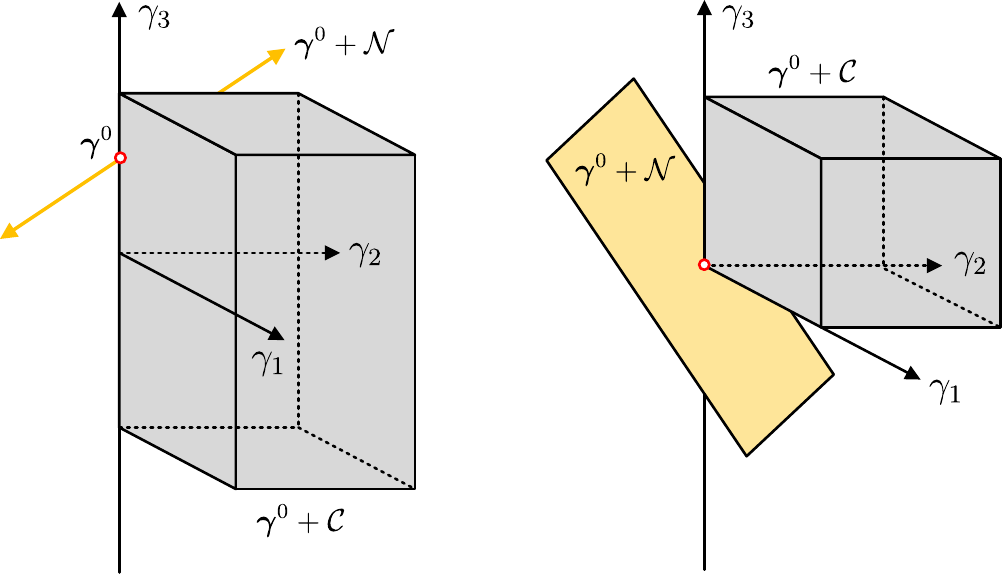,width=9cm}}
	\caption{{\color{black}Examples of $\mathcal{N}$ and $\mathcal{C}$ in $\mathbb{R}^{3}$. The red circle represents $\boldsymbol\gamma^0$, the shaded cube represents $\boldsymbol\gamma^0+\mathcal{C}$, and the yellow affine space represents $\boldsymbol\gamma^0+\mathcal{N}$. Left: $\mathcal{I}=\{1,2\}$, $\rm{dim}(\mathcal{N})=1$, and $\mathcal{N}\cap\mathcal{C}=\{\mathbf{0}\}$; right: $\mathcal{I}=\{1,2,3\}$, $\rm{dim}(\mathcal{N})=2$, and $\mathcal{N}\cap\mathcal{C}=\{\mathbf{0}\}$.
Note that a necessary condition for $\mathcal{N}\cap\mathcal{C}=\{\mathbf{0}\}$ is $\rm{dim}(\mathcal{N}) < |\mathcal{I}|$.}}
\label{fig.intersection}
\end{figure}
}

In general, there is no closed-form characterization of $\mathcal{N}\cap\mathcal{C}$. Thus, the condition $\mathcal{N}\cap\mathcal{C}=\{\mathbf{0}\}$ for a given $\mathbf{J}(\boldsymbol{\gamma}^0)$ cannot be verified analytically. However, by noting that {the sets} $\mathcal{N}$ and $\mathcal{C}$ are both convex, we can test whether the condition $\mathcal{N}\cap\mathcal{C}=\{\mathbf{0}\}$ holds by numerically searching for an $N$-dimensional nonzero vector in $\mathcal{N}\cap\mathcal{C}$. By further exploiting the positive semidefiniteness of the Fisher information matrix, the following theorem turns the verification of $\mathcal{N}\cap\mathcal{C}=\{\mathbf{0}\}$ into an LP in an $(N-K)$-dimensional space.

\begin{theorem}\label{prop.condition.mat}
Given $\mathbf{S}$, $\sigma_w^2$, and $\boldsymbol{\gamma}^0$, let $\mathbf{J}(\boldsymbol{\gamma}^0)$ be the Fisher information matrix in \eqref{eq.fim}.
Let $\mathbf{A}\in \mathbb{R}^{(N-K)\times (N-K)}$ be a submatrix of $\mathbf{J}(\boldsymbol{\gamma}^0)$ indexed by $\mathcal{I}$. Let $\mathbf{C}\in \mathbb{R}^{K\times K}$ be a submatrix of $\mathbf{J}(\boldsymbol{\gamma}^0)$ indexed by $\mathcal{I}^c$, where $\mathcal{I}^c$ is the complement of $\mathcal{I}$ with respect to $\{1,2,\ldots,N\}$. Let $\mathbf{B}\in \mathbb{R}^{(N-K)\times K}$ be a submatrix of $\mathbf{J}(\boldsymbol{\gamma}^0)$ with rows and columns indexed by $\mathcal{I}$ and $\mathcal{I}^c$, respectively. Then the condition $\mathcal{N}\cap\mathcal{C}=\{\mathbf{0}\}$ in Theorem~\ref{prop.condition} is equivalent to: (i) $\mathbf{C}$ is invertible; and (ii) the following problem is feasible
\begin{subequations}
\begin{alignat}{2}
&\quad \operatorname{find} &\quad& \mathbf{x} \label{eq.lp.1}\\
&\operatorname{subject\,to}&      & (\mathbf{A}-\mathbf{B}\mathbf{C}^{-1}\mathbf{B}^T)\mathbf{x} > \mathbf{0},\label{eq.lp.2}
\end{alignat}\label{eq.lp}%
\end{subequations}
where vector $\mathbf{x}\in \mathbb{R}^{N-K}$.
\end{theorem}
\begin{IEEEproof}
Please see Appendix~\ref{sec.apd.lp}.
\end{IEEEproof}

Theorem~\ref{prop.condition.mat} shows that if there exists a vector $\mathbf{x}$ in $\mathbb{R}^{N-K}$ such that $(\mathbf{A}-\mathbf{B}\mathbf{C}^{-1}\mathbf{B}^T)\mathbf{x}$ lies in the positive orthant, then $\mathcal{N}\cap\mathcal{C}=\{\mathbf{0}\}$ holds. Note that the feasibility problem in \eqref{eq.lp} depends only on the matrix $\mathbf{A}-\mathbf{B}\mathbf{C}^{-1}\mathbf{B}^T$.
The class of such matrices that satisfy the constraint in \eqref{eq.lp.2} is
referred to as $\mathcal{M}^+$, which is introduced in \cite{Bruckstein2008} in
the study of the NNLS problem, and also used in \cite{Fengler2019a} for the
performance analysis of device activity detection via the NNLS formulation. It is
interesting that, while we formulate the estimation of $\boldsymbol{\gamma}^0$ as an MLE
problem instead of an NNLS problem, the notion of $\mathcal{M}^+$ still appears.

The condition derived in Theorem~\ref{prop.condition.mat} can be efficiently tested
numerically by solving \eqref{eq.lp} for fixed problem parameters. Since
$\mathbf{A}-\mathbf{B}\mathbf{C}^{-1}\mathbf{B}^T$ is determined by
$\mathbf{J}(\boldsymbol{\gamma}^0)$, which depends on $\mathbf{S}$,
$\sigma_w^2$, and $\boldsymbol{\gamma}^0$, the solution to \eqref{eq.lp} could
also potentially depend on all of these parameters. 
However, we show later in Section~\ref{sec.subsec.match} that the solution
actually depends only on $\mathbf{S}$ and the index set $\mathcal{I}$ corresponding
to $\boldsymbol{\gamma}^0$.

Theorem~\ref{prop.condition.mat} gives us a way to identify the
phase transition of the MLE problem numerically.
Suppose that $\mathbf{S}$ and $\mathcal{I}$ are generated randomly for any fixed $N$, $L$, and $K$ (e.g., $\mathbf{S}$ is Gaussian and the elements in $\mathcal{I}$ are uniformly selected from $\{1,2,\ldots,N\}$), we can use \eqref{eq.lp} to test different realizations of $\mathbf{S}$ and $\mathcal{I}$.
This allows us to numerically characterize the region in the space of
$N$, $L$, and $K$ such that $\hat{\boldsymbol{\gamma}}^{\text{($M$)}}$ can approach
$\boldsymbol{\gamma}^0$ in the large $M$ limit.

\subsection{Distribution of Estimation Error ${\hat{\boldsymbol{\gamma}}^{\text{($M$)}}}-\boldsymbol{\gamma}^0$}
\label{sec.subsec.dist}
We now assume that the system parameters are in the operating regime where
the estimator {$\hat{\boldsymbol{\gamma}}^{\text{($M$)}}$} is consistent, i.e., it converges
to the true $\boldsymbol{\gamma}^0$ as $M\rightarrow\infty$, and aim to
characterize the distribution of the estimation error for finite $M$.
Specifically, we characterize the asymptotic distribution of
$\sqrt{M}({\hat{\boldsymbol{\gamma}}^{\text{($M$)}}}-\boldsymbol{\gamma}^0)$.

As mentioned before,
$\sqrt{M}({\hat{\boldsymbol{\gamma}}^{\text{($M$)}}}-\boldsymbol{\gamma}^0)$ does not tend to a Gaussian distribution,
because $\boldsymbol{\gamma}^0$ lies on the boundary of its feasible set.
In the following, we account for this boundary effect in the analysis of the solution to the MLE problem \eqref{eq.prob1}.
Specifically, note that {$\hat{\boldsymbol{\gamma}}^{\text{($M$)}}$} must converge to a small neighborhood of $\boldsymbol{\gamma}^0$ when $M$ is large. We can then quantify the deviation of {$\hat{\boldsymbol{\gamma}}^{\text{($M$)}}$} from $\boldsymbol{\gamma}^0$, by using a quadratic approximation of the log-likelihood function,
while constraining $\hat{\boldsymbol{\gamma}}^{\text{($M$)}}$ to be in the feasible set.

\eda{
\begin{theorem}\label{prop.dist}
Consider the maximum likelihood estimation of the device activity in \eqref{eq.prob1} with
given $\mathbf{S}$, $\sigma_w^2$, at finite $M$.
Let $\boldsymbol{\gamma}^0$ be the true activity pattern.
Let $\mathbf{J}(\boldsymbol{\gamma}^0)$ be the Fisher information matrix defined in \eqref{eq.fim}.
Let $\mathcal{N}$ and $\mathcal{C}$ be defined as in \eqref{eq.subspace} and \eqref{eq.cone}, respectively. Assume that $\mathcal{N}\cap\mathcal{C}=\{\mathbf{0}\}$.
Let {$\hat{\boldsymbol{\gamma}}^{\text{($M$)}}$} be a sequence of solutions to the problem \eqref{eq.prob1} with $M$ going to infinity. Let $\mathbf{x}\in\mathbb{R}^{N}$ be a random vector distributed as $\mathcal{N}\left(\mathbf{0},M\mathbf{J}^{\dagger}(\boldsymbol\gamma^0)\right)$.
Then, for each realization of $\mathbf{x}$, there exists a solution $\boldsymbol{\mu}^{(*)}$ to the following constrained QP:
\begin{subequations}
\begin{alignat}{2}
&\underset{\boldsymbol\mu}{\operatorname{minimize}}    &\quad& (\mathbf{x}-\boldsymbol\mu)^T \left( \frac{\mathbf{J}(\boldsymbol\gamma^0)}{M} \right) (\mathbf{x}-\boldsymbol\mu)\label{eq.prob2.1}\\
&\operatorname{subject\,to} &      & \boldsymbol{\mu} \in \mathcal{C}
\end{alignat}\label{eq.prob2}
\end{subequations}
such that $\sqrt{M}({\hat{\boldsymbol{\gamma}}^{\text{($M$)}}}-\boldsymbol\gamma^0)$ converges in distribution to the collection of $\boldsymbol{\mu}^{(*)}$'s.
\end{theorem}
\begin{IEEEproof}
Please see Appendix~\ref{sec.apd.qp}.
\end{IEEEproof}

An interpretation of Theorem \ref{prop.dist} is as follows.
We first draw a sample $\mathbf{x}$ from the Gaussian distribution
specified by the Fisher information matrix. We then project the sample to the cone
$\mathcal{C}$ under the distance metric defined by the quadratic function \eqref{eq.prob2.1}, so that the estimation error is consistent with the fact that the
true $\boldsymbol\gamma^0$ lies on the boundary.
These projected samples would have the same distribution as the limiting distribution
of the MLE error
$\sqrt{M}(\hat{\boldsymbol{\gamma}}^{\text{($M$)}}-\boldsymbol{\gamma}^0)$.

Since the QP \eqref{eq.prob2} does not admit a closed-form solution in general,
it is difficult to obtain the distribution of the estimation error analytically.
However, \eqref{eq.prob2} is still useful in the sense that it
reveals the connection between the Fisher information matrix and the error
distribution, and more importantly it enables us to numerically obtain the
distribution of the estimation error for the MLE problem for finite but large $M$.

Note that because the Fisher information matrix $\mathbf{J}(\boldsymbol{\gamma}^0)$
can be singular,
the MLE solution {$\hat{\boldsymbol{\gamma}}^{\text{($M$)}}$} may not be unique when $M$ is finite.
But, as $M$ goes to infinity, the estimation error of MLE does converge in distribution.
Further, for singular $\mathbf{J}(\boldsymbol{\gamma}^0)$, the solution to the QP is not guaranteed to be unique.
But for each realization of
$\mathbf{x} \sim \mathcal{N}\left(\mathbf{0},M\mathbf{J}^{\dagger}(\boldsymbol\gamma^0)\right)$,
there exists a $\boldsymbol{\mu}^{(*)}$, which is a solution of the QP,
such that collectively
these $\boldsymbol{\mu}^{(*)}$'s have the same distribution
as the limiting distribution of the MLE error.
}

\section{Phase Transition Analysis from a Covariance Matching Perspective}
\label{sec.matching}

The necessary and sufficient condition in Theorem~\ref{prop.condition} is based on the properties of the MLE and its associated Fisher information matrix. In this section, we provide an equivalent condition from a perspective of covariance matching by directly analyzing the optimization problem \eqref{eq.prob1}. This new perspective provides new insight into the phase transition analysis, and also shows the connection with a recent analytic scaling law derived in \cite{Fengler2019a}.

\subsection{Covariance Matching as $M\rightarrow \infty$}
\label{sec.subsec.match}

\begin{figure}
\centerline{\epsfig{figure=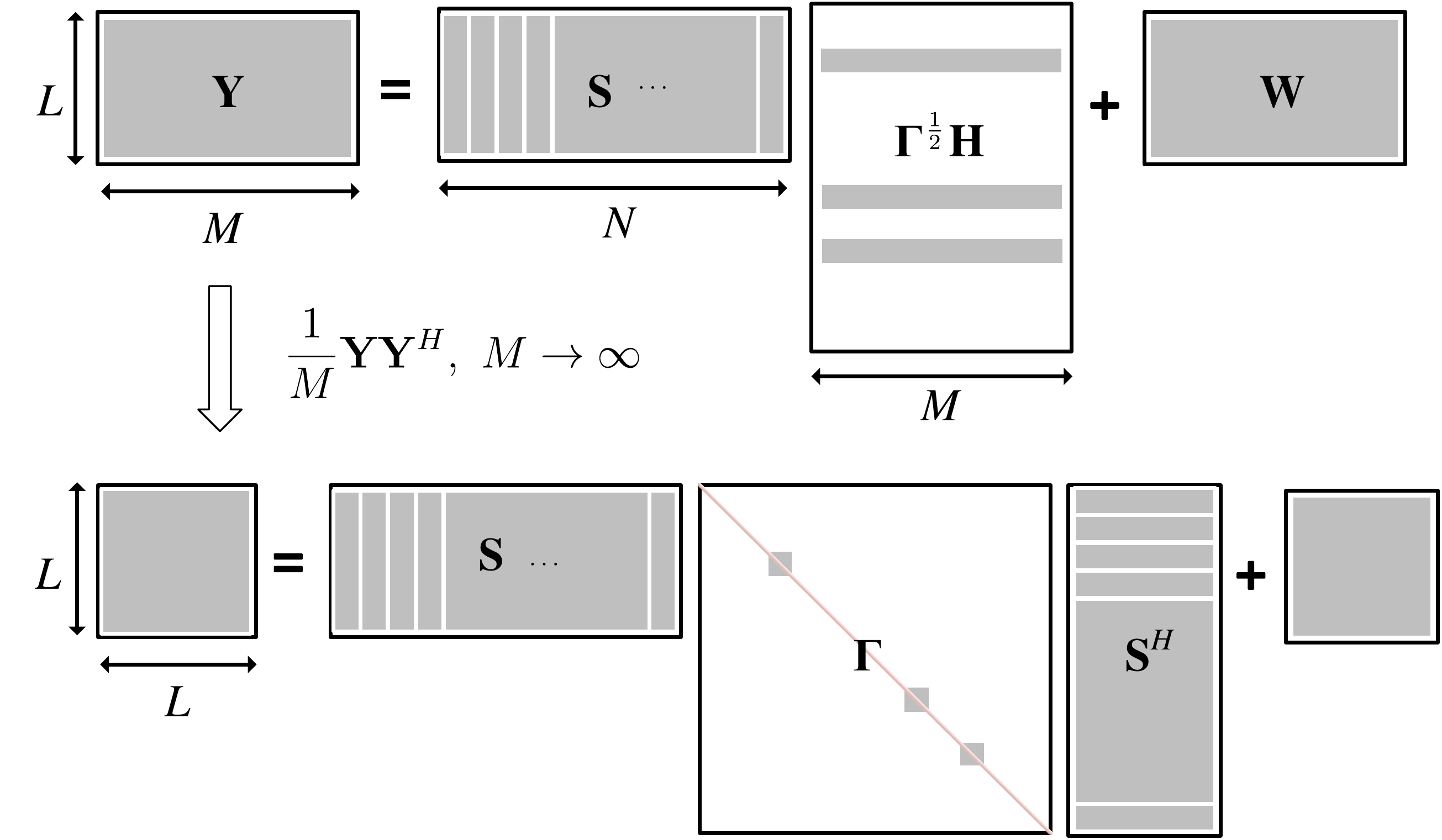,width=9cm}}
\caption{A visualization of the covariance matching as $M\rightarrow\infty$.}
\label{fig.matching}
\end{figure}

First, let us consider a relaxed version of \eqref{eq.prob1}, where the
optimization is performed over $\boldsymbol\Sigma$ instead of
$\boldsymbol\gamma$. A closed-form solution can be immediately obtained as
$\boldsymbol\Sigma=\hat{\boldsymbol\Sigma}$ by checking the
optimality condition of the objective
$\log|\boldsymbol\Sigma|+\mathrm{tr}({\boldsymbol\Sigma}^{-1}\hat{\boldsymbol\Sigma})$.
Unfortunately, under finite $M$, the closed-form solution
$\boldsymbol\Sigma=\hat{\boldsymbol\Sigma}$ may not lead to a feasible solution
for $\boldsymbol\gamma$ due to the fact that the sample covariance matrix may
not exhibit the structure that the true covariance matrix should have, i.e., it
may not be possible to express $\hat{\boldsymbol\Sigma} = \frac{1}{M}
\mathbf{Y} \mathbf{Y}^H$ as
$\mathbf{S}\boldsymbol\Gamma\mathbf{S}^{H}+\sigma_w^2\mathbf{I}$ for some
nonnegative diagonal matrix $\boldsymbol\Gamma$. Therefore, a solution for
$\boldsymbol\gamma$ cannot be obtained from $\hat{\boldsymbol\Sigma}$.
However, in the asymptotic regime of $M\rightarrow \infty$ where the sample
covariance matrix $\hat{\boldsymbol\Sigma}$ converges to the true covariance
matrix, i.e.,
$\mathbf{S}\boldsymbol\Gamma^0\mathbf{S}^{H}+\sigma_w^2\mathbf{I}$, where
$\boldsymbol\Gamma^0\triangleq\operatorname{diag}(\gamma_1^0,\ldots,\gamma_N^0)$,
a feasible solution for $\boldsymbol\gamma$ is guaranteed to exist, and it can
be found by solving for $\boldsymbol\Gamma$ in
\begin{align}\label{eq.mat.match}
\mathbf{S}\boldsymbol\Gamma\mathbf{S}^{H}+\sigma_w^2\mathbf{I} = \hat{\boldsymbol\Sigma}
\end{align}
under the constraint that $\boldsymbol\Gamma$ is a diagonal matrix with
nonnegative entries. An interpretation of \eqref{eq.mat.match}
in the limit of $M\rightarrow\infty$
is that \eqref{eq.mat.match} can be thought of as matching the sample
covariance matrix to the true covariance matrix. This can be visualized in
Fig.~\ref{fig.matching}.

Since the true parameter $\boldsymbol\gamma^0$, or equivalently $\boldsymbol\Gamma^0$, must be a solution to \eqref{eq.mat.match}, intuitively, to make ${\hat{\boldsymbol{\gamma}}^{\text{($M$)}}} \rightarrow \boldsymbol\gamma^0$ as $M\rightarrow \infty$, we need to make sure that $\boldsymbol\gamma^0$ is the unique solution to \eqref{eq.mat.match} in the regime of $M\rightarrow\infty$ under the nonnegative constraint. To guarantee this, we can analyze the null space
of the vectorized form of \eqref{eq.mat.match} as in
\begin{align}\label{eq.mat.match.cond}
\widehat{\mathbf{S}}(\boldsymbol\gamma -\boldsymbol\gamma^0) =\mathbf{0},
\end{align}
where $\widehat{\mathbf{S}}\in \mathbb{C}^{L^2\times N}$ is the column-wise Kronecker product (Khatri-Rao product) of $\mathbf{S}^*$ and $\mathbf{S}$ written as
\begin{align}\label{eq.col.wise.kron}
\widehat{\mathbf{S}}=[\mathbf{s}^{*}_{1}\otimes\mathbf{s}_{1}, \mathbf{s}^{*}_{2}\otimes\mathbf{s}_{2},\ldots, \mathbf{s}^{*}_{N}\otimes\mathbf{s}_{N}].
\end{align}
A necessary and sufficient condition to guarantee that the true parameter $\boldsymbol\gamma^0$ is the unique solution to \eqref{eq.mat.match.cond} under the nonnegative constraint in the limit of $M \rightarrow \infty$ can be obtained as follows:
\begin{theorem}\label{prop.alt.condition}
Consider the covariance matching problem (\ref{eq.mat.match}) with
$\hat{\boldsymbol \Sigma}$ as defined by (\ref{eq.sigma_hat}) with
$\boldsymbol\gamma^0$ being the true value of the activity pattern.
For the given signature sequence matrix $\mathbf{S}$, let $\widehat{\mathbf{S}}\in\mathbb{C}^{L^2\times N}$ be the column-wise Kronecker product of $\mathbf{S}^*$ and $\mathbf{S}$ as given in \eqref{eq.col.wise.kron}. We define the set $\widetilde{\mathcal{N}}$ in $\mathbb{R}^{N}$ as
\begin{align}\label{eq.alt.subspace}
\widetilde{\mathcal{N}}\triangleq\{\mathbf{x}\mid \widehat{\mathbf{S}}\mathbf{x}=\mathbf{0}, \mathbf{x}\in \mathbb{R}^{N}\}.
\end{align}
Then a necessary and sufficient condition for $\boldsymbol\Gamma^0 = {\rm diag}(\boldsymbol{\gamma}^0)$ to be the unique nonnegative solution to \eqref{eq.mat.match} in the limit $M\rightarrow\infty$ is $\widetilde{\mathcal{N}}\cap\mathcal{C}=\{\mathbf{0}\}$, where $\mathcal{C}$ is defined in \eqref{eq.cone}.
\end{theorem}

\begin{IEEEproof}
Please see Appendix~\ref{sec.apd.alt.condition}.
\end{IEEEproof}

The following result reveals the equivalence between the consistency of {$\hat{\boldsymbol{\gamma}}^{\text{($M$)}}$} and the uniqueness of the nonnegative solution to \eqref{eq.mat.match} in the regime $M\rightarrow\infty$, by showing that the condition $\mathcal{N}\cap\mathcal{C}=\{\mathbf{0}\}$ in Theorem~\ref{prop.condition} and $\widetilde{\mathcal{N}}\cap\mathcal{C}=\{\mathbf{0}\}$ in Theorem~\ref{prop.alt.condition} are actually equivalent.
\begin{theorem}\label{prop.condition.eq}
The sets $\widetilde{\mathcal{N}}$ defined in \eqref{eq.alt.subspace} and $\mathcal{N}$ defined in \eqref{eq.subspace} are identical, hence the condition $\widetilde{\mathcal{N}}\cap\mathcal{C}=\{\mathbf{0}\}$ is equivalent to $\mathcal{N}\cap\mathcal{C}=\{\mathbf{0}\}$.
\end{theorem}
\begin{IEEEproof}
Please see Appendix~\ref{sec.apd.condition.eq}.
\end{IEEEproof}

Note that $\mathcal{N}$ in \eqref{eq.subspace} is defined as the null space of $\mathbf{J}(\boldsymbol\gamma^0)$, which is determined by $\mathbf{S}$, $\sigma_w^2$, and $\boldsymbol\gamma^0$ as shown in \eqref{eq.fim}, whereas $\widetilde{\mathcal{N}}$ in \eqref{eq.alt.subspace} is defined as the null space of $\widehat{\mathbf{S}}$, which depends only on $\mathbf{S}$. The equivalence between $\widetilde{\mathcal{N}}$ and $\mathcal{N}$ indicates that $\sigma_w^2$ and $\boldsymbol\gamma^0$, although involved in the expression of $\mathbf{J}(\boldsymbol\gamma^0)$, have no impact on the null space of $\mathbf{J}(\boldsymbol\gamma^0)$.
By further noticing that $\mathcal{C}$ is determined by $\mathcal{I}$, we can conclude that the satisfiability of $\mathcal{N}\cap\mathcal{C}=\{\mathbf{0}\}$ in Theorem~\ref{prop.condition} only depends on
$\mathbf{S}$ and the support of $\boldsymbol\gamma^0$; it does not depend on $\sigma_w^2$ or the values of the nonzero entries of $\boldsymbol\gamma^0$.
This gives us a way of numerically analyzing the phase transition of both the MLE and the matrix matching approaches, as a function of only $K$, $L$, and $N$, in the massive MIMO regime.

Similar to Theorem~\ref{prop.condition.mat}, we can examine whether $\widetilde{\mathcal{N}} \cap \mathcal{C}=\{\mathbf{0}\}$ holds for given $\widehat{\mathbf{S}}$ and $\mathcal{I}$ by solving an LP. Since $\widehat{\mathbf{S}}$ is complex while $\widetilde{\mathcal{N}}$ is a real subspace, we need to separate the real and imaginary parts of $\widehat{\mathbf{S}}$. Let $\mathbf{r}_i^T=[s_{i1}, s_{i2},\ldots,s_{iN}]$ be the $i$-th row of $\mathbf{S}$. Based on $\mathbf{r}_i^T$, we construct two sets of row vectors to represent the real and imaginary parts of rows of $\widehat{\mathbf{S}}$:
\begin{equation}
\{\operatorname{Re}(\mathbf{r}_i^T)\odot \operatorname{Re}(\mathbf{r}_j^T)+\operatorname{Im}(\mathbf{r}_i^T)\odot \operatorname{Im}(\mathbf{r}_j^T), 1\leq i\leq j\leq L\}
\label{eq.set_one}
\end{equation}
and
\begin{equation}\{\operatorname{Re}(\mathbf{r}_i^T)\odot \operatorname{Im}(\mathbf{r}_j^T)-\operatorname{Im}(\mathbf{r}_i^T)\odot \operatorname{Re}(\mathbf{r}_j^T), 1\leq i< j\leq L\}.
\label{eq.set_two}
\end{equation}
\eda{
In total, these two sets consist of $L^2$ vectors in $\mathbb{R}^{1\times N}$.
Let $\mathbf{D}\in \mathbb{R}^{L^2\times N}$ be the matrix formed by all $L^2$ row vectors from the two sets, and let $\mathbf{D}_{\mathcal{I}^c}\in\mathbb{R}^{L^2\times K}$ be a sub-matrix of $\mathbf{D}$ constructed by the columns of $\mathbf{D}$ indexed by $\mathcal{I}^c$. Based on $\mathbf{D}$ and $\mathbf{D}_{\mathcal{I}^c}$, we can verify the condition $\widetilde{\mathcal{N}}\cap\mathcal{C}=\{\mathbf{0}\}$ as follows.

\begin{theorem}\label{prop.condition.mat.alt}
The condition $\widetilde{\mathcal{N}}\cap\mathcal{C}=\{\mathbf{0}\}$ is equivalent to:
(i) the rank of $\mathbf{D}_{\mathcal{I}^c}$ is $K$; and (ii) the following problem is infeasible
\begin{subequations}
\begin{alignat}{2}
&\quad \operatorname{find} &\quad& \mathbf{x}\label{eq.apd.lp.alt.1}\\
&\operatorname{subject\,to}&      & \mathbf{D}\mathbf{x} = \mathbf{0},\label{eq.apd.lp.alt.2}\\
& & &\mathbf{1}^T\mathbf{x}_{\mathcal{I}}=1,\label{eq.apd.lp.alt.3}\\
& & &{x}_i\geq 0, i\in\mathcal{I},\label{eq.apd.lp.alt.4}
\end{alignat}\label{eq.apd.lp.alt}%
\end{subequations}
where $\mathbf{x}\in\mathbb{R}^N$, and $\mathbf{x}_{\mathcal{I}}\in\mathbb{R}^{N-K}$ is a sub-vector of $\mathbf{x}$ with entries indexed by $\mathcal{I}$.
\end{theorem}
\begin{IEEEproof}
Please see Appendix~\ref{sec.apd.condition.mat.alt}.
\end{IEEEproof}
}

As compared to the LP in \eqref{eq.lp}, the LP in \eqref{eq.apd.lp.alt} does not include the true parameter $\boldsymbol\gamma^0$ and the noise variance $\sigma_w^2$; the solution to \eqref{eq.apd.lp.alt} depends on $\mathbf{S}$ and $\mathcal{I}$ only. There is also a difference in dimensionality. The LP in \eqref{eq.lp} aims to find an $(N-K)$-dimensional vector under $N-K$ inequality constraints, whereas the LP in \eqref{eq.apd.lp.alt} aims to find an $N$-dimensional vector under $L^2+1$ equality constraints and $N-K$ inequality constraints.

\subsection{Connection with the Scaling Law in \cite{Fengler2019a}}
\label{sec.subsec.conn}

The condition $\widetilde{\mathcal{N}}\cap\mathcal{C}=\{\mathbf{0}\}$ derived in this paper provides a precise criterion for any given $\mathbf{S}$ and $\mathcal{I}$ under any settings of $N$, $L$, and $K$ to ensure reliable activity detection as $M$ tends to infinity. The satisfiability of $\widetilde{\mathcal{N}}\cap\mathcal{C}=\{\mathbf{0}\}$ can be tested numerically for any finite $N$, $L$, and $K$.

A recent work in \cite{Fengler2019a} studies a similar problem but focuses on the NNLS formulation, and derives an analytic scaling law on $N$, $L$, $K$, and $M$ for a specific class of signature sequences that are drawn uniformly from a sphere in $\mathbb{C}^L$, such that the device activity can be reliably detected. Specifically, it is shown in \cite{Fengler2019a} that the number of identifiable active devices is $K=O(L^2)$, up to a logarithmic factor and a universal constant for sufficiently large $M$, under a covariance based NNLS formulation, which aims to solve the problem
\begin{subequations}
\begin{alignat}{2}\label{eq.NNLS.1}
&\underset{\boldsymbol\gamma}{\operatorname{minimize}}    &\quad& \|\boldsymbol\Sigma-\hat{\boldsymbol{\Sigma}}\|_{F}^2\\
&\operatorname{subject\,to} &      &\, \boldsymbol{\gamma} \geq 0.
\end{alignat}\label{eq.NNLS}%
\end{subequations}
Note that in the asymptotic regime $M\rightarrow\infty$ with fixed $N$, $K$, and $L$, NNLS becomes the covariance matching problem discussed in Section~\ref{sec.subsec.match}. Therefore, the results in Section~\ref{sec.subsec.match} should be related to the scaling law in \cite{Fengler2019a}.
To show the connection, we cite the following results derived in \cite[Theorem~2, Theorem~4]{Fengler2019a}, based on which the scaling law in \cite{Fengler2019a} is established.

\begin{theorem}[\hspace{1sp}\cite{Fengler2019a}]\label{scalinglaw}
Let $\mathbf{S}\in\mathbb{C}^{L\times N}$ be the signature sequence matrix
whose columns are uniformly drawn from the sphere of radius $\sqrt{L}$ in an
i.i.d. fashion. There exist some constants $c_1$, $c_2$, $c_3$, and $c_4$ whose
values do not depend on $K$, $L$, and $N$ such that if $K\leq c_1L^2/\log^2(eN/L^2)$,
then with probability at least $1-\exp(-c_2L)$, the following two statements are true:
\begin{enumerate}
\item
The matrix $\widehat{\mathbf{S}}$ defined in \eqref{eq.col.wise.kron} has
the $\ell_2$ robust null space property (NSP) of order $K$ with parameters $0<\rho<1$ and
$\tau>0$. More precisely, the following inequality
\begin{align}\label{eq.nsp}
\|\mathbf{x}_{\mathcal{K}}\|_2\leq \frac{\rho}{\sqrt{K}}\|\mathbf{x}_{\mathcal{K}^c}\|_1 + \tau\|\widehat{\mathbf{S}}\mathbf{x}\|_2
\end{align}
holds for any $\mathbf{x} \in \mathbb{R}^N$ and any index set
$\mathcal{K}\subseteq\{1,2,\ldots,N\}$ with $|\mathcal{K}|\leq K$, where
$\mathbf{x}_{\mathcal{K}}$ is a sub-vector of $\mathbf{x}$ with entries from
$\mathcal{K}$, and $\mathcal{K}^{c}$ is the complementary set of $\mathcal{K}$
with respect to $\{1,2,\ldots,N\}$.
\item
The solution of \eqref{eq.NNLS}, $\hat{\boldsymbol\gamma}^{\text{NNLS}}$, satisfies
\begin{align}\label{eq.errorbound}
\|\boldsymbol\gamma^0-\hat{\boldsymbol\gamma}^{\text{NNLS}}\|_2 \leq c_3\left(\sqrt{\frac{L}{K}}+c_4\right)\frac{\|\boldsymbol\Sigma^0-\hat{\boldsymbol\Sigma}\|_{F}}{L},
\end{align}
where $\boldsymbol\Sigma^0=\mathbf{S}\boldsymbol\Gamma^0\mathbf{S}^{H}+\sigma_w^2\mathbf{I}$.
\end{enumerate}
\end{theorem}
\begin{IEEEproof}
Please see \cite{Fengler2019a}.
\end{IEEEproof}

When $M\rightarrow\infty$, we note that the sample covariance matrix must
converge to the true covariance matrix. In this case, \eqref{eq.errorbound}
implies that as $N$, $K$, and $L$ go to infinity, the estimation error in NNLS
must vanish.

The following result shows that
$\widetilde{\mathcal{N}}\cap\mathcal{C}=\{\mathbf{0}\}$ can also be ensured
under the conditions in Theorem~\ref{scalinglaw}.

\begin{theorem}\label{prop.connection}
Under the same scaling law for $K$, $L$, $N$ and for the same randomly chosen $\mathbf{S}$ as specified in Theorem~\ref{scalinglaw}, $\widetilde{\mathcal{N}}\cap\mathcal{C}=\{\mathbf{0}\}$ in Theorem~\ref{prop.alt.condition} holds with probability at least $1-\exp(-c_2L)$.
\end{theorem}
\begin{IEEEproof}
Please see Appendix~\ref{sec.apd.connection}.
\end{IEEEproof}

Based on Theorem~\ref{prop.connection} and the equivalence between $\mathcal{N}$ and $\widetilde{\mathcal{N}}$, we can conclude that once the system parameters satisfy the scaling law, the condition $\mathcal{N}\cap\mathcal{C}=\{\mathbf{0}\}$ also holds with high probability. Therefore, with sufficiently large $M$, the activity pattern of the devices can be reliably detected by solving the MLE problem. Theorem~\ref{prop.connection} shows that the scaling law derived for the NNLS formulation also applies to the MLE formulation, which addresses the conjecture in \cite{Fengler2019a} in the asymptotic regime of $M$.
\eda{Note that in practice, MLE achieves a substantially lower error probability as compared to NNLS at finite $M$, as shown in the simulations in Section~\ref{sec.simu}. This is mainly due to the fact that MLE exploits both the distribution information of the observations and the nonnegativity of the parameters, whereas NNLS exploits the nonnegativity only.}
 
\subsection{Regularization}

It is worth mentioning that in both the MLE formulation and the NNLS formulation, $\boldsymbol\gamma$ is treated as a set of deterministic but unknown parameters. This means the fact that the true parameter $\boldsymbol\gamma^0$ is a sparse vector is not exploited. A straightforward way of incorporating such prior information is to add a regularization term to the objective functions in \eqref{eq.prob1} and \eqref{eq.NNLS} to promote the sparsity of the solution. For example, we can consider $l_1$ regularizer, i.e., $R(\boldsymbol\gamma)=\lambda\sum_{n=1}^N \gamma_n$, or log-sum regularizer, i.e., $R(\boldsymbol\gamma)=\lambda\sum_{n=1}^N \log(\epsilon + \gamma_n)$ with $\epsilon>0$, where $\lambda$ is a tunable parameter. With the regularization term, the new objective, based on \eqref{eq.prob1}, becomes
\begin{align}\label{eq.new.obj}
\min_{\boldsymbol\gamma\geq \mathbf{0}}\log\left|\boldsymbol\Sigma\right|+ \operatorname{tr}\left(\boldsymbol\Sigma^{-1}\hat{\boldsymbol{\Sigma}}\right)+\frac{1}{M}R(\boldsymbol\gamma).
\end{align}
However, such a regularization term may not be necessary. This can be justified by the identifiability of $\boldsymbol\gamma^0$ in the MLE formulation or the uniqueness of $\boldsymbol\gamma^0$ to the NNLS problem in the limit $M\rightarrow\infty$, provided that the condition $\widetilde{\mathcal{N}}\cap\mathcal{C}=\{\mathbf{0}\}$ is satisfied. Similar arguments have been discussed in \cite{Bruckstein2008} and \cite{Fengler2019a} for NNLS. Moreover, it is generally not easy to choose the parameter $\lambda$ properly. In the simulation part of this paper, we evaluate the impact of the regularization under finite $M$. The results show that, although the regularization cannot help improve the detection performance substantially as expected, it changes the trade-off between the two types of errors in the device activity detection.

\section{Joint Device Activity and Data Detection}
\label{sec.joint}

This section aims to show that the above analysis can also be applied to the
scenario in which each device is associated with multiple signature sequences
and can embed a few information bits in the random access phase.
\eda{
This data embedding scheme is first proposed in \cite{Senel2018} for grant-free
random access, where the AMP algorithm is employed for joint device activity
and data detection.
}
Here, we show that the joint detection problem can be formulated as
an optimization problem similar to \eqref{eq.prob1} via MLE,
and an asymptotic performance
analysis can be carried out using the approach discussed in
Sections~\ref{sec.analysis} and \ref{sec.matching}. It should be noted that the scheme considered in this section is different from the unsourced random access in \cite{Polyanskiy2017,Fengler2019a}, where all devices share the same set of sequences.

Suppose that each active device has $b$ bits to send. To encode the $b$-bit data as well as the device identification, we assume that each device is assigned a unique set of $Q\triangleq 2^b$ sequences with length $L$, which can be represented by a matrix as $\mathbf{S}_n=[\mathbf{s}_{n}^1,\mathbf{s}_{n}^2,\ldots,\mathbf{s}_{n}^Q]\in \mathbb{C}^{L \times Q}$, where $\mathbf{s}_{n}^q\in \mathbb{C}^{L}$ is the $q$-th sequence of device $n$. Each active device selects one sequence to transmit. Let $a_{n}^q\in\{1,0\}$ indicate whether or not sequence $q$ of device $n$ is transmitted. We have
that $\sum_{q=1}^Qa_{n}^q\in\{0,1\}$ for each $n$, where $\sum_{q=1}^Qa_{n}^q=0$ implies that device $n$ is inactive. Similar to \eqref{eq.sys}, the received signal at the BS is given by
\begin{align}\label{eq.multi.sys}
\widetilde{\mathbf{Y}}=\sum_{n=1}^{N}\mathbf{S}_n\mathbf{D}_n\mathbf{H}_n+\widetilde{\mathbf{W}}\triangleq\widetilde{\mathbf{S}}\widetilde{\boldsymbol{\Gamma}}^{\frac{1}{2}}\widetilde{\mathbf{H}}+\widetilde{\mathbf{W}},
\end{align}
where $\mathbf{D}_n\triangleq\operatorname{diag}(a_{n}^1g_n,\ldots,a_{n}^Qg_n)\in\mathbb{R}^{Q \times Q}$ is a diagonal matrix showing the sequence selection and the large-scale fading of device $n$, $\mathbf{H}_n\triangleq[\mathbf{h}_{n},\ldots,\mathbf{h}_{n}]^T\in\mathbb{C}^{Q \times M}$ is the channel matrix formed by repeated rows,
$\widetilde{\mathbf{W}}\in \mathbb{C}^{L\times M}$ is the effective
i.i.d.\ Gaussian noise with variance $\sigma_{w}^2$, $\widetilde{\mathbf{S}}\triangleq[\mathbf{S}_1,\ldots,\mathbf{S}_N]\in\mathbb{C}^{L\times NQ}$,  $\widetilde{\boldsymbol{\Gamma}}^{\frac{1}{2}}\triangleq\operatorname{diag}(\mathbf{D}_1,\ldots,\mathbf{D}_N)\in\mathbb{R}^{NQ \times NQ}$, and $\widetilde{\mathbf{H}}\triangleq[\mathbf{H}_{1}^T,\ldots,\mathbf{H}_{N}^T]^T\in\mathbb{C}^{NQ\times M}$. Note that \eqref{eq.multi.sys} differs from \eqref{eq.sys} in the extra block structure exhibited in $\widetilde{\boldsymbol{\Gamma}}$ and $\widetilde{\mathbf{H}}$.

The BS performs the joint device activity and data detection by estimating the diagonal matrix $\widetilde{\boldsymbol{\Gamma}}$ based on $\widetilde{\mathbf{Y}}$. Note that the columns of $\widetilde{\mathbf{Y}}$ can be seen as independent samples drawn from a complex Gaussian distribution with mean zero and covariance $\widetilde{\boldsymbol\Sigma}$, which can be computed from \eqref{eq.multi.sys} as
\begin{align}
\widetilde{\boldsymbol\Sigma}=\mathbb{E}\big[\widetilde{\mathbf{Y}}\widetilde{\mathbf{Y}}^H\big]=\widetilde{\mathbf{S}}\widetilde{\boldsymbol{\Gamma}}^{\frac{1}{2}}\boldsymbol{\Phi}\widetilde{\boldsymbol{\Gamma}}^{\frac{1}{2}}\widetilde{\mathbf{S}}^H+\sigma_w^2\mathbf{I},
\end{align}
where $\boldsymbol{\Phi}\triangleq \operatorname{diag}(\mathbf{E},\ldots,\mathbf{E})\in \mathbb{R}^{NQ\times NQ}$ is a block diagonal matrix with $\mathbf{E}\in \mathbb{R}^{Q\times Q}$ being the all-one matrix.
Since each diagonal block $\mathbf{D}_n$ in $\widetilde{\boldsymbol{\Gamma}}^{\frac{1}{2}}$ has at most one nonzero entry, the covariance matrix can be simplified as $\widetilde{\boldsymbol\Sigma}=\widetilde{\mathbf{S}}\widetilde{\boldsymbol{\Gamma}}\widetilde{\mathbf{S}}^H+\sigma_w^2\mathbf{I}$.

Let $\widetilde{\boldsymbol{\gamma}}\in\mathbb{R}^{NQ}$ be the diagonal entries of $\widetilde{\boldsymbol{\Gamma}}$, i.e., $\widetilde{\boldsymbol{\gamma}}=[\widetilde{\boldsymbol{\gamma}}_1^T,\ldots,\widetilde{\boldsymbol{\gamma}}_N^T]^T$ with $\widetilde{\boldsymbol{\gamma}}_n=[(a_n^1g_n)^2,\ldots,(a_n^Qg_n)^2]^T\in\mathbb{R}^{Q}$.
We use MLE to estimate $\widetilde{\boldsymbol{\gamma}}$. The maximization of $\log p(\widetilde{\mathbf{Y}}|\widetilde{\boldsymbol{\gamma}})$ can be cast as the following optimization problem
\begin{subequations}\label{eq.multi.prob1}
\begin{alignat}{2}
&\underset{\widetilde{\boldsymbol\gamma}}{\operatorname{minimize}}    &\quad& \log\big|\widetilde{\boldsymbol\Sigma}\big|+ \frac{1}{M}\operatorname{tr}\left(\widetilde{\boldsymbol\Sigma}^{-1}\widetilde{\mathbf{Y}}\widetilde{\mathbf{Y}}^H\right)\\
&\operatorname{subject\,to} &      & \widetilde{\boldsymbol{\gamma}} \geq 0,\\
&                           &      &  \|\widetilde{\boldsymbol{\gamma}}_n\|_0 \leq 1,\,\,n=1,2,\ldots,N, \label{eq.multi.prob1.3}
\end{alignat}%
\end{subequations}
where \eqref{eq.multi.prob1.3} comes from the fact that each active device only selects \edb{one} sequence \edb{from} its set of $Q$ sequences, i.e., $\sum_{q=1}^Qa_{n}^q\in\{0,1\}$.

As compared to \eqref{eq.prob1}, the extra constraints \eqref{eq.multi.prob1.3} on blocks of $\widetilde{\boldsymbol{\gamma}}_n$ make problem \eqref{eq.multi.prob1} difficult to solve. In this paper, we consider a heuristic method to deal with \eqref{eq.multi.prob1} by first dropping constraint \eqref{eq.multi.prob1.3}. The rationale is that, based on the analysis in Theorem~\ref{prop.condition}, if the Fisher information matrix associated with $\widetilde{\boldsymbol{\gamma}}$ satisfies the condition in Theorem~\ref{prop.condition}, it is guaranteed that the resulting estimate of $\widetilde{\boldsymbol{\gamma}}$ without considering \eqref{eq.multi.prob1.3} converges to its true value as $M\rightarrow\infty$, indicating that \eqref{eq.multi.prob1.3} is satisfied automatically due to the consistency. For large but finite $M$, since \eqref{eq.multi.prob1.3} may not be satisfied exactly, we then use a simple coordinate selection to enforce the constraint for each block.

Using such a method implies that the results in Theorem~\ref{prop.condition} as well as
Theorem~\ref{prop.alt.condition} and Theorem~\ref{prop.condition.eq} can be used to obtain a phase transition analysis on $N2^b$, $K$, and $L$. Moreover, the result in Section~\ref{sec.subsec.dist} can be used to characterize the error probability in the joint device activity and data detection.

\section{Simulation Results}
\label{sec.simu}
In this section, we validate the asymptotic results by simulations and demonstrate the detection performance of the covariance based method for \edb{massive random access}. We consider an mMTC system with one cell of radius $1000$m, where all devices are located at the cell edge for simplicity. Note that this scenario also corresponds to the case when all devices are distributed randomly in the cell but with a power control scheme in which the transmit power of each device is inversely proportional to its large-scale fading \edb{coefficient}. The power of the background noise is set to be $-169$dBm/Hz over $10$ MHz, and the transmit power of each device is set as $23$dBm. We assume that all sequences are generated from an i.i.d.\ complex Gaussian distribution with zero mean and unit variance, unless otherwise specified.

\subsection{Numerical Validation of the Phase Transition}
\begin{figure}
\centerline{\epsfig{figure=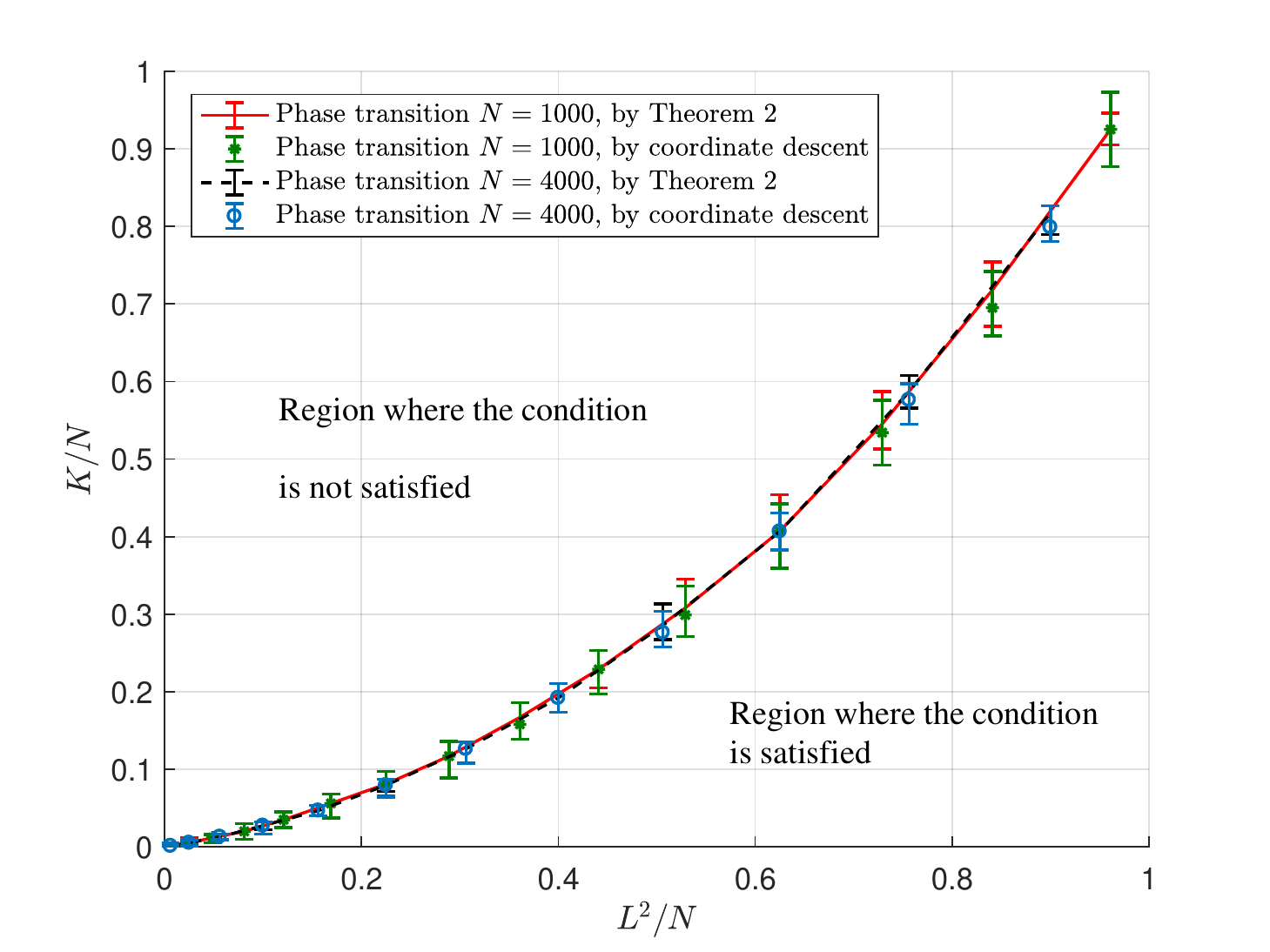,width=9cm}}
\caption{Phase transition of the covariance based method for device activity detection.}
\label{fig.phaseT}
\end{figure}

We consider the device activity detection problem, and numerically test the necessary and sufficient condition described in Theorem~\ref{prop.condition} under a variety of choices of $L$ and $K$, given $N=1000$ or $N=4000$. We draw the region of $L$, $K$ in which the condition is satisfied. Note that the satisfiability of the condition does not \edb{depend} on $\sigma_w^2$, as shown in Theorem~\ref{prop.condition.mat.alt}, thus we fix $\sigma_w^2$ in the simulations. We are interested in the case $L^2<N$ such that the Fisher information matrix $\mathbf{J}(\boldsymbol\gamma^0)$ is singular. Otherwise, the non-singular Fisher information matrix already guarantees that the condition is satisfied. Further, since the detection of $K$ active devices is based on effective $O(L^2)$ observations of the covariance matrix, we plot $L^2/N$ versus $K/N$ in Fig.~\ref{fig.phaseT}. Given $L$ and $K$, we generate $\mathbf{J}(\boldsymbol\gamma^0)$ based on random $\mathbf{S}$ and $\boldsymbol\gamma^0$, and identify the region where the condition can/cannot be satisfied. The result is obtained based on $100$ realizations of $\mathbf{S}$ and $\boldsymbol\gamma^0$ for each $K$ and $L$.
The error bars indicate the range beyond which either all $100$ realizations or zero realization satisfy the condition. Note that the error bar is due to the randomness of $\mathbf{S}$ and $\boldsymbol\gamma^0$.
\eda{
To validate the prediction from Theorem~\ref{prop.condition}, we also run the coordinate descent algorithm to solve the MLE problem in \eqref{eq.prob1} in the large $M$ limit by replacing the sample covariance matrix with the true covariance matrix. We then identify the region of $(L, K)$ in which the active devices can be perfectly detected, thus obtaining the phase transition curve empirically.  We observe that the curves obtained by Theorem~\ref{prop.condition} and by the coordinate descent algorithm match well. We also observe from Fig.~\ref{fig.phaseT} that the curves with different values of $N$ overlap, and the transition region becomes narrower with larger $N$, implying that the phase transition depends on $N$, $L$, and $K$ via the ratios $L^2/N$ and $K/N$.
}

\begin{figure}
\centerline{\epsfig{figure=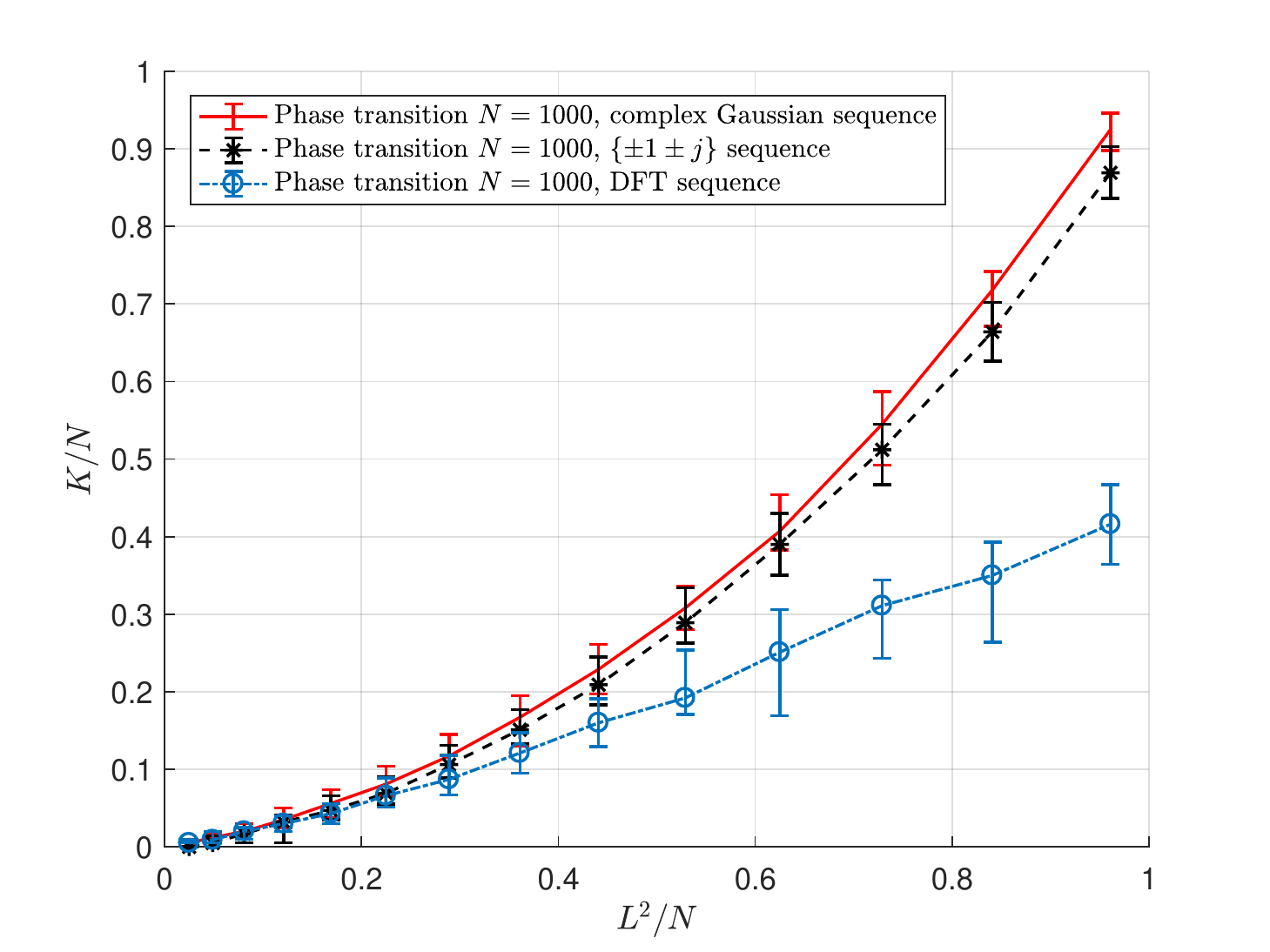,width=9cm}}
\caption{Phase transition comparison of the complex Gaussian signature sequences and the signature sequences whose entries are from $\{\pm1\pm j\}$.}
\label{fig.phaseT.seq}
\end{figure}

\eda{
Since the condition in Theorem~\ref{prop.condition} is applicable to any arbitrary sequence matrix $\mathbf{S}$, we can use the condition to evaluate the phase transitions for different types of signature sequence matrices. Fig.~\ref{fig.phaseT.seq} compares the complex Gaussian matrix with the (partial) DFT matrix and another random matrix whose elements are uniformly drawn from a finite alphabet, $\{\pm 1 \pm j\}$. We observe from Fig.~\ref{fig.phaseT.seq} that the Gaussian matrix slightly outperforms the matrix generated from $\{\pm 1 \pm j\}$ but is substantially better than the (partial) DFT matrix. From a practical point of view, it is easier to generate and store the sequence matrix with $\{\pm 1 \pm j\}$, as compared to the Gaussian matrix.
}

\subsection{Distribution of the Estimation Error}
\begin{figure}
\centerline{\epsfig{figure=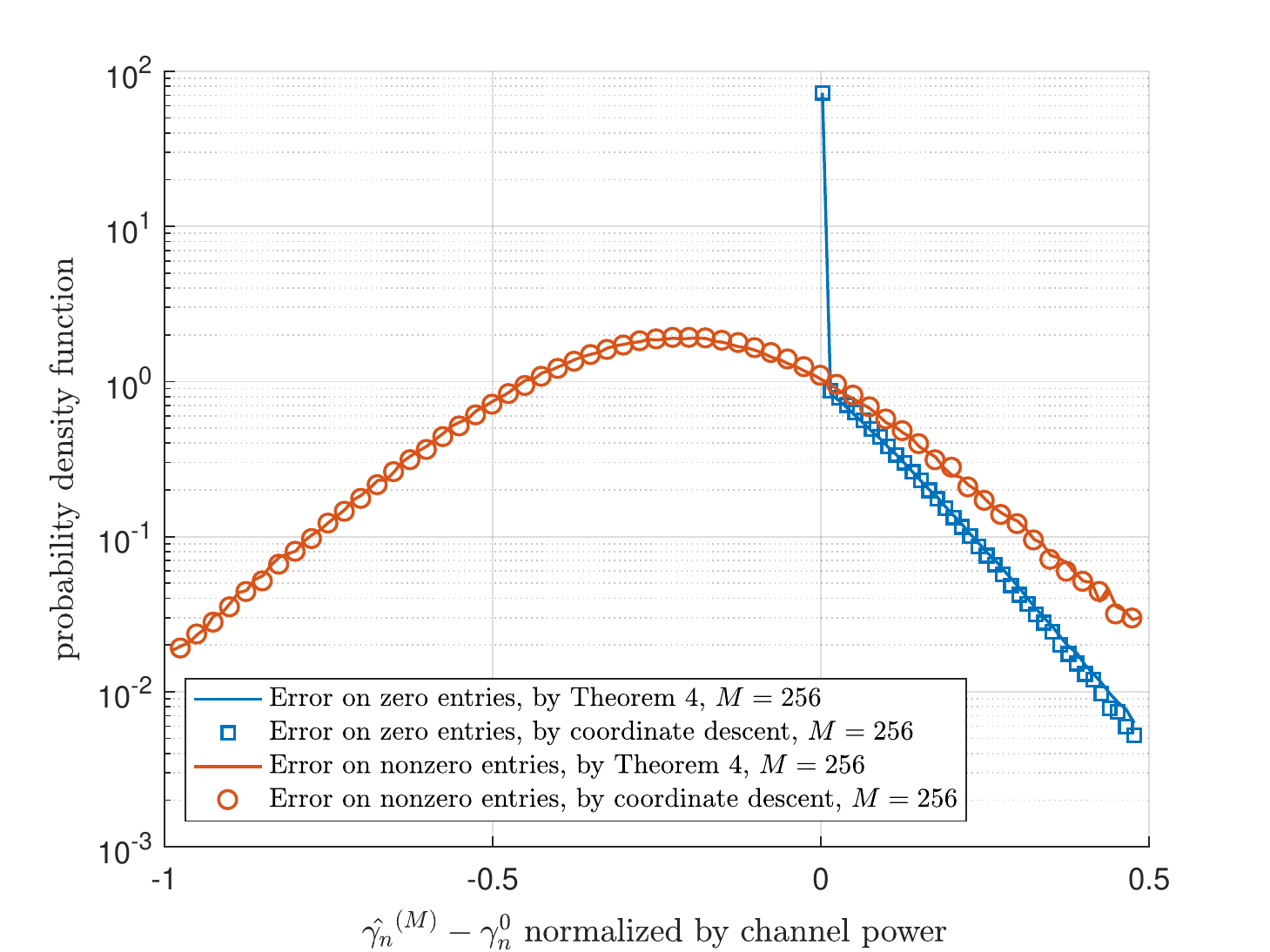,width=9cm}}
\caption{Probability density functions (PDFs) of the error on the zero entries and on the nonzero entries \edb{at $M=256$}.}
\label{fig.errorDistri}
\end{figure}

In Fig.~\ref{fig.errorDistri}, we validate the approximated distribution of $\hat{\boldsymbol\gamma}^{(M)}-\boldsymbol{\gamma}^0$ with $M=256$ from \edb{solving} the QP in \eqref{eq.prob2}, by comparing it with the result from running the coordinate descent algorithm to solve \eqref{eq.prob1}. We set $N=1000$, $K=50$, and $L=20$, which corresponds to $L^2/N=0.4$ and $K/N=0.05$ in Fig.~\ref{fig.phaseT}. We treat each coordinate of $\hat{\boldsymbol\gamma}^{(M)}-\boldsymbol{\gamma}^0$ as independent for simplicity and plot the empirical distribution of the coordinate-wise error. We consider two types of coordinates depending on whether or not the true value of device activity is zero, and plot their corresponding distributions separately. We observe that the curves obtained from \edb{solving} the QP in \eqref{eq.prob2} match those by solving \eqref{eq.prob1} with coordinate descent in both cases. We observe that there is a point mass in the distribution of the error for the zero entries. This is the probability that the inactive devices are correctly identified at finite $M=256$.

\begin{figure}
\centerline{\epsfig{figure=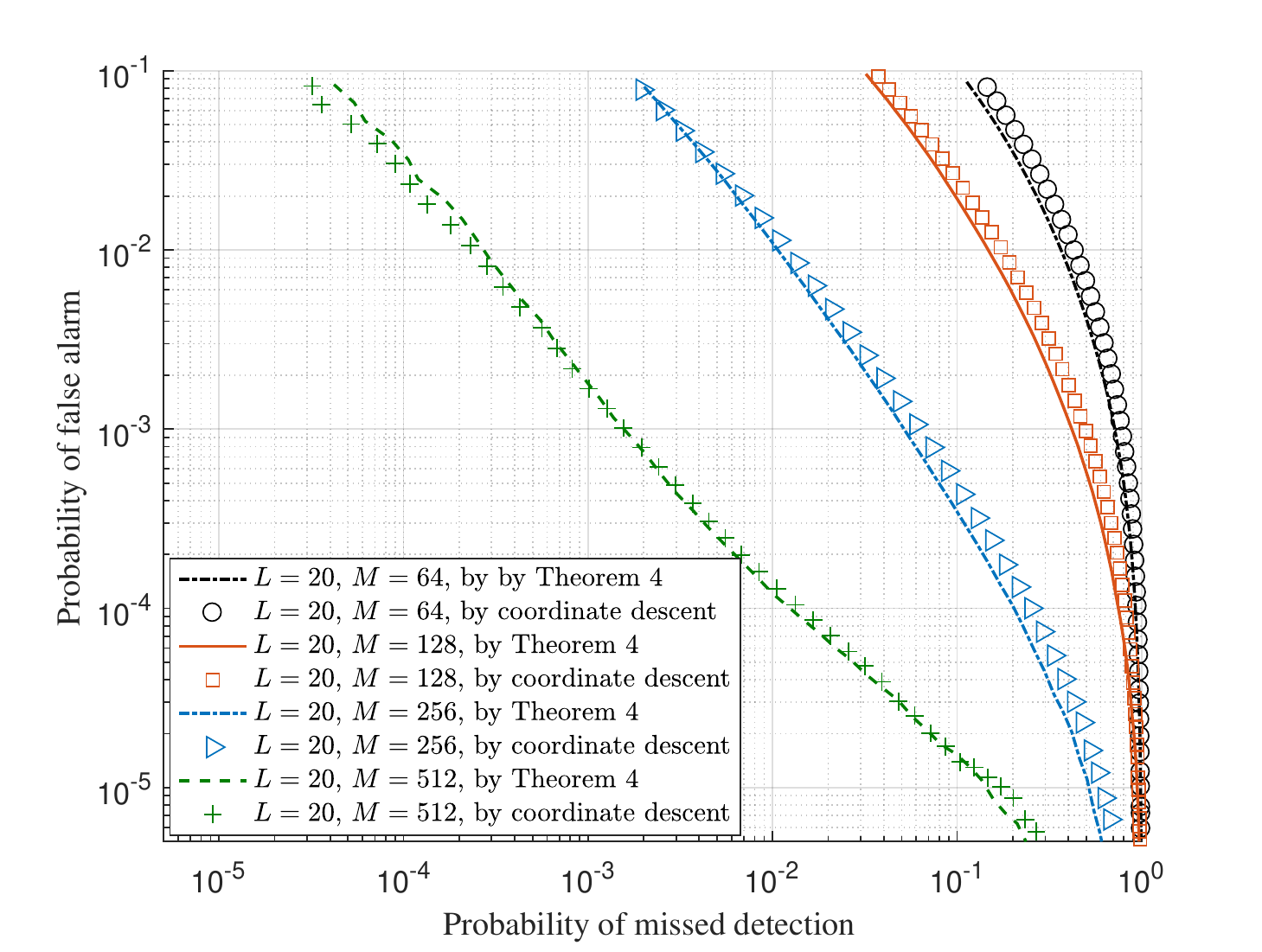,width=9cm}}
\caption{Comparison of the simulated results and the analysis in terms of probability of false alarm and probability of missed detection for device activity detection.}
\label{fig.vali}
\end{figure}

The distribution of the estimation error in Fig.~\ref{fig.errorDistri} helps characterize the probabilities of missed detection and false alarm for the device activity detection problem. A trade-off between missed detection and false alarm can be obtained by setting different thresholds in the last step of the activity detection. We compare the predicted result by the QP in \eqref{eq.prob2} and the simulated result by the coordinate descent algorithm in Fig.~\ref{fig.vali} with $N=1000$, $K=50$, and $L=20$. We observe that the simulated and theoretical curves match very well. The gap becomes even smaller when the number of antennas increases. 

\subsection{Joint Device Activity and Data Detection}
\begin{figure}
\centerline{\epsfig{figure=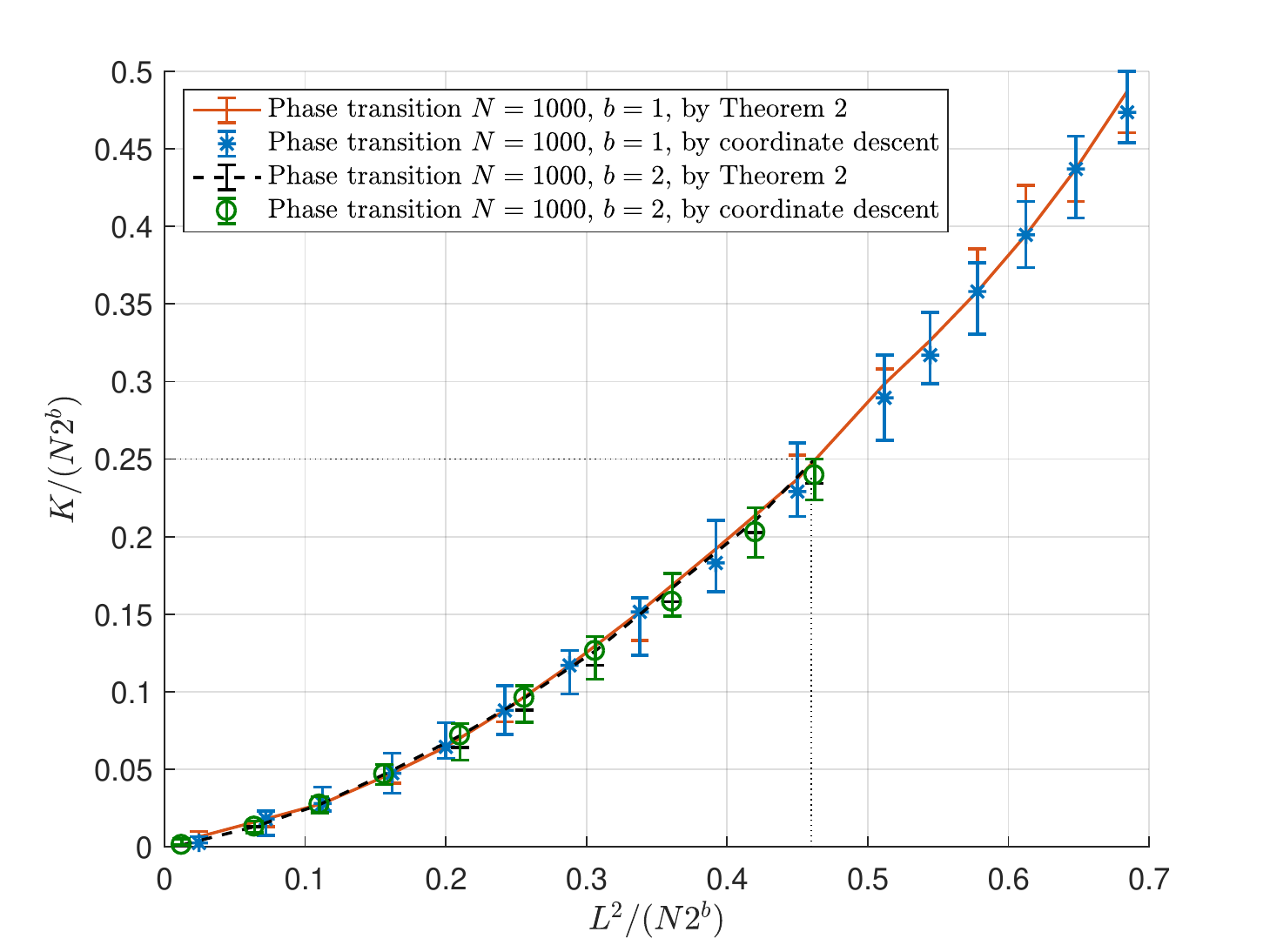,width=9cm}}
\caption{Phase transition of the covariance based method for joint device activity and data detection.}
\label{phaseT.joint}
\end{figure}

In this subsection, we validate the phase transition analysis and the characterization of the estimation error in MLE for joint device activity and data detection. The phase transition is shown in Fig.~\ref{phaseT.joint}, where $N=1000$ and $b$ is set as $1$ or $2$. We plot $K/(N2^b)$ versus $L^2/(N2^b)$, i.e., both $K/N$ and $L^2/N$ are normalized by an extra factor $2^b$. We observe from Fig.~\ref{phaseT.joint} that the curves obtained from Theorem~\ref{prop.condition} and from the coordinate descent algorithm match well. Moreover, we also observe that the curves with $b=1$ and $b=2$ partially overlap, indicating that the phase transition depends on $N$, $L$, $K$, and $b$ via the ratios $L^2/(N2^b)$ and $K/(N2^b)$.

\begin{figure}
\centerline{\epsfig{figure=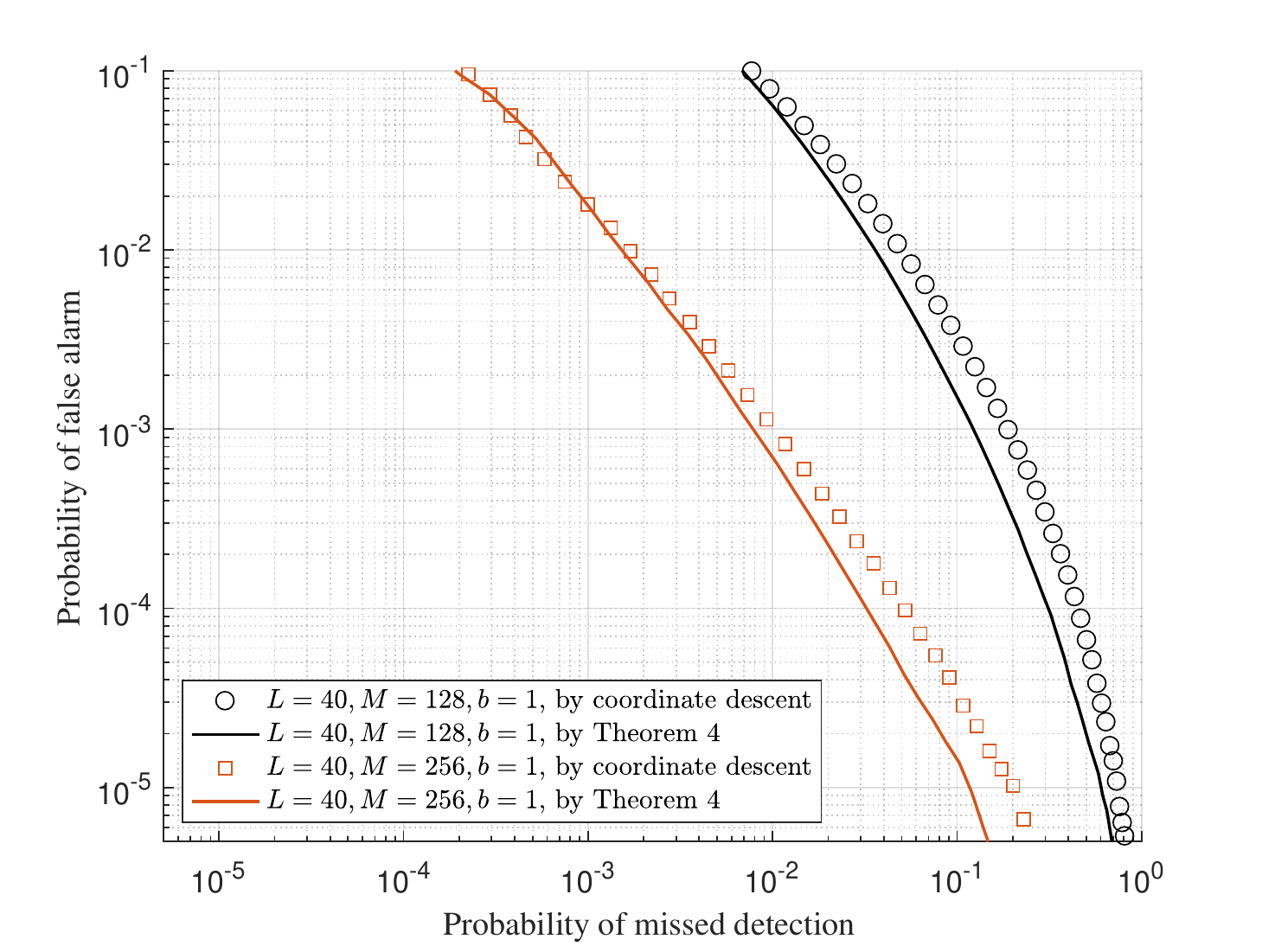,width=9cm}}
\caption{Comparison of the simulated results and the analysis in terms of probability of false alarm and probability of missed detection for joint device activity and data detection .}
\label{fig.vali.joint}
\end{figure}

Similar to Fig.~\ref{fig.vali}, the characterization of the estimation error in MLE can be used to predict the performance of joint device activity and data detection. Here, we still use the probability of false alarm and the probability of missed detection as the performance metrics. To take both device activity and data detection into consideration, we slightly modify the definitions of these two types of errors. Specifically, the probability of missed detection corresponds to two types of error events: a device is active but is declared to be inactive, or a device is active but the data is not correctly decoded although the device is declared active. The probability of false alarm corresponds to the event that a device is inactive but declared active no matter what the decoded data is. A trade-off between missed detection and false alarm can be obtained by setting different thresholds. Fig.~\ref{fig.vali.joint} shows the predicted and the simulated trade-off curves with $N=1000$, $K=100$, $b=1$, and $L=40$. As compared to Fig.~\ref{fig.vali}, the prediction becomes slightly less accurate, which might be due to the extra block-wise sparsity in $\widetilde{\boldsymbol{\gamma}}$ because of information embedding.

\subsection{MLE vs. NNLS vs. AMP}
In this subsection, we consider the joint device activity and data detection problem, and compare the covariance based method with the AMP based method that has been used to solve a similar problem for massive random access in \cite{Senel2018}. For the covariance based method, we consider both the MLE formulation employed in this paper and the NNLS formulation studied in \cite{Fengler2019a}. We fix $N=1000$, $K=100$, and consider various values for $L$ and $M$.  We set $b=1$ or $b=2$, i.e., each active device has $1$ or $2$ bits of information to transmit.

\begin{figure}
\centerline{\epsfig{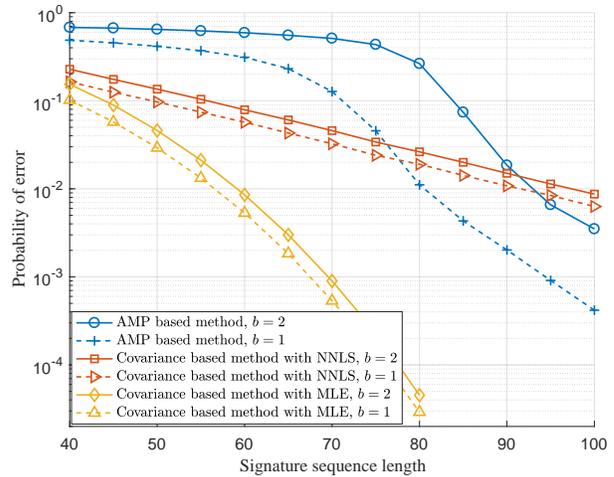}}
\caption{Performance comparison of the covariance based method with MLE, the covariance based method with NNLS, and the AMP based method under different $L$, where $M=64$.} 
\label{fig.length}
\end{figure}

In Fig.~\ref{fig.length}, we show the detection performance as the signature sequence length $L$ increases. Since there are two types of detection errors, to conveniently show the error behavior with $L$, we properly select the threshold to achieve a point where the probability of false alarm and the probability of missed detection are equal, which is represented as ``probability of error" in Fig.~\ref{fig.length}. We observe that increasing $L$ substantially decreases the error probability for the covariance based method with the MLE formulation. However, for the AMP based method, the benefit of increasing $L$ becomes obvious only when $L$ exceeds some point, e.g., $L=80$ when $b=2$. This can be explained by the phase transition in AMP \cite{Donoho2009}, which requires $L$ to be sufficiently large, depending on the problem size. We also observe from Fig.~\ref{fig.length} that the covariance based method with the MLE formulation consistently outperforms both the AMP based method and the covariance based method with the NNLS formulation. Moreover, by increasing the transmitted data from $1$ bit to $2$ bits, which doubles the size of the set of the non-orthogonal sequences, AMP suffers from far more severe performance degradation as compared to the covariance based methods.

\begin{figure}
\centerline{\epsfig{figure=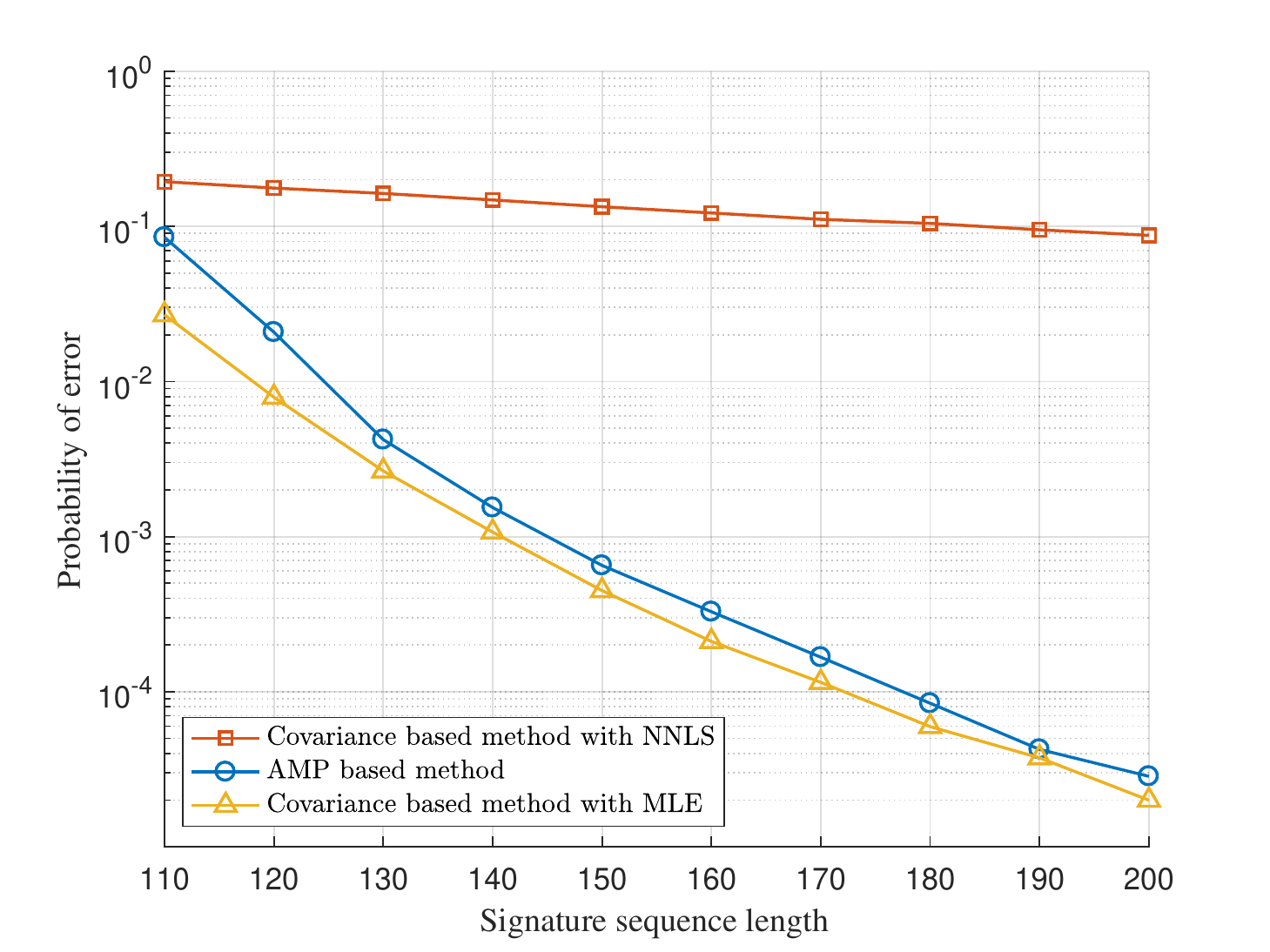,width=9cm}}
\caption{Performance comparison of the covariance based method with MLE, the covariance based method with NNLS, and the AMP based method under different $L$, where $M=8$ and $b=1$.} 
\label{fig.length.2}
\end{figure}

\begin{figure}
\centerline{\epsfig{figure=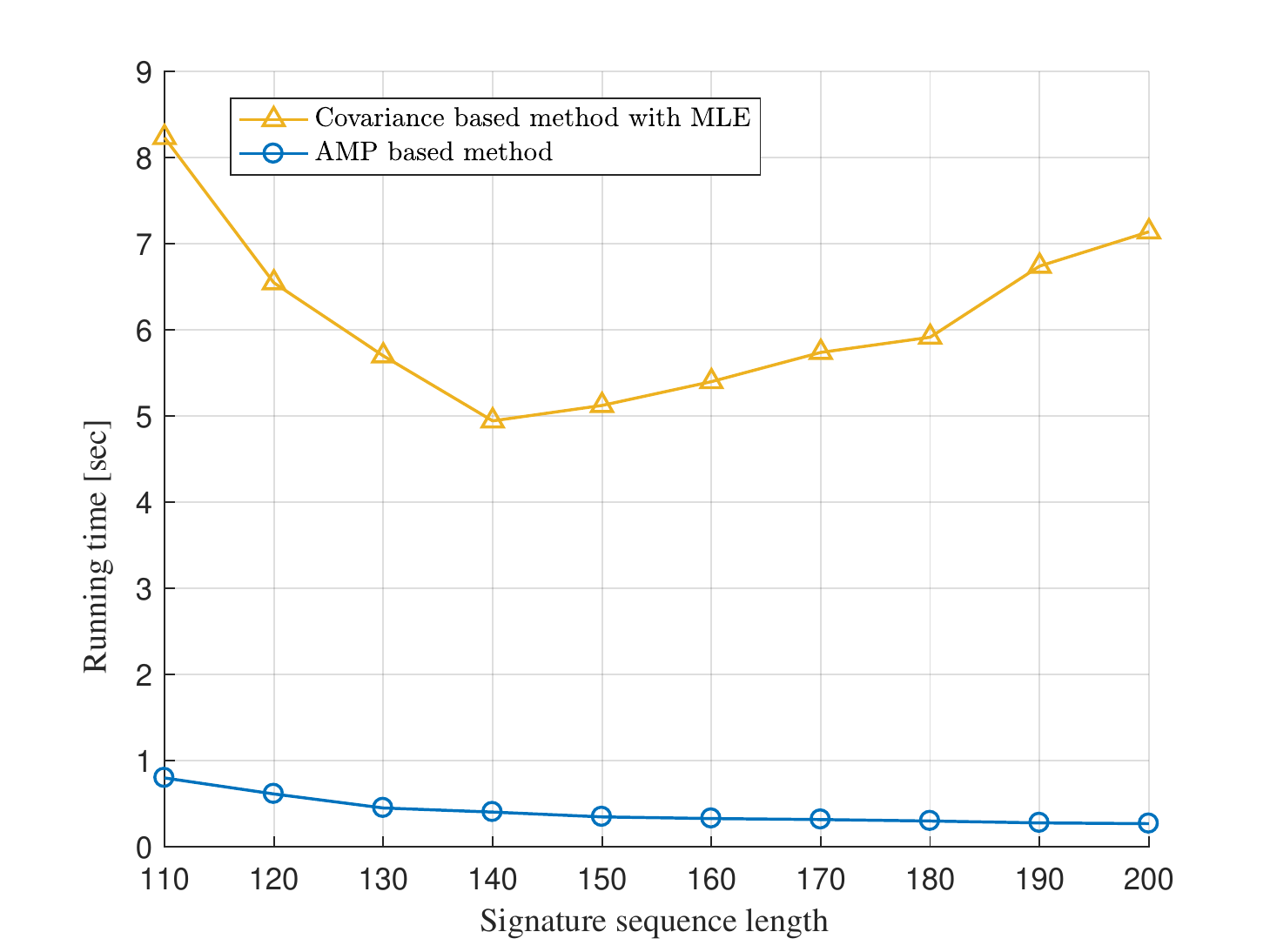,width=9cm}}
\caption{Computational time comparison of the covariance based method with MLE and the AMP based method under different $L$ for $M=8$ and $b=1$. \edb{In both algorithms, the termination condition is satisfied when the improvement between successive iterations falls below $10^{-4}$.}}
\label{fig.runtime}
\end{figure}

\edb{
Fig.~\ref{fig.length}
shows that MLE substantially outperforms the AMP based method when $M$ is large
and $L<K$, which is the preferred operating regime of the covariance based
method. However, it is worth mentioning that in the scenario where $M$ is small
and $L$ is relatively large as compared to $K$, these two methods can achieve
comparable performance, as illustrated in Fig.~\ref{fig.length.2}. 
In this case, the AMP based method has the
advantage of having lower computational complexity, which is largely attributed
to the fact that the complexity of AMP scales with $L$ linearly per iteration,
whereas the complexity of Algorithm~\ref{alg.cd} for solving the MLE problem scales with $L$
quadratically per iteration. For small $M$ and $L>K$, a comparison of the
overall computational time of the two algorithms (at comparable target
error tolerance) as implemented in Matlab on a computer with Intel Core
i5-5200U CPU and 8 GB of memory is shown in Fig.~\ref{fig.runtime}, from which
we observe that the AMP based method indeed has an overall lower complexity and
better scalability with $L$.
Interestingly, we also observe in Fig.~\ref{fig.runtime} that the
computational time of the MLE initially decreases and then increases with $L$.
This is because the overall computational time of MLE depends on both the complexity per
iteration, which is an increasing function of $L$, and the number of
iterations, which is a decreasing function of $L$ at fixed $N$ and $K$.
It should be emphasized that while Fig.~\ref{fig.runtime} shows
the complexity advantage of AMP over MLE for $M=8$, at larger $M$,
AMP becomes more difficult to converge, while the complexity of MLE
is not a strong function of $M$. In this regime, MLE would be perferred
over AMP.}

\subsection{Impact of Regularization}
\begin{figure}
\centerline{\epsfig{figure=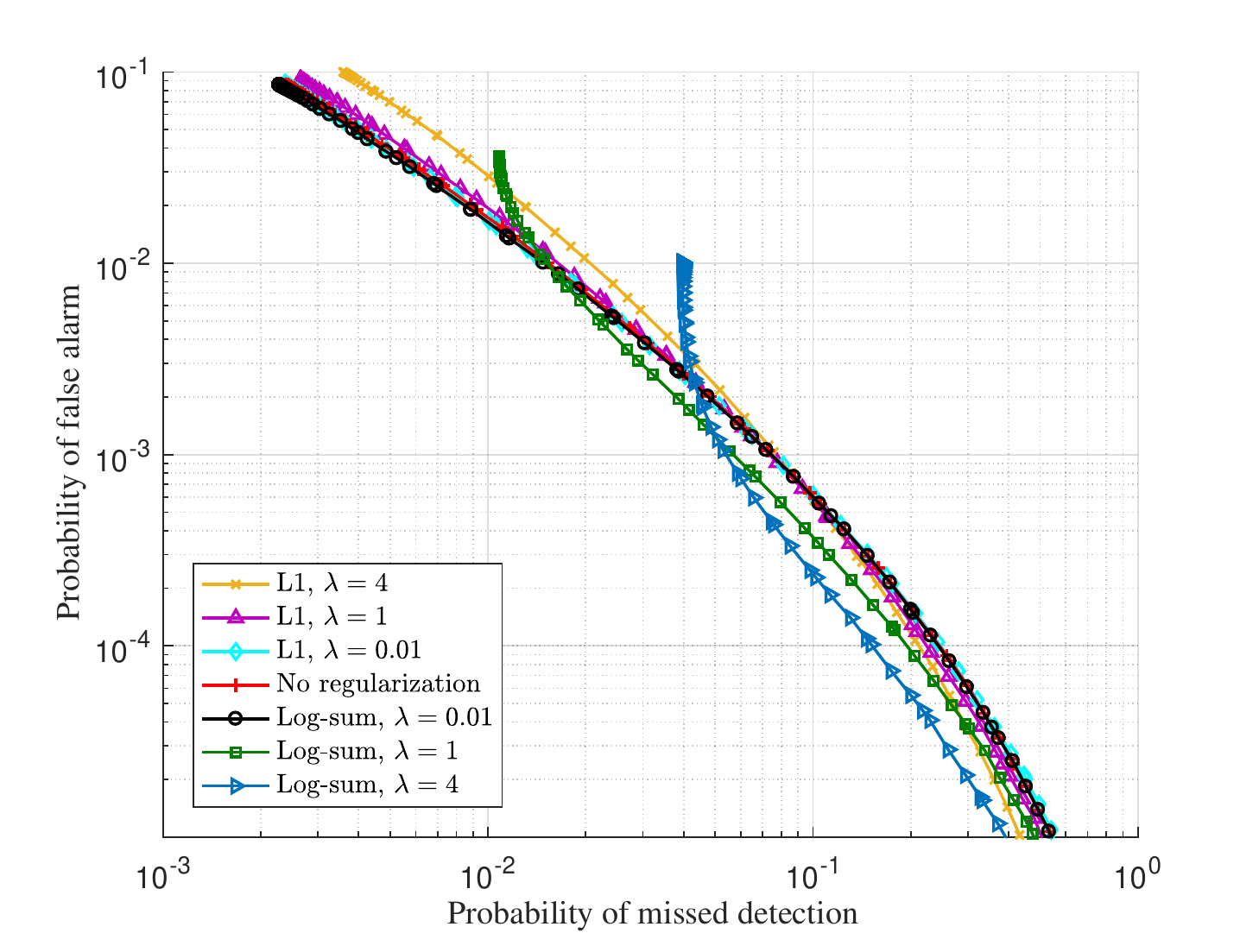,width=9cm}}
\caption{Probabilities of false alarm and missed detection with the regularization term in the log-likelihood function.}
\label{fig.prior}
\end{figure}

In the last part of the simulations, we investigate the impact of adding a regularization term to the objective in \eqref{eq.prob1} under finite $M$. We consider two different regularization terms, $l_1$ regularizer and log-sum regularizer, as discussed in Section~\ref{sec.subsec.conn}. We consider the device activity detection problem for a system with $N=1000$, $K=50$, $L=30$, and $M=64$. We use the coordinate descent algorithm to solve \eqref{eq.new.obj}. Similar to Algorithm~\ref{alg.cd}, closed-form expressions can be derived for the coordinate updates. In Fig.~\ref{fig.prior}, we plot the probability of missed detection versus the probability of false alarm under different choices of $\lambda$. We observe that the regularization terms, especially the log-sum regularizer, change the trade-off between the two types of detection errors. Specifically, for the log-sum regularizer, we observe that when the probability of missed detection is set to be larger than $0.02$ (or 0.05) for $\lambda=1$ (or $\lambda=4$), the regularization term leads to smaller probability of false alarm. In the meanwhile, with the log-sum regularizer it becomes harder to achieve a very low probability of missed detection, no matter what the probability of false alarm is. The change in the trade-off can be explained by the fact that the regularization term indeed promotes the sparsity of the solution, which makes the occurrence of the false alarm more unlikely; but it also increases the chance of missing one or two active devices among all active devices.

\section{Conclusion}
\label{sec.conclusion}
This paper studies the device activity detection problem for the random access that relies on the use of the non-orthogonal sequences in mMTC with massive MIMO. A covariance based approach is employed which formulates the activity detection problem as an MLE problem. By analyzing the asymptotic behavior of the MLE via its associated Fisher
information matrix, a necessary and sufficient condition on the Fisher
information matrix, under which a vanishing detection error is
guaranteed in the massive MIMO regime, is derived. This leads to a phase transition
analysis based on solving an LP that divides the space of the system
parameters into success and failure regions. When the condition is satisfied,
this paper further provides an approach based on solving a QP to accurately predict the probabilities
of detection error for device activity detection with a finite number of
antennas. This paper also considers a random access scheme consisting of joint
device activity and data detection, and shows that the joint detection problem
can be formulated in a similar way as the device activity detection problem
and admits a similar performance analysis.

\eda{
We conclude this paper with a brief discussion of future research directions.
First, this paper assumes that the pilot sequences are transmitted synchronously.
Imperfect synchronization in the pilot phase will require new methods
for dealing with the offset. Second, this paper considers single-cell only.
Extension to multi-cell systems with inter-cell interference would be of interest
in a cellular network setting. Finally, it is possible to develop more computationally
efficient algorithms than the coordinate descent method for solving the MLE problem
by exploiting the sparsity of the solution. Some recent {progresses} along these
directions {have} been reported in \cite{ZhilinFoadWei2021TSP,LiangYaFeng2021ICASSP,WangChen2021ICASSP}.
}

\appendix
\subsection{Proof of Theorem~\ref{prop.fim}}
\label{sec.apd.fisher}
We compute the $(i,j)$-th entry of the Fisher information matrix by using the following identity \cite[Eq.~(3.23)]{Kay1993}
\begin{align}\label{eq.fim.def.div}
\mathbb{E}\left[\frac{\partial \log p(\mathbf{Y}|\boldsymbol{\gamma})}{\partial \gamma_i}\frac{\partial \log p(\mathbf{Y}|\boldsymbol{\gamma})}{\partial \gamma_j}\right]=-\mathbb{E}\left[\frac{\partial^2 \log p(\mathbf{Y}|\boldsymbol{\gamma})}{\partial \gamma_i\partial \gamma_j}\right],
\end{align}
because it is easier to obtain an explicit expression from the right-hand side of \eqref{eq.fim.def.div} for the problem under consideration. Let $\mathcal{L}(\boldsymbol\gamma)\triangleq \log p(\mathbf{Y}|\boldsymbol\gamma)$. The first-order derivative of $\mathcal{L}(\boldsymbol\gamma)$ is given by
\begin{align}\label{eq.adp.first.deriv}
\frac{\partial\mathcal{L}(\boldsymbol\gamma)}{\partial\gamma_i}=
-M\operatorname{tr}\left(\boldsymbol\Sigma^{-1}\mathbf{s}_i\mathbf{s}_i^H\right)+M\operatorname{tr}\left(\boldsymbol\Sigma^{-1}\mathbf{s}_i\mathbf{s}_i^H
\boldsymbol\Sigma^{-1}\hat{\boldsymbol\Sigma}\right),
\end{align}
and its second-order partial derivative can be computed as
\begin{align}
\frac{\partial^2 \mathcal{L}(\boldsymbol\gamma)}{\partial\gamma_i\partial\gamma_j}
&=M\operatorname{tr}\left(\boldsymbol\Sigma^{-1}\mathbf{s}_j\mathbf{s}_j^H\boldsymbol\Sigma^{-1}\mathbf{s}_i\mathbf{s}_i^H\right)\nonumber\\
&\quad-M\operatorname{tr}\left(\boldsymbol\Sigma^{-1}\mathbf{s}_j\mathbf{s}_j^H\boldsymbol\Sigma^{-1}
\mathbf{s}_i\mathbf{s}_i^H
\boldsymbol\Sigma^{-1}\hat{\boldsymbol\Sigma}\right)\nonumber\\
&\quad-M\operatorname{tr}\left(\boldsymbol\Sigma^{-1}\mathbf{s}_i\mathbf{s}_i^H\boldsymbol\Sigma^{-1}
\mathbf{s}_j\mathbf{s}_j^H
\boldsymbol\Sigma^{-1}\hat{\boldsymbol\Sigma}\right).
\end{align}
By taking the expectation with respect to $\mathbf{Y}$ using $\mathbb{E}[\hat{\boldsymbol\Sigma}]=\boldsymbol\Sigma$,
we get the $(i,j)$-th entry of the Fisher information matrix as
\begin{align}
-\mathbb{E}\left[\frac{\partial^2 \mathcal{L}(\boldsymbol\gamma)}{\partial\gamma_i\partial\gamma_j}\right]&=M\operatorname{tr}\left(\boldsymbol\Sigma^{-1}
\mathbf{s}_i\mathbf{s}_i^H\boldsymbol\Sigma^{-1}\mathbf{s}_j\mathbf{s}_j^H\right)\nonumber\\
&=M\left(\mathbf{s}_i^H\boldsymbol\Sigma^{-1}\mathbf{s}_j\right)\left(\mathbf{s}_j^H\boldsymbol\Sigma^{-1}\mathbf{s}_i\right),
\end{align}
based on which $\mathbf{J}(\boldsymbol\gamma)$ can be written in a matrix form as \eqref{eq.fim}.

\subsection{Proof of Theorem~\ref{prop.condition}}
\label{sec.apd.condition}
We use the notion of identifiability in MLE, i.e., the true parameter
$\boldsymbol{\gamma}^0$ is (globally) identifiable if there exists no other
$\boldsymbol\gamma^\prime\neq \boldsymbol\gamma^0$ such that
$p(\mathbf{Y}|\boldsymbol{\gamma}^\prime)=p(\mathbf{Y}|\boldsymbol{\gamma}^0)$.
For the problem under consideration, it can be shown that the consistency of
\edb{the MLE} holds if and only if the true parameter is
identifiable based on the result in \cite[Theorem~14.1]{Veeravalli2018}.
Therefore, in this proof we aim to show that
$\mathcal{N}\cap\mathcal{C}=\{\mathbf{0}\}$ is necessary and sufficient for the
identifiability of $\boldsymbol\gamma^0$.

We start by introducing the notion of \emph{local} identifiability
\cite{Rothenberg1971}. As compared to the (global) identifiability, the local
identifiability of $\boldsymbol\gamma^0$ only requires that there exists a
neighborhood of $\boldsymbol\gamma^0$ such that it contains no other
$\boldsymbol\gamma^\prime\neq \boldsymbol\gamma^0$ with
$p(\mathbf{Y}|\boldsymbol{\gamma}^\prime)=p(\mathbf{Y}|\boldsymbol{\gamma}^0)$.
We first prove that $\mathcal{N}\cap\mathcal{C}=\{\mathbf{0}\}$ is necessary
and sufficient for the local identifiability of $\boldsymbol\gamma^0$. We then
show that the local identifiability is equivalent to the global identifiability
for the problem under consideration, which completes the proof.

First, we present two lemmas on the null space of $\mathbf{J}(\boldsymbol{\gamma})$.
\begin{lemma}\label{lemma.null.space}
Let $\boldsymbol\Sigma=\mathbf{U}\boldsymbol\Lambda\mathbf{U}^H$ be the
eigenvalue decomposition of the covariance matrix $\boldsymbol\Sigma =
\mathbf{S}\boldsymbol{\Gamma}\mathbf{S}^H+\sigma_w^2\mathbf{I}$ with
$\boldsymbol{\gamma}\geq \mathbf{0}$.
Let $\mathbf{V}\triangleq\mathbf{S}^H\mathbf{U}\in\mathbb{C}^{N \times L}$,
and denote its $i$-th column by $\mathbf{v}_i$. Then the set of
$\mathbf{x}\in\mathbb{R}^N$ satisfying
$\mathbf{x}^T\mathbf{J}(\boldsymbol{\gamma})\mathbf{x}=0$ is given by
\begin{align}\label{eq.adp.null.general.j}
\left\{\mathbf{x}\mid \mathbf{x}^T(\mathbf{v}_i\odot\mathbf{v}_j^*)=0,
\forall 1\leq i,j\leq L\right\}.
\end{align}
\end{lemma}
\begin{IEEEproof}
Let the eigenvalues of $\boldsymbol \Sigma$ be
$\boldsymbol\Lambda=\mathrm{diag}(\lambda_1,\lambda_2,\ldots,\lambda_L)$.
By plugging $\boldsymbol\Sigma=\mathbf{U}\boldsymbol\Lambda\mathbf{U}^H$ into
\eqref{eq.fim}, $\mathbf{J}(\boldsymbol{\gamma})$ can be expressed as
\begin{align}\label{eq.adp.fisher.evd}
\mathbf{J}(\boldsymbol{\gamma})
&= \left(\mathbf{S}^H\mathbf{U}\boldsymbol\Lambda^{-1}\mathbf{U}^H\mathbf{S}\right)\odot \left(\mathbf{S}^H\mathbf{U}\boldsymbol\Lambda^{-1}\mathbf{U}^H\mathbf{S}\right)^*\nonumber\\
&=\left(\mathbf{V}\boldsymbol\Lambda^{-1}\mathbf{V}^H\right)\odot \left(\mathbf{V}\boldsymbol\Lambda^{-1}\mathbf{V}^H\right)^*\nonumber\\
&=\left(\sum_{i=1}^L \lambda_i^{-1}\mathbf{v}_i\mathbf{v}^H_i\right)\odot\left(\sum_{j=1}^L \lambda_j^{-1}\mathbf{v}^*_j\mathbf{v}_j^T\right)\nonumber\\
&=\sum_{i=1}^L\sum_{j=1}^L\left(\lambda_i^{-1}\mathbf{v}_i\mathbf{v}_i^H\right)
\odot\left(\lambda_j^{-1}\mathbf{v}_j^*\mathbf{v}_j^T\right)\nonumber\\
&=\sum_{i=1}^L\sum_{j=1}^L\lambda_i^{-1}\lambda_j^{-1}\left(\mathbf{v}_i\odot\mathbf{v}_j^*\right)
\left(\mathbf{v}_i\odot\mathbf{v}_j^*\right)^H,
\end{align}
where the last step is due to the fact that $\left(\mathbf{v}_i\mathbf{v}_i^H\right)
\odot\left(\mathbf{v}_j^*\mathbf{v}_j^T\right)=\left(\mathbf{v}_i\odot\mathbf{v}_j^*\right)
\left(\mathbf{v}_i^H\odot\mathbf{v}_j^T\right)$. Note that $\boldsymbol\Sigma$ is positive definite when $\boldsymbol{\gamma}\geq \mathbf{0}$, which implies that $\lambda_i$'s are all positive. Moreover, $\left(\mathbf{v}_i\odot(\mathbf{v}_j)^*\right)
\left(\mathbf{v}_i\odot(\mathbf{v}_j)^*\right)^H, 1\leq i,j \leq L$, are all positive semidefinite. Therefore, any $\mathbf{x}\in\mathbb{R}^N$ satisfying $\mathbf{x}^T\mathbf{J}(\boldsymbol{\gamma}^0)\mathbf{x}=0$ must also satisfy
\begin{align}\label{eq.adp.null.j0}
\mathbf{x}^T\left(\mathbf{v}_i\odot\mathbf{v}_j^*\right)=0, \,\forall 1\leq i,j\leq L,
\end{align}
and vice versa, from which we obtain \eqref{eq.adp.null.general.j}.
\end{IEEEproof}
\begin{lemma}\label{lemma.eq}
If $\mathbf{x}$ is such that $\mathbf{x}^T\mathbf{J}(\boldsymbol{\gamma}^0)\mathbf{x}=0$, then $\mathbf{x}^T\mathbf{J}(\boldsymbol{\gamma})\mathbf{x}=0$ also holds for any $\boldsymbol{\gamma}\geq \mathbf{0}$.
\end{lemma}
\begin{IEEEproof}
The variables $\mathbf{U}$, $\boldsymbol\Lambda$, $\mathbf{V}$, and $\mathbf{v}_i$ in Lemma~\ref{lemma.null.space} all depend on $\boldsymbol{\gamma}$.
Let $\mathbf{U}_0$, $\boldsymbol\Lambda_0$,
$\mathbf{V}_0$, and $\mathbf{v}^0_i$ be the values of $\mathbf{U}$, $\boldsymbol\Lambda$,
$\mathbf{V}$, and $\mathbf{v}_i$, corresponding to
$\boldsymbol{\gamma}^0$. Since
$\mathbf{V}\triangleq\mathbf{S}^H\mathbf{U}$ and
$\mathbf{V}_0 \triangleq\mathbf{S}^H\mathbf{U}_0$, we have
\begin{align}\label{eq.adp.vv.relation}
\mathbf{V}=\mathbf{V}_0\mathbf{U}_0^H\mathbf{U}= \mathbf{V}_0\overline{\mathbf{U}},
\end{align}
where $\overline{\mathbf{U}}\triangleq \mathbf{U}_0^H\mathbf{U}$.
Let the $(i,j)$-th entry of $\overline{\mathbf{U}}$ be $\overline{u}_{ij}$.
By writing \eqref{eq.adp.vv.relation} explicitly as $\mathbf{v}_i=\sum_{l=1}^L \overline{u}_{li}\mathbf{v}_l^0$, we get
\begin{align}\label{eq.adp.vivj}
\mathbf{v}_i\odot\mathbf{v}_j^*&=\left(\sum_{l=1}^L \overline{u}_{li}\mathbf{v}_l^0\right)\odot\left(\sum_{k=1}^L \overline{u}_{kj}^*(\mathbf{v}_k^0)^*\right)\nonumber\\
&=\sum_{l=1}^L\sum_{k=1}^L\overline{u}_{li}\overline{u}_{kj}^*(\mathbf{v}_l^0\odot(\mathbf{v}_k^0)^*),
\end{align}
which indicates that $\mathbf{x}\in\mathbb{R}^N$ satisfying $\mathbf{x}^T\left(\mathbf{v}_i^0\odot(\mathbf{v}_j^0)^*\right)=0, \forall \,1\leq i,j\leq L,$ must also satisfy $\mathbf{x}^T\left(\mathbf{v}_i\odot\mathbf{v}_j^*\right)=0, \forall \,1\leq i,j\leq L,$ due to the linearity. By Lemma~\ref{lemma.null.space}, we can conclude that $\mathbf{x}^T\mathbf{J}(\boldsymbol{\gamma}^0)\mathbf{x}=0$ implies $\mathbf{x}^T\mathbf{J}(\boldsymbol{\gamma})\mathbf{x}=0$ for any $\boldsymbol{\gamma}\geq \mathbf{0}$.
\end{IEEEproof}

We now show the necessity of $\mathcal{N}\cap\mathcal{C}=\{\mathbf{0}\}$ for the
local identifiability, using contradiction.
Suppose that there exists a nonzero vector $\mathbf{x}\in \mathcal{N}\cap\mathcal{C}$.
Since $\mathbf{x}\in \mathcal{N}$, we must have
$\mathbf{x}^T\mathbf{J}(\boldsymbol{\gamma}^0)\mathbf{x}=0$.
By plugging \eqref{eq.fim.def} into $\mathbf{x}^T\mathbf{J}(\boldsymbol{\gamma}^0)\mathbf{x}$, we get
\begin{align}\label{eq.theorem1.expect}
\mathbf{x}^T\mathbf{J}(\boldsymbol{\gamma}^0)\mathbf{x}=\mathbb{E}\left[\sum_{i}\frac{\partial \log p(\mathbf{Y}|\boldsymbol{\gamma})}{\partial \gamma_i}x_i\right]^2_{\boldsymbol{\gamma}=\boldsymbol{\gamma}^0}=0.
\end{align}
By noting that the term inside the expectation is nonnegative, we get
\begin{align}\label{eq.theorem1.deriv}
\sum_{i}\left(\frac{\partial \log p(\mathbf{Y}|\boldsymbol{\gamma})}{\partial \gamma_i}\Big|_{\boldsymbol{\gamma}=\boldsymbol{\gamma}^0}\right)x_i=0.
\end{align}
Consider now $\boldsymbol{\gamma}$ in the neighborhood of $\boldsymbol{\gamma}^0$ along the direction $\mathbf{x}$.
Since $\mathbf{x}\in\mathcal{C}$, we must have that $\boldsymbol{\gamma}$ remains feasible.
Now by Lemma~\ref{lemma.eq},
$\mathbf{x}^T\mathbf{J}(\boldsymbol{\gamma}^0)\mathbf{x}= \mathbf{x}^T\mathbf{J}(\boldsymbol{\gamma})\mathbf{x} = 0$, we can repeat the same argument as in (\ref{eq.theorem1.expect})-(\ref{eq.theorem1.deriv}) to show that
$\sum_{i}\frac{\partial \log p(\mathbf{Y}|\boldsymbol{\gamma})}{\partial \gamma_i}x_i=0$.
This means that the directional derivative of $\log p(\mathbf{Y}|\boldsymbol{\gamma})$ along
$\mathbf{x}$ is zero, which implies that $\log p(\mathbf{Y}|\boldsymbol{\gamma})$
stays unchanged when $\boldsymbol{\gamma}$ moves from $\boldsymbol{\gamma}^0$
along the direction $\mathbf{x}$ in the neighborhood of $\boldsymbol{\gamma}^0$. This implies that
$\boldsymbol{\gamma}^0$ is not locally identifiable.

To show the sufficiency of $\mathcal{N}\cap\mathcal{C}=\{\mathbf{0}\}$ for the
local identifiability, we also use contradiction. Suppose that the local
identifiability is not satisfied. This implies that there exists a sequence
$\{\boldsymbol{\gamma}^1,\boldsymbol{\gamma}^2,\ldots\}$ approaching
$\boldsymbol{\gamma}^0$ in the feasible neighborhood of $\boldsymbol{\gamma}^0$
satisfying $p(\mathbf{Y}|\boldsymbol{\gamma}^1)=p(\mathbf{Y}|\boldsymbol{\gamma}^2)=\cdots=p(\mathbf{Y}|\boldsymbol{\gamma}^0)$
for all $\mathbf{Y}$. We can then construct an infinite sequence of unit vectors
$\{\frac{\boldsymbol{\gamma}^1-\boldsymbol{\gamma}^0}{\|\boldsymbol{\gamma}^1-\boldsymbol{\gamma}^0\|_2},
\frac{\boldsymbol{\gamma}^2-\boldsymbol{\gamma}^0}{\|\boldsymbol{\gamma}^2-\boldsymbol{\gamma}^0\|_2},\ldots\}$,
which must contain a limit point due to the fact that the sequence is
bounded. Let $\mathbf{x}$ denote this limit point. By the mean value theorem,
for all $n=1,2,\ldots$,
there exists $\overline{\boldsymbol{\gamma}^n}$ between $\boldsymbol{\gamma}^n$
and $\boldsymbol{\gamma}^0$, such that
\begin{multline}
\frac{\log p(\mathbf{Y}|\boldsymbol{\gamma}^n) - \log p(\mathbf{Y}|\boldsymbol{\gamma}^0)}{\|\boldsymbol{\gamma}^n-\boldsymbol{\gamma}^0\|_2}\nonumber\\
=
\sum_{i}\frac{\partial \log p(\mathbf{Y}|\boldsymbol{\gamma})}{\partial \gamma_i}\Big|_{\boldsymbol{\gamma}=\overline{\boldsymbol{\gamma}^n}}
\left(\frac{\gamma^n_i - \gamma^0_i}{\|\boldsymbol{\gamma}^n-\boldsymbol{\gamma}^0\|_2}\right).
\end{multline}
But $p(\mathbf{Y}|\boldsymbol{\gamma}^1)=p(\mathbf{Y}|\boldsymbol{\gamma}^2)=\cdots=p(\mathbf{Y}|\boldsymbol{\gamma}^0)$, which means that the above equation is actually zero for all $n$. Hence, for the limit point $\mathbf{x}$ of the sequence
$\{\frac{\boldsymbol{\gamma}^1-\boldsymbol{\gamma}^0}{\|\boldsymbol{\gamma}^1-\boldsymbol{\gamma}^0\|_2},
\frac{\boldsymbol{\gamma}^2-\boldsymbol{\gamma}^0}{\|\boldsymbol{\gamma}^2-\boldsymbol{\gamma}^0\|_2},\ldots\}$,
we must have
\begin{align}\label{eq.meanvalue_limit}
\sum_{i}\left(\frac{\partial \log p(\mathbf{Y}|\boldsymbol{\gamma})}{\partial \gamma_i}\Big|_{\boldsymbol{\gamma}=\boldsymbol{\gamma}^0}\right)x_i=0.
\end{align}
Note that \eqref{eq.meanvalue_limit} holds for all $\mathbf{Y}$. By taking the
expectation of its square, 
we get an equation identical to \eqref{eq.theorem1.expect}, which implies that
$\mathbf{J}(\boldsymbol{\gamma}^0)\mathbf{x}=\mathbf{0}$ by using the positive
semidefiniteness of $\mathbf{J}(\boldsymbol{\gamma}^0)$. Therefore,
$\mathbf{x}\in \mathcal{N}$.
In the meanwhile, since the sequence approaches $\boldsymbol{\gamma}^0$ in the
feasible neighborhood, we must have $\gamma_i^n-\gamma_i^0\geq 0$ for all
$i\in\mathcal{I}$, as $\gamma_i^0=0, i\in\mathcal{I}$, and $\gamma_i^n\geq
0, i\in\mathcal{I}$. Thus, the vectors in the sequence
$\{\frac{\boldsymbol{\gamma}^1-\boldsymbol{\gamma}^0}{\|\boldsymbol{\gamma}^1-\boldsymbol{\gamma}^0\|_2},
\frac{\boldsymbol{\gamma}^2-\boldsymbol{\gamma}^0}{\|\boldsymbol{\gamma}^2-\boldsymbol{\gamma}^0\|_2},\ldots\}$
are all unit vectors in $\mathcal{C}$. This means that the limit point $\mathbf{x}$ must also be a unit vector in $\mathcal{C}$, because
the intersection of the unit sphere and $\mathcal{C}$ is
a closed set. Thus we have that $\mathbf{x} \in \mathcal{N}\cap\mathcal{C}$, and
therefore $\mathcal{N}\cap\mathcal{C}\neq\{\mathbf{0}\}$.

Finally, we show that the local identifiability of the true parameter $\boldsymbol{\gamma}^0$ is equivalent to the global identifiability of $\boldsymbol{\gamma}^0$ for the problem under consideration. Since the global identifiability already implies the local identifiability, we only need to prove that the local identifiability also implies the global identifiability. In the following, we use contradiction to show that if $\boldsymbol{\gamma}^0$ is not globally identifiable, then $\boldsymbol{\gamma}^0$ is not locally identifiable. Suppose that there exists another $\boldsymbol\gamma^\prime\neq \boldsymbol\gamma^0$ in $[0,+\infty)^{N}$ such that $p(\mathbf{Y}|\boldsymbol{\gamma}^\prime)=p(\mathbf{Y}|\boldsymbol{\gamma}^0)$. Since both $p(\mathbf{Y}|\boldsymbol{\gamma}^\prime)$ and $p(\mathbf{Y}|\boldsymbol{\gamma}^0)$ are zero-mean multivariate Gaussian distributions, the corresponding covariance matrices, denoted by $\boldsymbol{\Sigma}^\prime\triangleq \sum_{n=1}^N \gamma^\prime_n\mathbf{s}_n\mathbf{s}_n^H+\sigma_w^2\mathbf{I}$ and $\boldsymbol{\Sigma}^0\triangleq \sum_{n=1}^N \gamma^0_n\mathbf{s}_n\mathbf{s}_n^H+\sigma_w^2\mathbf{I}$ from \eqref{eq.gauss} must be identical if their distribution functions are the same, implying
\begin{align}
\sum_{n=1}^N (\gamma^{\prime}_n - \gamma^0_n)\mathbf{s}_n\mathbf{s}_n^H = \mathbf{0}.
\end{align}
Then, we can construct another
$\boldsymbol\gamma^{\prime\prime}\triangleq \boldsymbol\gamma^0 +
t(\boldsymbol\gamma^\prime - \boldsymbol\gamma^0)$ with $t\in (0,1)$ such that
$p(\mathbf{Y}|\boldsymbol{\gamma}^{\prime\prime})=p(\mathbf{Y}|\boldsymbol{\gamma}^0)$,
since its corresponding mean would be zero and its covariance matrix would also be identical.
Note that the positive scalar $t$ can be arbitrarily small, which implies that
we can construct such $\boldsymbol\gamma^{\prime\prime}$ in any neighborhood of
$\boldsymbol\gamma^0$, and thus $\boldsymbol\gamma^0$ is not locally
identifiable. This completes the proof of Theorem~\ref{prop.condition}.

\subsection{Proof of Proposition~\ref{prop.dim_necessary}}
\label{sec.apd.dim_necessary}
Let $\mathbf{x}^{(1)},\ldots,\mathbf{x}^{(S)}$ be a basis of $\mathcal{N}$, where $S$ is the {dimension} of $\mathcal{N}$.
If $\mathcal{N}\cap\mathcal{C}=\{\mathbf{0}\}$, then the following $S$ vectors
\begin{align}
\mathbf{x}^{(1)}_{\mathcal{I}},\ldots,\mathbf{x}^{(S)}_{\mathcal{I}}
\end{align}
must be linearly independent, where $\mathbf{x}_{\mathcal{I}}^{(s)}$ is a sub-vector of $\mathbf{x}^{(s)}$ indexed by $\mathcal{I}$. This can be proved by contradiction. Suppose that $\mathbf{x}^{(1)}_{\mathcal{I}},\ldots,\mathbf{x}^{(S)}_{\mathcal{I}}$ are linearly dependent, then there must exist a nonzero vector $\boldsymbol\alpha\triangleq [\alpha_1,\ldots,\alpha_S ]$ such that
\begin{align}\label{eq1}
\mathbf{x}_{\mathcal{I}}^\alpha \triangleq \alpha_1\mathbf{x}^{(1)}_{\mathcal{I}}+\cdots+\alpha_S\mathbf{x}^{(S)}_{\mathcal{I}} = \mathbf{0},
\end{align}
and consequently we get
\begin{align}\label{eq2}
    \mathbf{x}^\alpha\triangleq \alpha_1\mathbf{x}^{(1)}+\cdots+\alpha_S\mathbf{x}^{(S)}\in \mathcal{C},
\end{align}
by using the definition of $\mathcal{C}$ in \eqref{eq.cone} and recognizing that $\mathbf{x}_{\mathcal{I}}^\alpha$ in \eqref{eq1} is a sub-vector of $\mathbf{x}^\alpha$ indexed by $\mathcal{I}$. Therefore, $\mathbf{x}^\alpha \in \mathcal{N}\cap\mathcal{C}$. By further noting that $\mathbf{x}^\alpha\neq \mathbf{0}$ since $\mathbf{x}^{(1)},\ldots,\mathbf{x}^{(S)}$ are a basis of $\mathcal{N}$ and $\boldsymbol\alpha$ is nonzero, we immediately have $\mathcal{N}\cap\mathcal{C}\neq\{\mathbf{0}\}$, which contradicts with $\mathcal{N}\cap\mathcal{C}=\{\mathbf{0}\}$.
With $\mathbf{x}^{(1)}_{\mathcal{I}},\ldots,\mathbf{x}^{(S)}_{\mathcal{I}}$ being linearly independent, we get $S \leq |\mathcal{I}|$ by noting that $\mathbf{x}^{(s)}_{\mathcal{I}}\in \mathbb{R}^{|\mathcal{I}|}$.

Finally, we show that $S\neq|\mathcal{I}|$. We also use contradiction. Suppose $S=|\mathcal{I}|$. Then $\mathbf{x}^{(1)}_{\mathcal{I}},\ldots,\mathbf{x}^{(S)}_{\mathcal{I}}$ spans $\mathbb{R}^{|\mathcal{I}|}$, and there must exist a nonzero vector $\boldsymbol\beta\triangleq [\beta_1,\ldots,\beta_S ]$ such that
\begin{align}\label{eq4}
\mathbf{x}_{\mathcal{I}}^\beta \triangleq \beta_1\mathbf{x}^{(1)}_{\mathcal{I}}+\cdots+\beta_S\mathbf{x}^{(S)}_{\mathcal{I}} \geq \mathbf{0},
\end{align}
with $\mathbf{x}_{\mathcal{I}}^\beta\neq \mathbf{0}$. We then get
\begin{align}\label{eq5}
\mathbf{x}^\beta \triangleq\beta_1\mathbf{x}^{(1)}+\cdots+\beta_S\mathbf{x}^{(S)}\in\mathcal{C},
\end{align}
and $\mathbf{x}^\beta\neq \mathbf{0}$ by noticing that $\mathbf{x}_{\mathcal{I}}^\beta$ is a sub-vector of $\mathbf{x}^\beta$. We then have $\mathbf{x}^\beta \in \mathcal{N}\cap\mathcal{C}$ and $\mathcal{N}\cap\mathcal{C}\neq\{\mathbf{0}\}$, which contradicts with $\mathcal{N}\cap\mathcal{C}=\{\mathbf{0}\}$. Therefore, $S < |\mathcal{I}|$ must hold.

\subsection{Proof of Theorem~\ref{prop.condition.mat}}
\label{sec.apd.lp}
Let $\mathbf{z}\in \mathbb{R}^{N}$ be a nonzero vector in the null space of $\mathbf{J}(\boldsymbol{\gamma}^0)$, i.e., $\mathbf{J}(\boldsymbol{\gamma}^0)\mathbf{z}=\mathbf{0}$. Since $\mathbf{J}(\boldsymbol{\gamma}^0)$ is symmetric, by rearranging the columns and rows of $\mathbf{J}(\boldsymbol{\gamma}^0)$ and the entries of $\mathbf{z}$ according to the index sets $\mathcal{I}$ and $\mathcal{I}^c$, the equation $\mathbf{J}(\boldsymbol{\gamma}^0)\mathbf{z}=\mathbf{0}$ can be rewritten in a block-wise form as
\begin{align}\label{eq.apd.mateq}
\left[
\begin{array}{cc}
\mathbf{A} & \mathbf{B} \\
\mathbf{B}^T & \mathbf{C}
\end{array}
\right]
\left[
\begin{array}{c}
\mathbf{z}_{\mathcal{I}} \\
\mathbf{z}_{\mathcal{I}^{c}}
\end{array}
\right]=
\left[
\begin{array}{c}
\mathbf{0} \\
\mathbf{0}
\end{array}
\right],
\end{align}
where $\mathbf{A}$, $\mathbf{B}$, and $\mathbf{C}$ are submatrices of $\mathbf{J}(\boldsymbol{\gamma}^0)$ defined in the theorem, and $\mathbf{z}_{\mathcal{I}}\in \mathbb{R}^{N-K}$, $\mathbf{z}_{\mathcal{I}^c}\in \mathbb{R}^{K}$ are sub-vectors of $\mathbf{z}$ with indices from $\mathcal{I}$ and $\mathcal{I}^c$, respectively. We rewrite \eqref{eq.apd.mateq} as
\begin{align}
\mathbf{A}\mathbf{z}_{\mathcal{I}} + \mathbf{B}\mathbf{z}_{\mathcal{I}^c} &= \mathbf{0},\label{eq.adp.lp.3}\\
\mathbf{B}^T\mathbf{z}_{\mathcal{I}} + \mathbf{C}\mathbf{z}_{\mathcal{I}^c} &= \mathbf{0}.\label{eq.adp.lp.4}
\end{align}

We first show that $\mathbf{C}$ is invertible if $\mathcal{N}\cap\mathcal{C}=\{\mathbf{0}\}$. Suppose that $\mathbf{C}$ is singular, i.e., there exists a nonzero vector $\mathbf{v}\in\mathbb{R}^K$ such that $\mathbf{C}\mathbf{v}=\mathbf{0}$. We then construct a nonzero vector $\mathbf{z}$ with $\mathbf{z}_{\mathcal{I}^c}=\mathbf{v}$ and $\mathbf{z}_{\mathcal{I}}=\mathbf{0}$. It can be verified from \eqref{eq.apd.mateq} that $\mathbf{z}$ satisfies $\mathbf{z}^T\mathbf{J}(\boldsymbol{\gamma}^0)\mathbf{z}=0$, based on which we get $\mathbf{J}(\boldsymbol{\gamma}^0)\mathbf{z}=\mathbf{0}$ by using the positive semidefiniteness of $\mathbf{J}(\boldsymbol{\gamma}^0)$. Therefore, $\mathbf{z} \in \mathcal{N}$. Moreover, the constructed $\mathbf{z}$ is also in the cone $\mathcal{C}$ since $z_i= 0, i\in\mathcal{I}$. Therefore, $\mathbf{z}\in\mathcal{N}\cap\mathcal{C}$. The condition $\mathcal{N}\cap\mathcal{C}=\{\mathbf{0}\}$ is not satisfied, since $\mathbf{z} \neq \mathbf{0}$.

With invertible $\mathbf{C}$, we eliminate $\mathbf{z}_{\mathcal{I}^c}$ in \eqref{eq.adp.lp.3} and \eqref{eq.adp.lp.4}, and obtain the following equation
\begin{align}\label{eq.adp.lp.5}
(\mathbf{A}-\mathbf{B}\mathbf{C}^{-1}\mathbf{B}^T)\mathbf{z}_{\mathcal{I}}=\mathbf{0}.
\end{align}
Since the cone constraints are on the coordinates indexed by $\mathcal{I}$, to check whether $\mathcal{N}\cap\mathcal{C}=\{\mathbf{0}\}$ holds, we only need to examine if there exists a nonzero vector $\mathbf{z}_{\mathcal{I}}$ with nonnegative entries that satisfies \eqref{eq.adp.lp.5}. Note that we require $\mathbf{z}_{\mathcal{I}}\neq \mathbf{0}$ to make $\mathbf{z}$ nonzero due to \eqref{eq.adp.lp.4} and the invertibility of $\mathbf{C}$. Based on \eqref{eq.adp.lp.5}, the existence of a nonzero vector $\mathbf{z}_{\mathcal{I}}$ can be formulated as a feasibility problem as follows
\begin{subequations}
\begin{alignat}{2}
&\quad \operatorname{find} &\quad& \mathbf{x}\label{eq.apd.lp.1}\\
&\operatorname{subject\,to}&      & (\mathbf{A}-\mathbf{B}\mathbf{C}^{-1}\mathbf{B}^T)\mathbf{x} = \mathbf{0}\label{eq.apd.lp.2}\\
& & &\mathbf{x}\geq \mathbf{0}, \mathbf{x}\neq\mathbf{0}.
\end{alignat}\label{eq.apd.lp}%
\end{subequations}
To get \eqref{eq.lp} from \eqref{eq.apd.lp}, we use the following lemma from \cite{Ben-Israel1964}.
\begin{lemma}\label{lemma}
Let $\mathbf{M}$ be any matrix over some field. Then, the following statements are equivalent: (i) $\mathbf{Mx}=\mathbf{0}$ has no solution for $\mathbf{x}\geq \mathbf{0}$ and $\mathbf{x}\neq \mathbf{0}$; (ii) $\mathbf{M}^T\mathbf{v}>\mathbf{0}$ has solutions.
\end{lemma}

By using Lemma~\ref{lemma} and noting that $\mathbf{A}-\mathbf{B}\mathbf{C}^{-1}\mathbf{B}^T$ is symmetric, the infeasibility of \eqref{eq.apd.lp} is equivalent to
the feasibility of \eqref{eq.lp}.

\subsection{Proof of Theorem~\ref{prop.dist}}
\label{sec.apd.qp}
The derivation of the QP is based on \cite{Self1987}, which considers the case of non-singular Fisher information matrix. Here, we consider the case where the Fisher information matrix may be singular. Let $F_M(\boldsymbol\gamma)$ denote the log-likelihood function normalized by $M$, i.e.,
\begin{align}
F_M(\boldsymbol\gamma)& \triangleq \frac{1}{M}\log p(\mathbf{Y}|\boldsymbol\gamma)=\frac{1}{M}\sum_{m=1}^M \log p(\mathbf{y}_m|\boldsymbol\gamma).
\end{align}
Since $\hat{\boldsymbol\gamma}^{(M)}$ is obtained by maximizing $F_M(\boldsymbol\gamma)$, and $\hat{\boldsymbol\gamma}^{(M)}$ converges to the true parameter $\boldsymbol\gamma^0$ as $M\rightarrow\infty$, we study the function $\Delta F_M(\boldsymbol\gamma)\triangleq F_M(\boldsymbol\gamma) - F_M(\boldsymbol\gamma^0)$ for large $M$ in the neighborhood of $\boldsymbol\gamma^0$. Let $\Delta\boldsymbol\gamma \triangleq \boldsymbol\gamma -\boldsymbol\gamma^0$. We consider the quadratic approximation of $\Delta F_M(\boldsymbol\gamma)$ at $\boldsymbol\gamma^0$ as
\begin{align}\label{eq.adp.diff}
\Delta F_M(\boldsymbol\gamma) \approx
\Delta\boldsymbol\gamma^T\nabla F_M(\boldsymbol\gamma^0)
+\frac{1}{2}\Delta\boldsymbol\gamma^T\nabla^2 F_M(\boldsymbol\gamma^0)\Delta\boldsymbol\gamma, 
\end{align}
where $\nabla F_M(\boldsymbol\gamma^0)$ and $\nabla^2 F_M(\boldsymbol\gamma^0)$ represent the gradient and the Hessian of $F_M(\boldsymbol\gamma)$ at $\boldsymbol\gamma^0$, respectively.

We now aim to relate the gradient and the Hessian to the associated Fisher information matrix. For the gradient term in \eqref{eq.adp.diff}, the $i$-th entry of $\nabla F_M(\boldsymbol\gamma^0)$ can be written as
\begin{align}\label{eq.adp.sum}
[\nabla F_M(\boldsymbol\gamma^0)]_i =\frac{1}{M}\sum_{m=1}^M\frac{\partial\log p(\mathbf{y}_m|\boldsymbol\gamma)}{\partial\gamma_i}\bigg|_{\boldsymbol\gamma=\boldsymbol\gamma^0},
\end{align}
where each term in the summation can be seen as a random variable with the mean and variance, respectively, as
\begin{align}
\mathbb{E}\left[\frac{\partial\log p(\mathbf{y}_m|\boldsymbol\gamma)}{\partial\gamma_i}\bigg|_{\boldsymbol\gamma=\boldsymbol\gamma^0}\right] &= 0,\label{eq.apd.mean}\\
\mathrm{var}\left[\frac{\partial\log p(\mathbf{y}_m|\boldsymbol\gamma)}{\partial\gamma_i}\bigg|_{\boldsymbol\gamma=\boldsymbol\gamma^0}\right]&=\frac{1}{M}\left[\mathbf{J}(\boldsymbol\gamma_0)\right]_{ii}.\label{eq.apd.var}
\end{align}
The mean \eqref{eq.apd.mean} is obtained by taking the expectation of \eqref{eq.adp.first.deriv} using $\mathbb{E}[\hat{\boldsymbol\Sigma}]=\boldsymbol\Sigma$, and the variance \eqref{eq.apd.var} is obtained based on \eqref{eq.fim.def} and \eqref{eq.apd.mean}.
In particular, notice from \eqref{eq.fim.def} that
\begin{align}\label{eq.apd.fim.def}
[\mathbf{J}(\boldsymbol\gamma)]_{ii}&=\mathbb{E}\left[\left(\sum^{M}_{m=1}\frac{\partial \log p(\mathbf{y}_m|\boldsymbol{\gamma})}{\partial \gamma_i}\right)^2\right]\nonumber\\
&= M \mathbb{E}\left[\left(\frac{\partial \log p(\mathbf{y}_m|\boldsymbol{\gamma})}{\partial \gamma_i}\right)^2\right],
\end{align}
where the last step is due to \eqref{eq.apd.mean} and the fact that $\mathbf{y}_m$'s are i.i.d. Gaussian random variables conditioned on $\boldsymbol{\gamma}$. This shows that \eqref{eq.apd.var} holds. Similarly, the covariance
can be computed as
\begin{align}\label{eq.adp.cov}
\mathbb{E}\left[\left(\frac{\partial\log p(\mathbf{y}_m|\boldsymbol\gamma)}{\partial\gamma_i}\right)
\left(\frac{\partial\log p(\mathbf{y}_m|\boldsymbol\gamma)}{\partial\gamma_j}\right)\bigg|_{\boldsymbol\gamma=\boldsymbol\gamma^0}\right]
=\frac{\left[\mathbf{J}(\boldsymbol\gamma_0)\right]_{ij}}{M}.
\end{align}
Thus, $\nabla F_M(\boldsymbol\gamma^0)$ in \eqref{eq.adp.sum}
is the sample average of $M$ i.i.d.\ random vectors, whose mean, variance, and covariance are given in \eqref{eq.apd.mean}, \eqref{eq.apd.var}, and \eqref{eq.adp.cov}, respectively. By the central limit theorem, we have that
\begin{align}
\sqrt{M}\nabla F_M(\boldsymbol\gamma^0)\overset{D}{\rightarrow} \mathcal{N}(\mathbf{0},\mathbf{J}(\boldsymbol\gamma^0)/M), \quad\operatorname{as}~ M\rightarrow \infty.
\end{align}

For the Hessian term in \eqref{eq.adp.diff}, based on \eqref{eq.fim.def.div} and by the law of large numbers, we immediately have that
\begin{align}
\nabla^2 F_M(\boldsymbol\gamma^0)\overset{P}{\rightarrow} -\mathbf{J}(\boldsymbol\gamma^0)/M, \quad\operatorname{as}~ M\rightarrow \infty.
\end{align}

Therefore, the \edb{right-hand} side of \eqref{eq.adp.diff} converges in distribution to the following random variable
\begin{align}\label{eq.adp.qp.0}
\frac{1}{\sqrt{M}}\Delta\boldsymbol\gamma^T\mathbf{z}
- \frac{1}{2M}\Delta\boldsymbol\gamma^T\mathbf{J}(\boldsymbol\gamma^0)\Delta\boldsymbol\gamma,
\end{align}
where $\mathbf{z}\in\mathbb{R}^{N}$ is a random vector following $\mathcal{N}(\mathbf{0},\mathbf{J}(\boldsymbol\gamma^0)/M)$. We further replace $\mathbf{z}$ in \eqref{eq.adp.qp.0} by $\mathbf{J}(\boldsymbol\gamma^0)\mathbf{x}/M$, in which $\mathbf{x}\in\mathbb{R}^{N}$ is a random vector following
$\mathcal{N}(\mathbf{0},M\mathbf{J}^{\dagger}(\boldsymbol\gamma^0))$, and \eqref{eq.adp.qp.0} can be rewritten as
\begin{multline}\label{eq.adp.qp}
\frac{1}{M\sqrt{M}}\Delta\boldsymbol\gamma^T\mathbf{J}(\boldsymbol\gamma^0)\mathbf{x}
- \frac{1}{2M}\Delta\boldsymbol\gamma^T\mathbf{J}(\boldsymbol\gamma^0)\Delta\boldsymbol\gamma \\
= -\frac{1}{2M}\left(\Delta \boldsymbol\gamma - \frac{\mathbf{x}}{\sqrt{M}}\right)^T\mathbf{J}(\boldsymbol\gamma^0)
\left(\Delta \boldsymbol\gamma - \frac{\mathbf{x}}{\sqrt{M}}\right) \\
+\frac{1}{2M^2}\mathbf{x}^T\mathbf{J}(\boldsymbol\gamma^0)
\mathbf{x},
\end{multline}
where the last step is obtained by completing a square.

Finally, the maximization of $F_M(\boldsymbol\gamma)$ is equivalent to the maximization of $\Delta F(\boldsymbol\gamma)$ since $F_M(\boldsymbol\gamma^0)$ does not depend on $\boldsymbol\gamma$. Based on \eqref{eq.adp.qp}, the optimization problem can be cast as
\begin{subequations}
\begin{alignat}{2}
&\underset{\Delta \boldsymbol\gamma}{\operatorname{minimize}}    &\,& \left(\frac{\mathbf{x}}{\sqrt{M}}-\Delta\boldsymbol\gamma \right)^T\frac{\mathbf{J}(\boldsymbol\gamma^0)}{M}
\left(\frac{\mathbf{x}}{\sqrt{M}} - \Delta \boldsymbol\gamma\right)\\
&\operatorname{subject\,to} &      & \,\,\,\,\Delta \boldsymbol\gamma \in \mathcal{C},
\end{alignat}\label{eq.apd.opt}%
\end{subequations}
where $\Delta \boldsymbol\gamma \in \mathcal{C}$ comes from the fact that $\boldsymbol\gamma$ should be nonnegative.
Replacing $\Delta \boldsymbol\gamma$ by $\boldsymbol\mu/\sqrt{M}$ gives the QP
in \eqref{eq.prob2}, meaning that the MLE error has a limiting distribution, which
is the same as the distribution of a solution to the QP with $\mathbf{x} \sim \mathcal{N}(\mathbf{0},M\mathbf{J}^{\dagger}(\boldsymbol\gamma^0))$.

\subsection{Proof of Theorem~\ref{prop.alt.condition}}
\label{sec.apd.alt.condition}
In the limit $M\rightarrow\infty$, \eqref{eq.mat.match} can be written as \eqref{eq.mat.match.cond} via vectorization and noting that the sample covariance matrix $\hat{\boldsymbol\Sigma}$ converges to the true covariance matrix $\mathbf{S}\boldsymbol\Gamma^0\mathbf{S}^{H}+\sigma_w^2\mathbf{I}$. We prove the necessity of $\widetilde{\mathcal{N}}\cap\mathcal{C}=\{\mathbf{0}\}$ by contradiction. We assume that there exists
a nonzero vector $\mathbf{x} \in \widetilde{\mathcal{N}}\cap\mathcal{C}$. Then we can construct a nonnegative vector
$\boldsymbol\gamma^1\triangleq \boldsymbol\gamma^0 + t\mathbf{x}$ with $t=\min_{n\in\mathcal{I}^c}(\gamma^0_{n}/|x_{n}|)$. Since $\mathbf{x}\in\mathcal{C}$ and $\mathbf{x}$ is nonzero, it can be verified that $\boldsymbol\gamma^1\geq \mathbf{0}$ and $\boldsymbol\gamma^1\neq\boldsymbol\gamma^0$. Moreover, since $\mathbf{x}\in\mathcal{N}$, we have that $\boldsymbol\gamma^1$ is also a solution to \eqref{eq.mat.match.cond}, implying that the condition $\widetilde{\mathcal{N}}\cap\mathcal{C}=\{\mathbf{0}\}$ must be necessary.

To show the sufficiency, we also use contradiction. Suppose that there exists a nonnegative vector $\boldsymbol{\gamma}^1\neq \boldsymbol{\gamma}^0$ such that \eqref{eq.mat.match.cond} holds at $\boldsymbol{\gamma}^1$, i.e., $\widehat{\mathbf{S}}(\boldsymbol\gamma^1 -\boldsymbol\gamma^0)=\mathbf{0}$. Let $\mathbf{x}\triangleq\boldsymbol\gamma^1 -\boldsymbol\gamma^0$. We immediately have $\mathbf{x}\in\widetilde{\mathcal{N}}$. In the meanwhile, since $\gamma^1_n, n\in{\mathcal{I}},$ are nonnegative whereas $\gamma^0_n=0, n\in\mathcal{I}$, we have $x_n=\gamma^1_n-\gamma^0_n \geq 0, n\in\mathcal{I}$, indicating that $\mathbf{x}\in \mathcal{C}$. Therefore, there exist a nonzero vector $\mathbf{x}\in\widetilde{\mathcal{N}}\cap\mathcal{C}$  which contradicts with $\widetilde{\mathcal{N}}\cap\mathcal{C}=\{\mathbf{0}\}$, implying that $\widetilde{\mathcal{N}}\cap\mathcal{C}=\{\mathbf{0}\}$ is sufficient.

\subsection{Proof of Theorem~\ref{prop.condition.eq}}
\label{sec.apd.condition.eq}
First, $\mathcal N$ is as characterized in Lemma~\ref{lemma.null.space} in
Appendix \ref{sec.apd.condition}. Let $\mathbf{U}_0$, $\mathbf{V}_0$, and
$\mathbf{v}_i^0$ denote the values of $\mathbf{U}$, $\mathbf{V}$, and
$\mathbf{v}_i$ at $\boldsymbol{\gamma}^0$, respectively, as defined in
Lemma~\ref{lemma.null.space}. Note that $\mathbf{V}_0=\mathbf{S}^H\mathbf{U}_0$.
Then, the null set $\mathcal N$ is the set of $\mathbf{x}\in\mathbb{R}^{N}$
that satisfies
$\mathbf{x}^T\mathbf{J}(\boldsymbol\gamma^0)\mathbf{x}=0$, which is given by
\begin{align}\label{eq.adp.null.1}
\mathcal{N}=\{\mathbf{x}\mid \mathbf{x}^T(\mathbf{v}_i^0\odot(\mathbf{v}_j^0)^*)=0, \mathbf{x}\in \mathbb{R}^{N}, 1\leq i,j\leq L\}.
\end{align}

Next, we express $\widetilde{\mathcal{N}}$ in a form similar to \eqref{eq.adp.null.1}. We write
$\widehat{\mathbf{S}}=[\mathbf{s}^{*}_{1}\otimes\mathbf{s}_{1}, \mathbf{s}^{*}_{2}\otimes\mathbf{s}_{2},\ldots, \mathbf{s}^{*}_{N}\otimes\mathbf{s}_{N}]\in \mathbb{C}^{L^2\times N}$ explicitly as
\begin{align}\label{eq.apd.explicit.s}
\widehat{\mathbf{S}} &=
\left[
\begin{array}{cccc}
s_{11}^*\mathbf{s}_1 & s_{12}^*\mathbf{s}_2&\ldots& s_{1N}^*\mathbf{s}_N\\
s_{21}^*\mathbf{s}_1 & s_{22}^*\mathbf{s}_2&\ldots& s_{2N}^*\mathbf{s}_N\\
\vdots & \vdots & \ddots & \vdots\\
s_{L1}^*\mathbf{s}_1 & s_{L2}^*\mathbf{s}_2&\ldots& s_{LN}^*\mathbf{s}_N
\end{array}
\right],
\end{align}
from which we observe that the $L^2$ rows of $\widehat{\mathbf{S}}$ can be expressed in the form of
$\mathbf{r}_i^T\odot\mathbf{r}_j^H$ for $1\leq i,j \leq L$, where $\mathbf{r}_i^T\triangleq[s_{i1}, s_{i2},\ldots,s_{iN}]$ is the $i$-th row of $\mathbf{S}$. Therefore, the null space $\widetilde{\mathcal{N}}$ can be expressed using $\mathbf{r}_i^T\odot\mathbf{r}_j^H$ as follows
\begin{align}\label{eq.adp.null.2}
\widetilde{\mathcal{N}}=\{\mathbf{x}\mid \mathbf{x}^T(\mathbf{r}_i\odot\mathbf{r}_j^*)=0, \mathbf{x}\in \mathbb{R}^{N}, 1\leq i,j\leq L\}.
\end{align}

We now relate $\widetilde{\mathcal{N}}$ in \eqref{eq.adp.null.1} and $\mathcal{N}$ in \eqref{eq.adp.null.2} by noticing that $\mathbf{v}_i^0$ in \eqref{eq.adp.null.1} and $\mathbf{r}_i$ in \eqref{eq.adp.null.2} are connected via $\mathbf{V}_0=\mathbf{S}^H\mathbf{U}_0$. Let $u_{ij}^0$ denote the $(i,j)$-th entry of $\mathbf{U}_0$, based on which $\mathbf{v}_i^0$ can be written as $\mathbf{v}_i^0=\sum_{l=1}^L u_{li}^0\mathbf{r}_l^*$. We then have
\begin{align}\label{eq.adp.null.linear1}
\mathbf{v}_i^0\odot(\mathbf{v}_j^0)^*&=\left(\sum_{l=1}^L u_{li}^0\mathbf{r}_l^*\right)\odot\left(\sum_{k=1}^L (u_{kj}^0)^*\mathbf{r}_k\right)\nonumber\\
&=\sum_{l=1}^L\sum_{k=1}^Lu_{li}^0(u_{kj}^0)^*(\mathbf{r}_l^*\odot\mathbf{r}_k).
\end{align}
Similarly, we have $\mathbf{S}^H=\mathbf{V}_0\mathbf{U}_0^H$ since $\mathbf{U}_0$ is unitary, and $\mathbf{r}_i$ can be written as $\mathbf{r}_i=\sum_{l=1}^Lu_{il}^0\mathbf{v}_l^*$, based on which we get
\begin{align}\label{eq.adp.null.linear2}
\mathbf{r}_i\odot\mathbf{r}_j^*=\sum_{l=1}^L\sum_{k=1}^Lu_{il}^0(u_{jk}^0)^*\left((\mathbf{v}_l^0)^*\odot\mathbf{v}_k^0\right).
\end{align}
We observe from \eqref{eq.adp.null.linear1} and \eqref{eq.adp.null.linear2} that any vector $\mathbf{x}$ that satisfies $\mathbf{x}^T\left(\mathbf{v}_i^0\odot(\mathbf{v}_j^0)^*\right)=0$ for all $1\leq i,j\leq L$ should also satisfy $\mathbf{x}^T(\mathbf{r}_i\odot\mathbf{r}_j^*)=0, 1\leq i,j\leq L$, and vice versa. Therefore, the two sets
$\mathcal{N}$ and $\widetilde{\mathcal{N}}$ are identical, implying that $\widetilde{\mathcal{N}}\cap\mathcal{C}=\{\mathbf{0}\}$ and $\mathcal{N}\cap\mathcal{C}=\{\mathbf{0}\}$ are equivalent.

\subsection{Proof of Theorem~\ref{prop.condition.mat.alt}}
\label{sec.apd.condition.mat.alt}
First, note that $\mathbf{x}\in\widetilde{\mathcal{N}}$ can be equivalently expressed as $\mathbf{D}\mathbf{x}=\mathbf{0}$ since $\mathbf{D}$ is formed by the real and imaginary parts of rows of $\widehat{\mathbf{S}}$.

We now prove that $\widetilde{\mathcal{N}}\cap\mathcal{C}=\{\mathbf{0}\}$ implies these two conditions: (i) $\mathbf{D}_{\mathcal{I}^c}\in\mathbb{R}^{L^2\times K}$ is rank $K$; and (ii) the problem \eqref{eq.apd.lp.alt} is infeasible. We use contradiction. Suppose that $\mathbf{D}_{\mathcal{I}^c}$ is not rank $K$, then $\mathbf{D}_{\mathcal{I}^c}\mathbf{x}_{\mathcal{I}^c}=\mathbf{0}$ must admit nonzero solutions, and we can construct a vector $\mathbf{x}^{(*)}\in\mathbb{R}^{N}$ with its sub-vectors $\mathbf{x}^{(*)}_{\mathcal{I}}=\mathbf{0}$ and $\mathbf{x}^{(*)}_{\mathcal{I}^c}$ being a nonzero solution to $\mathbf{D}_{\mathcal{I}^c}\mathbf{x}_{\mathcal{I}^c}=\mathbf{0}$. For such an $\mathbf{x}^{(*)}$, we have $\mathbf{D}\mathbf{x}^{(*)}=\mathbf{0}$, and thus $\mathbf{x}^{(*)}\in\widetilde{\mathcal{N}}$. In the meanwhile, $\mathbf{x}^{(*)}\in\mathcal{C}$ since $\mathbf{x}^{(*)}_{\mathcal{I}}=\mathbf{0}$. Therefore, we have $\mathbf{x}^{(*)}\in\widetilde{\mathcal{N}}\cap\mathcal{C}$. Now suppose the problem \eqref{eq.apd.lp.alt} is feasible, with a slight abuse of notation, let $\mathbf{x}^{(*)}$ be a solution to \eqref{eq.apd.lp.alt}, we immediately have that $\mathbf{x}^{(*)}$ is nonzero and $\mathbf{x}^{(*)}\in\widetilde{\mathcal{N}}\cap\mathcal{C}$. Therefore, in both cases we can find a nonzero vector $\mathbf{x}^{(*)}$ such that $\mathbf{x}^{(*)}\in\widetilde{\mathcal{N}}\cap\mathcal{C}$, which contradicts with $\widetilde{\mathcal{N}}\cap\mathcal{C}=\{\mathbf{0}\}$.

We then prove that these two conditions imply $\widetilde{\mathcal{N}}\cap\mathcal{C}=\{\mathbf{0}\}$. We still use contradiction. Suppose that there exists a nonzero vector $\mathbf{x}^{(*)}$ such that $\mathbf{x}^{(*)}\in\widetilde{\mathcal{N}}\cap\mathcal{C}$. The vector $\mathbf{x}^{(*)}$ can be categorized into one of the following two cases: (i) $\mathbf{x}^{(*)}_{\mathcal{I}}=\mathbf{0}$; or (ii) $\mathbf{x}^{(*)}_{\mathcal{I}}\neq\mathbf{0}$. In the case where $\mathbf{x}^{(*)}_{\mathcal{I}}=\mathbf{0}$, $\mathbf{x}^{(*)}_{\mathcal{I}^c}$ must be nonzero, and $\mathbf{x}^{(*)}_{\mathcal{I}^c}$ must be a solution to $\mathbf{D}_{\mathcal{I}^c}\mathbf{x}_{\mathcal{I}^c}=\mathbf{0}$ since $\mathbf{x}^{(*)}\in\widetilde{\mathcal{N}}$. Then the rank of $\mathbf{D}_{\mathcal{I}^c}$ cannot be $K$, which contradicts with the first condition. In the case where $\mathbf{x}^{(*)}_{\mathcal{I}}\neq\mathbf{0}$, $\mathbf{x}^{(*)}$ must satisfy constraints \eqref{eq.apd.lp.alt.2} and \eqref{eq.apd.lp.alt.4} since $\mathbf{x}^{(*)}\in\widetilde{\mathcal{N}}\cap\mathcal{C}$. Then we can find a scalar $t$ such that $t\mathbf{x}$ satisfies all \eqref{eq.apd.lp.alt.2}, \eqref{eq.apd.lp.alt.3}, and \eqref{eq.apd.lp.alt.4}, which contradicts with the infeasibility of the problem \eqref{eq.apd.lp.alt}. Therefore, in both cases at least one of these two conditions does not hold.

\subsection{Proof of Theorem~\ref{prop.connection}}
\label{sec.apd.connection}

The proof is based on the robust $\ell_2$ NSP of $\widehat{\mathbf{S}}$
established in Theorem~\ref{scalinglaw}. Suppose that $\widehat{\mathbf{S}}$ has
the robust $\ell_2$ NSP of order $K$ with parameters $0<\rho<1$ and $\tau$ under
the conditions specified in Theorem~\ref{scalinglaw}.  As indicated in \cite[Sec.
4.3]{Foucart2013}, $\widehat{\mathbf{S}}$ also satisfies the robust $\ell_1$
NSP of order $K$, expressed as
\begin{align}\label{eq.adp.nsp}
\|\mathbf{x}_{\mathcal{K}}\|_1\leq \rho\|\mathbf{x}_{\mathcal{K}^c}\|_1 + \sqrt{K}\tau\|\widehat{\mathbf{S}}\mathbf{x}\|_2,
\end{align}
by using $\|\mathbf{x}_{\mathcal{K}}\|_1\leq \sqrt{K}\|\mathbf{x}_{\mathcal{K}}\|_2$ on \eqref{eq.nsp}.
Consider an $\mathbf{x}$ in the null space of $\widehat{\mathbf{S}}$, i.e.,
$\mathbf{x}\in\widetilde{\mathcal{N}}$. Based on \eqref{eq.adp.nsp}, we get
\begin{align}\label{eq.adp.nsp.2}
\|\mathbf{x}_{\mathcal{K}}\|_1\leq \rho\|\mathbf{x}_{\mathcal{K}^{c}}\|_1.
\end{align}
This condition must be satisfied for any $\mathbf{x}\in\widetilde{\mathcal{N}}$
and any index set $\mathcal{K}$ with $|\mathcal{K}|\leq K$.

Now, suppose that $\mathbf{x}\in\widetilde{\mathcal{N}}\cap\mathcal{C}$.
First, we observe from \eqref{eq.apd.explicit.s} that the $L^2$
rows of $\widehat{\mathbf{S}}$ can be expressed in the form of
$\mathbf{r}_i^T\odot\mathbf{r}_j^H$ for $1\leq i,j \leq L$, and therefore
$\sum_{i=1}^L\mathbf{r}_i^T\odot\mathbf{r}_i^H$ is in the row space of
$\widehat{\mathbf{S}}$. So, any $\mathbf{x}\in\widetilde{\mathcal{N}}$ should satisfy
\begin{align}\label{eq.adp.zerosum}
0&=\left(\sum_{i=1}^L\mathbf{r}_i^T\odot\mathbf{r}_i^H\right)\mathbf{x}=\sum_{i=1}^L\sum_{j=1}^N x_js_{ij}s_{ij}^{*}=L\sum_{j=1}^N x_j,
\end{align}
where the last step is obtained by swapping the summation and
noticing that the columns of $\widehat{\mathbf{S}}$ share identical $\ell_2$
norm since all columns are drawn from a sphere in $\mathbb{C}^L$. By breaking
the summation in the \edb{right-hand} side of \eqref{eq.adp.zerosum} into two parts
according to $\mathcal{I}$ and $\mathcal{I}^c$, we have
\begin{equation}
\sum_{i\in\mathcal{I}}x_i=-\sum_{i\in\mathcal{I}^c}x_i\leq \|\mathbf{x}_{\mathcal{I}^c}\|_1.
\end{equation}
Since $\mathbf{x}\in \mathcal{C}$ also holds, this means that $x_i$'s, $i\in\mathcal{I}$, are
nonnegative. So, $\sum_{i\in\mathcal{I}}x_i=\|\mathbf{x}_{\mathcal{I}}\|_1$, and therefore
\begin{align}\label{eq.adp.nsp.ineq.1}
\|\mathbf{x}_{\mathcal{I}}\|_1\leq \|\mathbf{x}_{\mathcal{I}^c}\|_1.
\end{align}
This above condition should be satisfied for any $\mathbf{x}\in\widetilde{\mathcal{N}}\cap\mathcal{C}$.

Finally, we show that the only $\mathbf{x} \in \widetilde{\mathcal{N}}\cap\mathcal{C}$
that can satisfy \eqref{eq.adp.nsp.2} is the zero vector. This is because we
can choose $\mathcal{K}$ in \eqref{eq.adp.nsp.2} to be $\mathcal{I}^c$ since
$|\mathcal{I}^c|=K$. In this case, \eqref{eq.adp.nsp.2} implies
\begin{equation}\label{eq.adp.nsp.ineq.2}
\|\mathbf{x}_{\mathcal{I}^c}\|_1 \leq \rho\|\mathbf{x}_{\mathcal{I}}\|_1,
\end{equation}
while $\mathbf{x} \in \widetilde{\mathcal{N}}\cap\mathcal{C}$ implies \eqref{eq.adp.nsp.ineq.1}.
Noting that $\rho<1$, the only $\mathbf{x}$ that can satisfy both \eqref{eq.adp.nsp.ineq.1} and \eqref{eq.adp.nsp.ineq.2} is $\mathbf{x}=\mathbf{0}$.
This shows that if $\widehat{\mathbf{S}}$ has the robust $\ell_2$ NSP, then
$\widetilde{\mathcal{N}}\cap\mathcal{C} = \{ \mathbf{0} \}$.

\bibliography{chenbib}
\end{document}